\documentclass[reprint,twocolumn,superscriptaddress,preprintnumbers,amsmath,amssymb,aps,prd,floatfix]{revtex4-2}
\usepackage[utf8]{inputenc}

\usepackage{mathtools}
\usepackage{amsfonts}
\usepackage{mathrsfs}
\usepackage{bm}
\usepackage[normalem]{ulem}
\usepackage{bbold}
\usepackage{braket}
\usepackage{slashed}
\usepackage{dsfont}
\usepackage{float}
\usepackage{dcolumn}
\usepackage{amssymb,amsmath}
\usepackage{soul}  

\usepackage{graphicx}
\usepackage{array}

\usepackage{placeins}
\usepackage{booktabs}

\usepackage{xspace}
\usepackage{hyperref}
\usepackage[nameinlink]{cleveref}
\usepackage{bookmark}
\usepackage{units}

\setkeys{Gin}{width=0.48\textwidth}

\usepackage{multirow}

\newcolumntype{C}{>{$}c<{$}} \AtBeginDocument{ \heavyrulewidth=.08em \lightrulewidth=.05em \cmidrulewidth=.03em \belowrulesep=.65ex \belowbottomsep=0pt \aboverulesep=.4ex \abovetopsep=0pt \cmidrulesep=\doublerulesep \cmidrulekern=.5em \defaultaddspace=.5em }

\graphicspath{{./figures/}{./figures/3Dphi4/}{./figures/celeba/}}

\usepackage{xifthen}
\usepackage{xcolor}

\def\eq#1{\eqref{#1}}
\def\Eq#1{Eq.~\labelcref{#1}} 

\def\fig#1{Fig.~\labelcref{#1}} \def\Fig#1{Fig.~\labelcref{#1}}   \def\Tab#1{Tab.~\labelcref{#1}}  \def\Sec#1{Sec.~\labelcref{#1}}
\def\App#1{App.~\labelcref{#1}}

\newcommand{\gettitle}{Diffusion Models for Sampling Near Criticality in Lattice Field Theories}

\hypersetup{ colorlinks, linkcolor={blue!75!black}, citecolor={blue!75!black}, urlcolor={blue!75!black},
pdftitle={\gettitle}, pdfauthor={}, pdfkeywords={}, bookmarksopen=true, bookmarksopenlevel=2, bookmarksnumbered=true }

\begin{document}
\preprint{RIKEN-iTHEMS-Report-26}
\title{\gettitle}
\author{Yang-yang Tan}
\email{yangyang-tan@foxmail.com}
\affiliation{Institute for Physics of Intelligence, Graduate School of Science, The University of Tokyo, Bunkyo-ku, Tokyo 113-0033, Japan}
\affiliation{RIKEN Center for Interdisciplinary Theoretical and Mathematical Sciences (iTHEMS), Wako, Saitama 351-0198, Japan}

\author{Gert Aarts}
\email{g.aarts@swansea.ac.uk}
\affiliation{Centre for Quantum Fields and Gravity, Department of Physics, Swansea University, Swansea SA2 8PP, United Kingdom}

\author{Diaa E.~Habibi}
\email{n.e.habibi@swansea.ac.uk}
\affiliation{Centre for Quantum Fields and Gravity, Department of Physics, Swansea University, Swansea SA2 8PP, United Kingdom}

\author{Biagio Lucini}
\email{b.lucini@qmul.ac.uk}
\affiliation{School of Mathematical Sciences, Queen Mary University of London, Mile End Road, London, E1 4NS, UK}

\author{Lingxiao Wang}
\email{lingxiao.wang@riken.jp}
\affiliation{RIKEN Center for Interdisciplinary Theoretical and Mathematical Sciences (iTHEMS), Wako, Saitama 351-0198, Japan}
\affiliation{Institute for Physics of Intelligence, Graduate School of Science, The University of Tokyo, Bunkyo-ku, Tokyo 113-0033, Japan}

\date{\today}

\begin{abstract}
We investigate generative diffusion models as denoising samplers for two- and three-dimensional lattice $\phi^4$ theory across the symmetric, near-critical, and broken phases. Validated against ensembles generated by Fourier-accelerated HMC combined with Wolff cluster updates, the reverse-SDE sampler reproduces scalar observables and the momentum-space propagator $G(|k|)$, with residual bias concentrated in the zero-mode and, in three dimensions, the action density. We introduce two local diagnostics and an HMC-referenced effective sample size (ESS), which probe the learned drift directly, through a Metropolis-adjusted Langevin acceptance rate, and through observable-level bias and variance. Exploiting a fully convolutional architecture with weights shared across different volumes ($V=L^d$), we show that cross-volume training transfers to unseen sizes, matching or slightly improving in-distribution training in the two-dimensional symmetric and broken phases. A three-dimensional model trained on $L \in \{4, 8, 16, 32\}$ reproduces the propagator and most scalar observables at the unseen lattice size $L = 64$ across the phase diagram, with the residual susceptibility excess in the broken phase as the main exception, and improves several critical observables relative to in-distribution $L = 64$ training. This establishes cross-volume generalization as a viable mechanism for large-volume sampling, and the score learned from many cheap small-lattice configurations transfers to the target volume without retraining.
\end{abstract}

\maketitle

\section{Introduction}

Lattice field theory provides the standard non-perturbative formulation of Euclidean quantum field theory, reducing expectation values to high-dimensional integrals over field configurations~\cite{Montvay:1994cy,Smit:2002ug}. The central numerical task is therefore to sample configurations from the Boltzmann distribution
$p(\phi)\sim e^{-S[\phi]}$.
Local Metropolis updates and hybrid Monte Carlo (HMC)~\cite{Duane:1987de} construct such samples through a Markov chain, but the integrated autocorrelation time typically scales as $\tau_{\rm int}\sim \xi^z$ with the correlation length $\xi$ and critical exponent $z>1$. Close to a continuous phase transition, where $\xi$ grows with the system size, this critical slowing down becomes the dominant cost. Cluster algorithms can remove much of this bottleneck in special $\mathbb{Z}_2$ systems~\cite{Swendsen:1987ce,Wolff:1988uh}, and Fourier acceleration can improve the decorrelation of approximately Gaussian modes~\cite{Davies:1987vs}. However, neither strategy gives a general-purpose solution for interacting lattice field theories across phases, couplings and lattice sizes.

The real scalar $\phi^4$ theory on a hypercubic lattice is an ideal controlled setting in which to study this problem. It is the simplest Euclidean field theory with a second-order phase transition~\cite{Luscher:1988uq,Montvay:1994cy}. In two and three dimensions the transition lies in the Ising universality class~\cite{Pelissetto:2000ek}, and accurate determinations of the critical hopping parameter $\kappa_c(\lambda)$ are available from conventional simulations~\cite{Schaich:2009jk,Wozar:2011gu}. At the same time, the model contains the main algorithmic features relevant for more complicated lattice theories, including a symmetric phase, a broken phase with two $\mathbb{Z}_2$-related vacua, long-wavelength critical modes, and finite-size effects governed by universal scaling. We use this theory as a benchmark in which the target distribution is non-trivial but sufficiently well understood to permit stringent validation.

The difficulty of sampling close to criticality has motivated a broader program of learned samplers for lattice and statistical field theories~\cite{Pawlowski:2018qxs}. Normalizing flows, optimal-transport constructions and renormalization-group-inspired maps provide one route, in which the sampler is built as an explicit transformation between simple and interacting measures~\cite{Albergo:2019eim,DelDebbio:2021qwf,Singha:2022icw,Albergo:2022qfi,Caselle:2022acb,Finkenrath:2022ogg,Cotler:2022fze,Cranmer:2023xbe,Ihssen:2024ihp,Mate:2024huc,Bauer:2024byr,Singha:2025lsd,Masuki:2025pfx}. Physics-informed kernels and related flow constructions further aim to encode locality, symmetries and field-theoretic structure directly into the generative dynamics, including applications to generic lattice sampling and sign-problem settings~\cite{Ihssen:2025ybn,Ihssen:2026njd,Aarts:2026uiu}, and the solution of functional renormalization group equations~\cite{Tan:2026bqn}. These developments show that machine-learned samplers should not be regarded merely as black-box density estimators, since their performance is tied to physical structure such as locality, scale separation and symmetry.

Generative diffusion models~\cite{SongErmon2019,Ho2020DDPM,Song2021SDE} provide a complementary approach. A neural network is trained to approximate the score $\nabla_\phi\log p_t(\phi)$ along a forward noising process, and the learned score is then used in the reverse-time stochastic dynamics~\cite{Anderson1982} to transport Gaussian noise back to the target distribution. In a lattice-field-theory context this construction has a direct connection with stochastic quantization~\cite{Parisi:1980ys,Damgaard:1987rr}, where the score is the drift field of a Langevin-type process on configuration space. This viewpoint has recently been developed for scalar theories~\cite{Wang:2023exq,Wang:2023sry,Aarts:2024rsl,Hirono:2024zyg,Fukushima:2024oij,Chen:2026bqw},
extended to Abelian and non-Abelian gauge-field settings with physics conditioning and equivariant architectures~\cite{Zhu:2024kiu,Zhu:2025pmw,Aarts:2026zzr,Alharazin:2026lcb,Komijani:2026lan}, and connected to complex Langevin dynamics~\cite{Aarts:2025lpi} and inverse modeling in quantum chromodynamics~\cite{Aarts:2025gyp}. After training, independent noise realizations generate independent reverse trajectories, suggesting a possible way to bypass Markov-chain autocorrelations.

This promise must nevertheless be manifested quantitatively. A learned sampler can fail through finite training statistics, discretization of the reverse process, imperfect score approximation, insufficient treatment of long-wavelength modes, or poor transfer across lattice sizes. Moreover, recent work indicates that generative models themselves can exhibit forms of criticality and critical slowing down in training and generation unless the architecture and score representation exploit physical locality and scale structure~\cite{2605.12597,Ruan:2026mdn}. The relevant question is therefore not simply whether a diffusion model can reproduce a few low-dimensional observables at one lattice size, but whether it can generate statistically reliable field ensembles across phases, diagnose its own failures without always requiring a reference ensemble, and generalize to lattice sizes outside the training set.

In this paper we address these questions for variance-exploding score-based diffusion models trained on two- and three-dimensional lattice $\phi^4$ theory. First, we validate generated ensembles at the trained couplings against 
reference ensembles generated using Fourier-accelerated HMC combined with Wolff cluster updates, which we refer to as FA-HMC--Wolff ensembles. The comparison includes scalar observables, the joint distribution of magnetization and action density, and the momentum-space propagator $G(|k|)$, and covers the symmetric, near-critical and broken regimes. Second, we introduce two local diagnostics, a score-network self-consistency test on clean configurations and the acceptance rate of a Metropolis-adjusted Langevin update constructed from the learned score, together with an HMC-referenced effective sample size that combines bias and variance into a single performance measure. Third, we test cross-$L$ generalization by training on multiple small lattice sizes and sampling at a larger unseen lattice size. In three dimensions, a model trained on $L\in\{4,8,16,32\}$ reproduces the $L=64$ propagator from ultraviolet to infrared momenta, and at criticality performs better than the in-distribution $L=64$ model. Single-size extrapolation baselines, each trained on one lattice size and extrapolated to $L=64$, show that this advantage is not set by the largest training size alone. A plausible interpretation is that the smaller lattices expose the score network to finite-volume infrared fluctuations, including the zero-mode distribution, at several volumes, and that this multi-volume information constrains the long-wavelength part of the score more effectively than training at any single lattice size.

The paper is organized as follows. \Sec{sec:phi4_setup} fixes notation for lattice $\phi^4$ theory. \Sec{sec:dm_framework} reviews the score-based diffusion model. \Sec{sec:indist} validates the diffusion model generated ensembles at the trained conditions through observables and propagators. \Sec{sec:diagnostics} introduces two local diagnostics and an HMC-referenced effective sample size. \Sec{sec:crossL} presents the cross-$L$ generalization results. \Sec{sec:summary} summarizes the findings and outlines future directions. Appendices collect the detailed numerical tables (\App{app:numerical_tables}), the FA-HMC--Wolff reference ensemble construction (\App{app:fahmc_wolff}), the determination of $\kappa_c$ (\App{app:kappac}), the calibration of the MALA step size (\App{app:mala_calibration}), the score-network architecture (\App{app:ncsnpp}), the HMC-referenced effective sample size computation (\App{app:ess_details}), and the non-conservative part of the learned score (\App{app:nonconservative}).

\section{Lattice $\phi^4$ theory}
\label{sec:phi4_setup}

Real scalar $\phi^4$ theory  serves as a classic testing ground for non-perturbative methods. In Euclidean signature, on a $D$-dimensional hypercubic lattice of spacing $a$, after discretizing the kinetic term and rescaling the field and bare parameters to dimensionless combinations, the action reads,
\begin{equation}
S[\phi] = \sum_{x} \Bigl[
  -2\kappa \sum_{\mu=1}^{D} \phi_x\,\phi_{x+\hat\mu}
  + \phi_x^2 + \lambda\bigl(\phi_x^2 - 1\bigr)^2
\Bigr],
\label{eq:phi4_action}
\end{equation}
on an $L^D$ lattice with periodic boundary conditions. The Boltzmann weight $e^{-S[\phi]}$ defines the lattice path integral as a classical statistical-mechanics system in $D$ Euclidean dimensions, with $\kappa$ the hopping parameter and $\lambda$ the dimensionless self-coupling.
Observables are expectation values under this weight,
\begin{equation}
\langle\mathcal{O}\rangle = \frac{1}{Z}\int\!\mathcal{D}\phi\,\mathcal{O}[\phi]\,e^{-S[\phi]},
\qquad Z=\int\!\mathcal{D}\phi\,e^{-S[\phi]}.
\label{eq:path_integral}
\end{equation}
Lattice Monte Carlo methods estimate $\langle\mathcal{O}\rangle$ as a configurational average over a finite set of samples produced by a Markov chain. Producing statistically independent samples efficiently is the central computational task.

The action~\Eq{eq:phi4_action} is invariant under the global $\mathbb{Z}_2$ transformation $\phi\to -\phi$. At sufficiently large $\kappa$ this symmetry is spontaneously broken and the field develops a non-zero expectation value. As $\kappa$ decreases through a critical value $\kappa_c(\lambda)$, the system undergoes a second-order phase transition into the symmetric phase.
The transition belongs to the Ising universality class in the corresponding dimension. The renormalization-group statement of universality is that the long-distance behavior of $\phi^4$ near $\kappa_c$ flows to the same fixed point as the Ising spin model, so the critical exponents, scaling functions and amplitude ratios are independent of the microscopic details of the action.
In this work we fix $\lambda=0.022$ for $D=2$ and $\lambda=0.9$ for $D=3$, and the determination of $\kappa_c$ using standard methodology can be found in~\App{app:kappac}.

Approaching $\kappa_c$ from either phase, the correlation length $\xi$ diverges as,
\begin{equation}
\xi \;\sim\; |\kappa-\kappa_c|^{-\nu},
\label{eq:xi_scaling}
\end{equation}
with $\nu$ a universal critical exponent ($\nu=1$ for $D=2$ and $\nu\approx 0.63$ for $D=3$)~\cite{Kleinert:2001ax}. This divergence has two consequences for any lattice simulation near $\kappa_c$. First, on a finite lattice $\xi$ saturates at the system size $L$, smoothing the phase transition into a finite-volume crossover. The residual transition signatures, such as the rounding of order-parameter curves and the crossing of dimensionless cumulants, scale as definite powers of $L$, captured by the standard machinery of finite-size scaling and used to locate $\kappa_c$ in \App{app:kappac}. Second, the autocorrelation time of any local Markov-chain update grows as $\tau\sim\xi^z$ with dynamical exponent $z\approx 2$, an effect known as \emph{critical slowing down}. Local Metropolis or HMC simulations therefore lose statistical efficiency rapidly as the lattice volume grows near $\kappa_c$. This bottleneck is the principal motivation for non-local cluster algorithms and accelerated Markov-chain schemes in conventional lattice simulations, and is also the regime in which generative neural-network models, such as the diffusion models studied here, are expected to provide a benefit.

\section{Diffusion models}
\label{sec:dm_framework}

This section sets up the variance-exploding score-based diffusion model used in this work. We introduce the forward stochastic differential equation that evolves a field configuration into a broad, nearly Gaussian field, the reverse-time samplers that transport noise back to the target distribution, and the denoising score-matching objective used to train the score network.

\subsection{Forward stochastic process}
\label{sec:dm_forward}

The forward process can be formulated as an It\^o stochastic differential equation~\cite{Song2021SDE},
\begin{equation}
\mathrm{d}\phi \;=\; f(\phi,t)\,\mathrm{d}t \;+\; g(t)\,\mathrm{d}w ,
\label{eq:forward_sde}
\end{equation}
where $f(\phi,t)$ is the drift term, $g(t)$ is the diffusion coefficient, and $w$ is a Wiener process with independent components for different field degrees of freedom. This SDE form is the unified continuous-time limit of the earlier discrete diffusion~\cite{Ho2020DDPM} and score-matching~\cite{SongErmon2019} constructions. We can equivalently express this process as a Langevin equation,
\begin{equation}
\begin{aligned}
\frac{\partial \phi(t,x)}{\partial t} &\;=\; f\!\bigl(\phi(t,x),t\bigr) \;+\; g(t)\,\eta(t,x),\\
\bigl\langle \eta(t,x)\,\eta(t',x')\bigr\rangle &\;=\; \delta(x-x')\,\delta(t-t'),
\end{aligned}
\label{eq:forward_langevin}
\end{equation}
where $\eta$ is Gaussian white noise uncorrelated in both diffusion time and space. Different choices of $f$ and $g$ correspond to different ways of evolving the data distribution toward a tractable terminal distribution. We adopt the variance-exploding (VE) form throughout this work, detailed comparisons with alternative SDE choices on the lattice can be found in Ref.~\cite{Aarts:2024rsl}.

The variance-exploding process is defined by removing the drift term,
\begin{equation}
f(\phi,t)=0,
\qquad
\mathrm{d}\phi \;=\; g(t)\,\mathrm{d}w .
\label{eq:ve_sde_explicit}
\end{equation}
In this case, there is no deterministic drift that pulls the field toward the origin. The forward process only adds Gaussian noise, and the amount of added noise increases with the diffusion time.

Solving \Eq{eq:ve_sde_explicit} from the initial condition $\phi_0$ gives the conditional distribution,
\begin{equation}
p_{0t}\bigl(\phi_t\,\bigl|\,\phi_0\bigr)
\;=\;
\mathcal{N}\!\Bigl(
\phi_t;\,
\phi_0,\,
\Sigma^2(t)\mathbb{1}
\Bigr),
\label{eq:ve_sde_kernel}
\end{equation}
with,
\begin{equation}
\Sigma^2(t)
\;=\;
\int_0^t g^2(s)\,\mathrm{d}s .
\label{eq:ve_integrated_noise_variance}
\end{equation}
\Eq{eq:ve_sde_kernel} keeps the conditional mean fixed at $\phi_0$ while the variance $\Sigma^2(t)$ accumulates monotonically from $\Sigma^2(0)=0$ via \Eq{eq:ve_integrated_noise_variance} without saturation. This unbounded growth gives the name \emph{variance-exploding}.

The marginal distribution of $\phi_t$ is obtained by averaging this conditional distribution over initial configurations sampled from $p_{\mathrm{data}}$,
\begin{equation}
p_t(\phi)
\;=\;
\int
p_{\mathrm{data}}(\phi')
\,
\mathcal{N}\!\Bigl(
\phi;\,
\phi',
\Sigma^2(t)\mathbb{1}
\Bigr)
\,\mathrm{d}\phi' .
\label{eq:ve_marginal_distribution}
\end{equation}
At the final diffusion time $t_{\max}=T$ we write $\Sigma_{\max}\equiv\Sigma(T)$ for the terminal noise scale. When $\Sigma_{\max}$ greatly exceeds the spread of $p_{\mathrm{data}}$, the Gaussian distribution in \Eq{eq:ve_marginal_distribution} varies on a scale much larger than $p_{\mathrm{data}}$ itself, and the integral is well approximated by a single broad Gaussian of width $\Sigma_{\max}$, with the detailed structure of $p_{\mathrm{data}}$ suppressed, i.e., the cumulants due to $p_{\mathrm{data}}$ are much smaller than the variance~\cite{Aarts:2024rsl}.

A common practical prescription~\cite{SongErmon2020} sets $\Sigma_{\max}$ at the scale of the maximal pairwise distance between training samples,
\begin{equation}
\Sigma_{\max}
\;\gtrsim\;
\max_{i,j}\|\phi_i-\phi_j\|_2 .
\label{eq:ve_sigma_max_prescription}
\end{equation}
The reverse process can therefore be initialized by drawing a Gaussian random field.

We choose a geometric diffusion coefficient with $T=1$,
\begin{equation}
g(t)=\sigma^{t/T}=\sigma^t,
\qquad
t\in[0,T],
\qquad
\sigma>1 .
\label{eq:ve_geometric_g}
\end{equation}
For this schedule, the integrated noise variance can be evaluated explicitly,
\begin{equation}
\Sigma^2(t)
\;=\;
\int_0^t \sigma^{2s}\,\mathrm{d}s
\;=\;
\frac{\sigma^{2t}-1}{2\ln\sigma}.
\label{eq:ve_sigma_schedule}
\end{equation}
The terminal noise scale is,
\begin{equation}
\Sigma_{\max}
\;=\;
\Sigma(1)
\;=\;
\sqrt{\frac{\sigma^2-1}{2\ln\sigma}} .
\label{eq:ve_sigma_max}
\end{equation}

\subsection{Reverse denoising generation}
\label{sec:dm_reverse}

In \Sec{sec:dm_forward} we showed that the forward SDE~\Eq{eq:forward_sde} drives the marginal $p_t$ from $p_0=p_{\text{data}}$ to a broad, approximately Gaussian $p_{T}$ of width $\Sigma_{\max}$. To generate samples from $p_{\text{data}}$ we reverse this evolution. While drawing $\phi_{T}$ from the Gaussian is trivial, the remaining ingredient is a backward dynamics from $T$ to $0$ that yields $\phi_0\sim p_{\text{data}}$. Anderson~\cite{Anderson1982} showed that the time-reversal of an It\^o SDE is itself an SDE by matching the bidirectional Fokker-Planck equations. Applied to~\Eq{eq:forward_sde}, this yields~\cite{Song2021SDE},
\begin{equation}
\mathrm{d}\phi \;=\; \bigl[f(\phi,t)-g^{2}(t)\,\nabla_{\phi}\log p_{t}(\phi)\bigr]\,\mathrm{d}t \;+\; g(t)\,\mathrm{d}\bar{w}.
\label{eq:reverse_sde_exact}
\end{equation}
Here $t$ runs from $T$ down to $0$ ($\mathrm{d}t<0$), and $\bar{w}$ is a standard Wiener process independent of the forward $w$.

By sharing the same marginal distributions in the Fokker-Planck equation, one can establish a deterministic probability-flow ODE,
\begin{equation}
\frac{\mathrm{d}\phi}{\mathrm{d}t} \;=\; f(\phi,t)-\tfrac{1}{2}\,g^{2}(t)\,\nabla_{\phi}\log p_{t}(\phi),
\label{eq:prob_flow_ode_exact}
\end{equation}
whose marginal density at every $t\in[0,T]$ coincides with that of the forward SDE~\Eq{eq:forward_sde} and the reverse SDE~\Eq{eq:reverse_sde_exact}. The three processes thus share the same marginal $p_{t}(\phi)$ at every intermediate $t$, although they generate distinct trajectories. The probability-flow ODE provides a deterministic alternative to the reverse SDE for generation and to estimate the likelihood~\cite{Song2021SDE,Wang:2023exq}, integrated from $\phi_{T}\sim p_{T}$ in the reverse direction down to $t=0$.

Both reverse dynamics depend on $p_{\text{data}}$ only through the time-dependent \emph{score function},
\begin{equation}
s(\phi,t) \;=\; \nabla_{\phi}\log p_{t}(\phi).
\label{eq:score_def}
\end{equation}
Replacing $s(\phi,t)$ in~\Eq{eq:reverse_sde_exact} and~\Eq{eq:prob_flow_ode_exact} with a neural-network approximation $s_{\theta}(\phi,t)$ yields the practical samplers
\begin{align}
\mathrm{d}\phi &\;=\; \bigl[f(\phi,t)-g^{2}(t)\,s_{\theta}(\phi,t)\bigr]\,\mathrm{d}t \;+\; g(t)\,\mathrm{d}\bar{w},
\label{eq:reverse_sde}\\
\frac{\mathrm{d}\phi}{\mathrm{d}t} &\;=\; f(\phi,t)-\tfrac{1}{2}\,g^{2}(t)\,s_{\theta}(\phi,t).
\label{eq:prob_flow_ode}
\end{align}
The training of $s_{\theta}$ is the subject of \Sec{sec:dm_score_matching}.

Although~\Eq{eq:reverse_sde_exact} and~\Eq{eq:prob_flow_ode_exact} have identical exact marginals, their discretizations with an imperfect $s_{\theta}$ behave differently at the trajectory level. The reverse SDE~\Eq{eq:reverse_sde} is stochastic, and its injected noise acts as a Langevin-like corrector that damps accumulated score errors at each step~\cite{Karras2022EDM}. The probability-flow ODE~\Eq{eq:prob_flow_ode} is deterministic and invertible, and admits higher-order numerical solvers such as DPM-Solver~\cite{LuDPMSolver2022} or the Heun integrator of EDM~\cite{Karras2022EDM} that require only $10$--$35$ function evaluations.

The two samplers therefore involve a trade-off. A deterministic ODE integrator reproduces any systematic bias in $s_{\theta}$, whereas the stochastic reverse SDE provides a relaxation mechanism toward the true marginal at every $t$. We found empirically that for diagnostics based on the two-point function $G(|k|)$ (as in the propagator study of \Sec{sec:propagator_results}), the reverse SDE, or a predictor-corrector hybrid that periodically injects Langevin noise into an ODE trajectory~\cite{Song2021SDE}, tracks the true marginal more closely.

\subsection{Score matching}
\label{sec:dm_score_matching}

\begin{figure*}[!tbp]
\centering
\includegraphics[width=\textwidth]{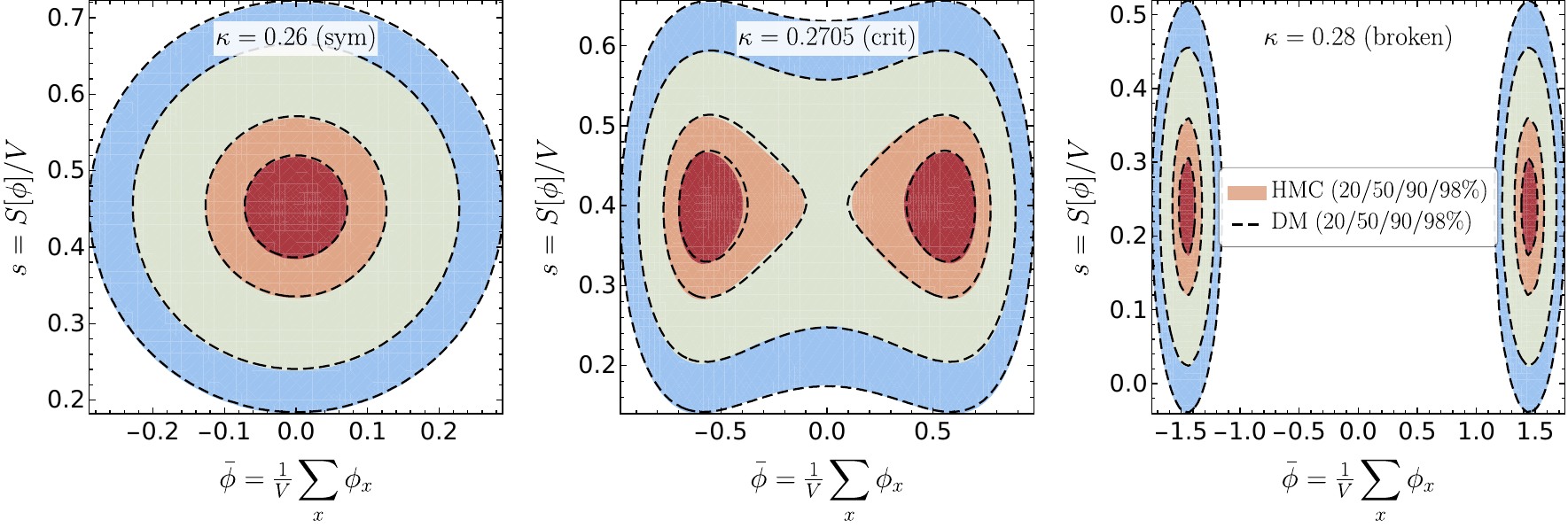}
\caption{Joint distribution of $\bar\phi=V^{-1}\sum_x\phi_x$ and the action density $s=S[\phi]/V$ at $D=2$, $L=128$, $\lambda=0.022$ for three values of $\kappa$. Filled bands are HMC iso-density contours enclosing $20\%$, $50\%$, $90\%$ and $98\%$ of the probability mass. Dashed black contours are the corresponding mass-enclosed contours of the diffusion-model ensemble at the same physical point. From left to right, the panels show the symmetric phase ($\kappa=0.26$), the near-critical regime ($\kappa=0.2705\simeq\kappa_c$), and the broken phase ($\kappa=0.28$). The diffusion-model contours around the two broken-phase peaks are slightly broader along the $\bar\phi$ direction. The effect is too small to be resolved by eye on the scale of the figure and is quantified by the susceptibility excess reported in \Tab{tab:indist_scalar_L128}.}
\label{fig:mS_joint_L128}
\end{figure*}

The reverse samplers~\Eq{eq:reverse_sde} and~\Eq{eq:prob_flow_ode} require the score $s(\phi,t)=\nabla_{\phi}\log p_{t}(\phi)$ at any diffusion time $t$. Since $p_t$ is a high-dimensional integral over $p_{\text{data}}$, it cannot be evaluated in closed form, so the score $\nabla_{\phi}\log p_{t}$ is not available as a direct training target.

Denoising score matching (DSM)~\cite{Vincent2011} sidesteps this problem by replacing the unavailable marginal score with the score of the conditional and noised distribution $p_{0t}(\phi_{t}\mid\phi_{0})$, which is analytically known once the forward process is fixed. The training objective is
\begin{multline}
\mathcal{L}(\theta) \;=\; \Bigl\langle\omega(t)\,\bigl\|s_{\theta}(\phi_{t},t)\\[-1pt]
-\nabla_{\phi_{t}}\log p_{0t}\bigl(\phi_{t}\,\bigl|\,\phi_{0}\bigr)\bigr\|^{2}\Bigr\rangle_{t,\,\phi_{0},\,\phi_{t}\mid\phi_{0}},
\label{eq:dsm}
\end{multline}
where the expectation runs over $t\sim\mathrm{Uniform}(0,T)$, with a small lower cutoff in practice (\App{app:ncsnpp}), $\phi_{0}\sim p_{\text{data}}$ and $\phi_{t}\sim p_{0t}(\cdot\mid\phi_{0})$, and $\omega(t)>0$ is a weighting function across diffusion times.

The conditional and marginal scores are different objects, but an identity due to Vincent~\cite{Vincent2011} shows that~\Eq{eq:dsm} and the (intractable) explicit score-matching loss $\bigl\langle\omega(t)\|s_{\theta}(\phi_{t},t)-\nabla_{\phi_{t}}\log p_{t}(\phi_{t})\|^{2}\bigr\rangle_{t,\phi_{t}}$ differ only by a $\theta$-independent constant. Their gradients in $\theta$ therefore coincide, and the global minimizer of $\mathcal{L}(\theta)$ is the marginal score $\nabla_{\phi_{t}}\log p_{t}$, even though the per-sample target in~\Eq{eq:dsm} is only the conditional score. Because $p_{0t}(\phi_{t}\mid\phi_{0})$ is Gaussian with the mean and variance fixed by the VE forward kernel~\Eq{eq:ve_sde_kernel}, this conditional score has a closed form, and~\Eq{eq:dsm} is evaluated by Monte Carlo sampling.

For the VE schedule adopted in this work, the conditional score reads,
\begin{equation}
\nabla_{\phi_{t}}\log p_{0t}\bigl(\phi_{t}\,\bigl|\,\phi_{0}\bigr) \;=\; -\frac{\phi_{t}-\phi_{0}}{\Sigma^{2}(t)} \;=\; -\frac{\eta}{\Sigma(t)},
\label{eq:ve_cond_score}
\end{equation}
having used the reparameterization $\phi_{t}=\phi_{0}+\Sigma(t)\,\eta$ with $\eta\sim\mathcal{N}(0,\mathbb{1})$. The standard weighting $\omega(t)=\Sigma^{2}(t)$ equalizes the signal-to-noise ratio along the diffusion trajectory. Substituting~\Eq{eq:ve_cond_score} into~\Eq{eq:dsm} with this weighting reduces the objective to,
\begin{equation}
\mathcal{L}(\theta) \;=\; \left\langle\bigl\|\Sigma(t)\,s_{\theta}(\phi_{t},t)+\eta\bigr\|^{2}\right\rangle_{t,\,\phi_{0},\,\eta},
\label{eq:dsm_loss_ve}
\end{equation}
the loss minimized in all subsequent training runs.

The diffusion-model score network $s_{\theta}$ is a convolutional U-Net with circular padding enforcing periodic boundary conditions. Its architecture and training schedule are given in \App{app:ncsnpp}.

Since $p_{0}=p_{\text{data}}$ by construction, a well-trained score network $s_{\theta}$ of sufficient capacity satisfies, in the limit $t\to 0$,
\begin{equation}
s_{\theta}(\phi,t)\;\xrightarrow[t\to 0]{}\;\nabla_{\phi}\log p_{\text{data}}(\phi)\;=\;-\,\nabla_{\phi}S[\phi],
\label{eq:score_small_t}
\end{equation}
where the final equality uses the Boltzmann form $p_{\text{data}}(\phi)\propto e^{-S[\phi]}$. In this limit the learned score therefore reproduces the negative action gradient $-\nabla_{\phi}S[\phi]$. In practice this identification is only approximate, since a finite score network need not be a conservative field~\cite{Aarts:2025lpi}, a deviation quantified for our trained networks in \App{app:nonconservative}, and further deviations arise from finite training statistics and the discretization of the reverse process. Such residual deviations can be removed by a Metropolis-adjusted Langevin correction, which restores exact sampling from $e^{-S}$~\cite{Zhu:2025pmw,lam2026metropolis}. The precision of this limit, and the sensitivity of the reverse samplers~\Eq{eq:reverse_sde} and~\Eq{eq:prob_flow_ode} to its residual error, underlie the diagnostics developed in \Sec{sec:score_quality} and \Sec{sec:acceptance}.

\section{Generated ensembles at the trained couplings}
\label{sec:indist}

After training, the diffusion model defines a sampler for the Boltzmann distribution of the lattice theory. The relevant question is therefore whether the generated ensemble reproduces the statistical properties of the reference Monte Carlo ensemble at the same couplings. This in-distribution test is a necessary baseline before addressing more demanding questions such as generalization across lattice sizes or changes of the physical parameters.

We perform this validation at representative couplings across different phases of the theory. In two dimensions we use $L=128$ and $\lambda=0.022$, with $\kappa=0.26$, $\kappa=0.2705\simeq\kappa_c$, and $\kappa=0.28$, corresponding respectively to the symmetric, near-critical, and broken phases. In three dimensions we use $L=64$ and $\lambda=0.9$, with $\kappa=0.18$, $\kappa=0.1923\simeq\kappa_c$, and $\kappa=0.2$, again spanning the symmetric, near-critical, and broken regimes. The reference ensembles are generated with the FA-HMC--Wolff Markov chain described in \App{app:fahmc_wolff}, which combines Fourier-accelerated HMC trajectories with Wolff cluster reflections. The diffusion-model ensembles are generated by integrating the reverse SDE with an Euler--Maruyama discretization. The 2D samples use a logarithmic time grid and the 3D samples use a linear time grid, with the detailed sampling choices collected in \App{app:numerical_tables} and \App{app:ncsnpp}.

The comparison is organized into two complementary tests. Scalar observables are reported in \Sec{sec:indist_scalar} and the momentum-space propagator $G(|k|)$ in \Sec{sec:propagator_results}. More quantitative diagnostics of score quality are deferred to \Sec{sec:diagnostics}.

\begin{figure*}[!tbp]
\centering
\includegraphics[width=\textwidth]{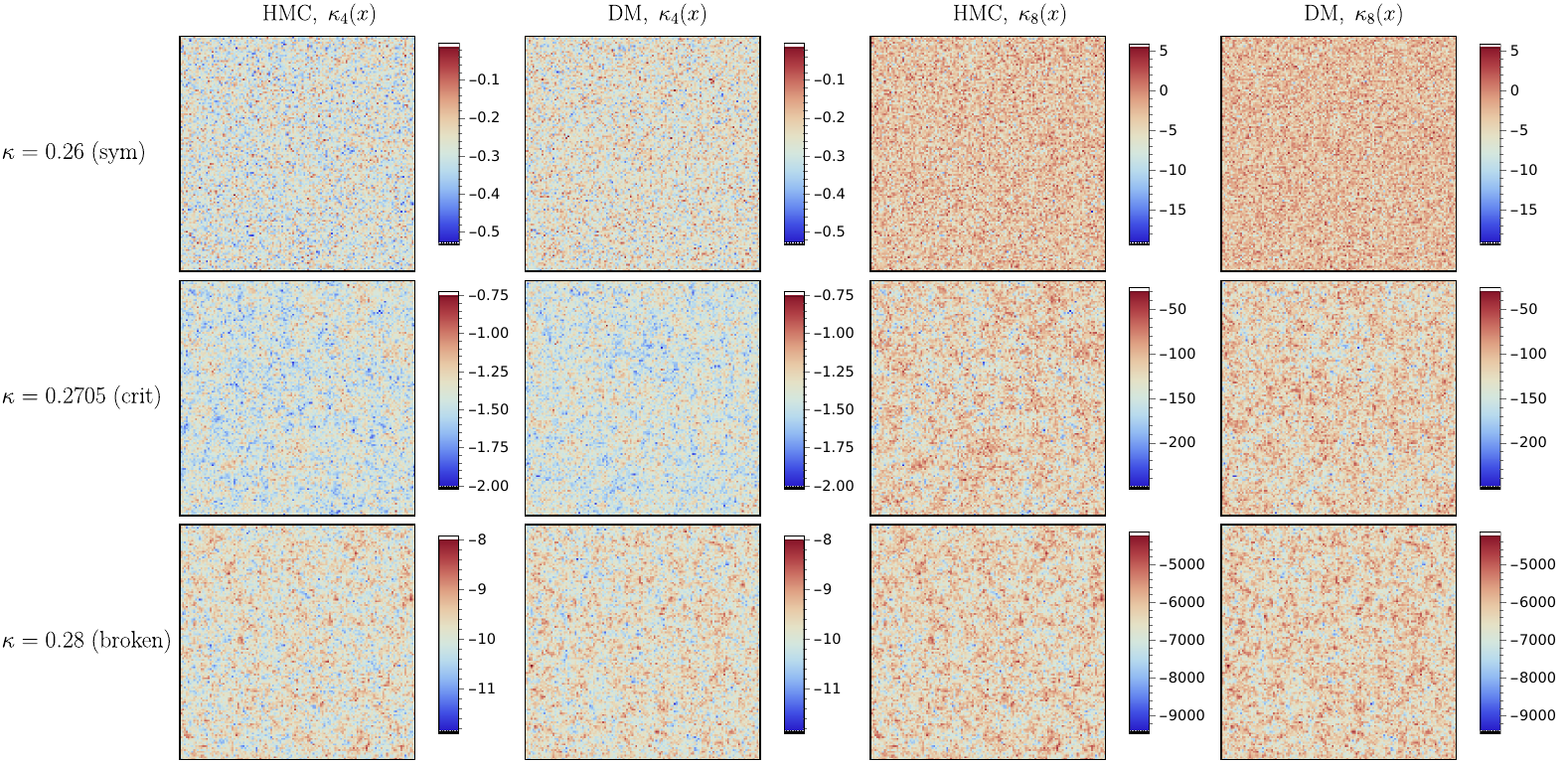}
\caption{Single-site cumulants $\kappa^{(\mathrm{site})}_4(x)$ and $\kappa^{(\mathrm{site})}_8(x)$ on the $L=128$ lattice at $\lambda=0.022$. Rows label the three values of $\kappa$ (symmetric, near-critical, broken). The four columns compare HMC and DM for $n=4$ and $n=8$. Within each row and each value of $n$, HMC and DM share the same color range, so that residual spatial structure can be compared directly. The heatmaps are homogeneous within statistical fluctuations and show no visible boundary, corner, or seam artifact. The scalar-observable tables report the corresponding spatial averages over $x$.}
\label{fig:cumulant_heatmaps_L128}
\end{figure*}

\subsection{Scalar observables}
\label{sec:indist_scalar}

We first test the diffusion-model ensemble through scalar observables, which provide low-dimensional projections of the lattice-field probability measure. These observables cannot by themselves establish equality of the full distribution, but they probe physically distinct aspects of the generated ensemble, including the order parameter, the zero-mode fluctuations, the shape of the magnetization distribution, the single-site field statistics, and the typical action range. They therefore constitute a necessary first validation before the more differential momentum-space comparison in \Sec{sec:propagator_results}.

For each configuration we define the spatially averaged field,
\begin{equation}
  \bar\phi = \frac{1}{V}\sum_x \phi_x .
\end{equation}
The corresponding global observables are the absolute magnetization $\langle |\bar\phi| \rangle$, the finite-volume susceptibility,
\begin{equation}
  \chi =
  V\left(
  \langle \bar\phi^2\rangle
  - \langle |\bar\phi|\rangle^2
  \right),
\end{equation}
and the Binder cumulant,
\begin{equation}
  U_L =
  1 -
  \frac{\langle \bar\phi^4\rangle}
       {3\langle \bar\phi^2\rangle^2}.
\end{equation}
We denote the connected cumulants of the zero-mode by,
\begin{equation}
  \kappa^{(0)}_n = \langle \bar\phi^n\rangle_c,
  \qquad n=2,4,6,8 .
\end{equation}
Here $\langle |\bar\phi| \rangle$ is the finite-volume order parameter. The susceptibility $\chi$ measures the width of the magnetization distribution around this order parameter and is therefore especially sensitive to the zero-mode. It is related to, but not identical with, $\kappa^{(0)}_2$ because it subtracts $\langle|\bar\phi|\rangle^2$. The Binder cumulant and the higher zero-mode cumulants provide additional information on the non-Gaussianity and tail structure of the magnetization distribution.

Detailed numerical values for the scalar-observable comparisons are collected in \App{app:numerical_tables}, \Tab{tab:indist_scalar_L128}. The main text summarizes the phase-dependent trends.

We also monitor the action density,
\begin{equation}
  s[\phi] \equiv \frac{S[\phi]}{V}.
  \label{eq:action_density}
\end{equation}
The field-independent constant $\lambda V$ in \Eq{eq:phi4_action} is omitted from all reported action densities. Since the target distribution is proportional to $e^{-S[\phi]}=e^{-Vs}$, the action density directly probes whether the diffusion model samples configurations in the typical action range of the Boltzmann measure. Agreement in $\langle s\rangle$ alone, however, is only a one-dimensional test. To retain information on the correlation between the action and the order parameter, we further consider the joint distribution,
\begin{equation}
P(\bar\phi,s)
=
\frac{1}{Z}
\int\!\mathcal{D}\phi\,
e^{-S[\phi]}\,
\delta\!\left(
\bar\phi-\frac{1}{V}\sum_x\phi_x
\right)
\delta\!\left(
s-\frac{S[\phi]}{V}
\right).
\label{eq:joint_ms}
\end{equation}
The joint distribution $P(\bar\phi,s)$ contains information beyond the separate marginals $P(\bar\phi)$ and $P(s)$. It tests whether the diffusion model reproduces the correlation between the magnetization and the action density, i.e., whether configurations with a given $\bar\phi$ occur at the correct action scale.

\Fig{fig:mS_joint_L128} compares the HMC and DM joint distributions at the three values of $\kappa$ in 2D, $L=128$. The DM iso-density contours track the HMC ones across all three regimes. In the broken phase, $\kappa=0.28$, the two $\mathbb{Z}_2$-related peaks sit at the correct magnetizations and at the same typical action density, with a slight broadening of the DM peaks along the $\bar\phi$ direction. This broadening is too small to be seen directly in \Fig{fig:mS_joint_L128} and is quantified by the susceptibility excess reported in \Tab{tab:indist_scalar_L128}.

This susceptibility excess admits a simple two-peak interpretation. Let $v$ denote the position of either magnetization peak and let $\sigma$ denote its intra-sector width. Since $\sigma\ll v$ in the broken phase, the absolute magnetization is controlled mainly by the peak position,
\begin{equation}
  \langle |\bar\phi| \rangle \simeq v,
\end{equation}
whereas the susceptibility defined above subtracts the order-parameter contribution and is therefore dominated by the residual peak width,
\begin{equation}
  \frac{\chi}{V}
  =
  \langle \bar\phi^2\rangle
  -
  \langle |\bar\phi|\rangle^2
  \simeq
  \sigma^2 .
\end{equation}
The remaining zero-mode observables behave in the opposite way. Because the mean vanishes by the $\mathbb{Z}_2$ symmetry, they are controlled by the peak position rather than by its width,
\begin{equation}
\kappa^{(0)}_2 = \langle\bar\phi^2\rangle \simeq v^2,
\qquad
U_L \simeq \frac{2}{3} - \frac{4}{3}\Big(\frac{\sigma}{v}\Big)^2 ,
\end{equation}
with the higher even cumulants similarly set by powers of $v$. The width $\sigma$ enters these observables only through the suppressed ratio $(\sigma/v)^2$, while it fixes $\chi$ at leading order.

A diffusion-model ensemble may therefore reproduce the peak locations, the Binder cumulant, and the higher even zero-mode cumulants rather accurately, while still showing a measurable excess in $\chi$ if the individual peaks are slightly too broad. We interpret the broken-phase $\chi$ excess as an overestimate of the intra-sector zero-mode fluctuations, rather than as a sector-weight imbalance or an error in the local action scale.

We also examine observables constructed from the single-site field distribution. To test translation invariance explicitly, we define the single-site cumulants,
\begin{equation}
  \kappa^{(\mathrm{site})}_n(x)
  =
  \langle \phi_x^n\rangle_c,
  \qquad n=2,4,6,8 ,
  \label{eq:site_cumulants_x}
\end{equation}
where the connected moments are taken over the ensemble at fixed lattice site $x$. The single-site Binder cumulant is,
\begin{equation}
  U^{(\mathrm{site})}_4(x)
  =
  1 -
  \frac{\langle \phi_x^4\rangle}
       {3\langle \phi_x^2\rangle^2}.
  \label{eq:site_binder_x}
\end{equation}
For an exactly translation-invariant ensemble these quantities are independent of $x$. In a finite generated ensemble, their residual spatial dependence is a direct diagnostic of translation-symmetry breaking and of possible boundary artifacts from the convolutional architecture.

The quantities reported in the scalar-observable tables are the spatial averages,
\begin{equation}
  \kappa^{(\mathrm{site})}_n
  =
  \frac{1}{V}\sum_x \kappa^{(\mathrm{site})}_n(x),
  \qquad
  U^{(\mathrm{site})}_4
  =
  \frac{1}{V}\sum_x U^{(\mathrm{site})}_4(x).
  \label{eq:site_cumulants_avg}
\end{equation}
Their spatial dependence is shown in \Fig{fig:cumulant_heatmaps_L128}. The heatmaps indicate that the circular-padded score network preserves the periodic translation symmetry of the lattice measure at the level probed by these single-site cumulants. They are complementary to the zero-mode cumulants, probing the single-site field distribution with high statistical precision while observables constructed from $\bar\phi$ are sensitive mainly to long-wavelength fluctuations.

The HMC rows in the scalar tables use the raw reference ensembles. DM ensembles are $\mathbb{Z}_2$-symmetrized where stated by augmenting generated configurations with their reflected partners, $\{\phi,-\phi\}$. The tables report $\mathbb{Z}_2$-even observables. Observables constructed from $\bar\phi$ are estimated with block jackknife errors. The single-site cumulants are evaluated at each site and then averaged over sites, following \Eq{eq:site_cumulants_avg}. Where shown, the matched-$N$ HMC rows thin the same chains to the size of the DM ensembles.

For the two-dimensional ensembles at $L=128$ and $\lambda=0.022$, \Tab{tab:indist_scalar_L128} shows that the diffusion model reproduces the HMC scalar observables well across the symmetric, near-critical, and broken regimes. In the symmetric phase, $\kappa=0.26$, the order parameter and susceptibility agree with HMC within the statistical precision of the present ensembles, while the Binder cumulant and higher zero-mode cumulants are small and noise dominated, as expected for an approximately Gaussian distribution of $\bar\phi$. The leading single-site cumulant is reproduced at the percent level, with larger relative deviations only in higher single-site cumulants, which are more sensitive to the tails of the single-site field distribution.

At $\kappa=0.2705\simeq\kappa_c$, the agreement remains good despite the enhanced infrared fluctuations. The order parameter, susceptibility and Binder cumulant agree at the percent level, and the single-site cumulants agree at the sub-percent to percent level. The higher zero-mode cumulants show larger relative deviations, reflecting their sensitivity to the tails of the finite-volume magnetization distribution and to finite-sample fluctuations.

In the broken phase, $\kappa=0.28$, the diffusion model reproduces $\langle|\bar\phi|\rangle$, the Binder cumulant, the higher zero-mode cumulants, the single-site cumulants and the action density at the percent level. The main residual discrepancy is the susceptibility, $\chi=5.44(18)$ against the HMC value $\chi=4.22(6)$.

The corresponding three-dimensional comparison at $L=64$ and $\lambda=0.9$ is summarized in \Tab{tab:indist_scalar_L64_3d}. In the symmetric phase, $\kappa=0.18$, the diffusion model overestimates $\langle|\bar\phi|\rangle$ by about $9\%$ and $\chi$ by about $26\%$, while the higher zero-mode cumulants are dominated by statistical noise on the present sample size. The single-site cumulants deviate at the few- to fifteen-percent level, of similar character to two dimensions.

At the near-critical point, $\kappa=0.1923\simeq\kappa_c$, the scalar observables show the largest residual bias. The diffusion model overestimates $\langle|\bar\phi|\rangle$ and $\chi$, and the higher zero-mode cumulants are substantially larger in magnitude than in the HMC ensemble. This indicates that the generated distribution is biased toward the magnetized side of the finite-volume critical distribution. The single-site cumulants, by contrast, deviate by at most about twenty percent, far below the factor-of-several discrepancy in the zero-mode cumulants, showing that the dominant mismatch is infrared rather than local.

In the broken phase, $\kappa=0.2$, the same pattern seen in two dimensions reappears more strongly. The diffusion model reproduces the peak position, Binder cumulant, and higher zero-mode cumulants to within a few percent, but overestimates the susceptibility, $\chi=7.8(5)$ against the HMC value $4.9(1)$. Through $\chi/V\simeq\sigma^2$ this $59\%$ excess is equivalent to an intra-sector width about $26\%$ larger in the diffusion model, $\sigma_{\rm DM}\simeq0.0055$ against $\sigma_{\rm HMC}\simeq0.0043$.

The three-dimensional ensembles also show a systematic shift of the action density. The diffusion model overestimates $\langle S\rangle/V$ by about $19\%$, $16\%$, and $6\%$ at $\kappa=0.18$, $0.1923$, and $0.2$, respectively, far outside the quoted statistical errors, while the corresponding two-dimensional deviations remain below one percent. The origin of this shift is identified by splitting the action density of \Eq{eq:phi4_action} into its non-local hopping and on-site parts. By translation invariance,
\begin{equation}
\begin{aligned}
\frac{\langle S\rangle}{V} &= s_{\rm hop}+s_{\rm on},\\
s_{\rm hop} &= -2\kappa\sum_{\mu=1}^{D}\bigl\langle\phi_x\,\phi_{x+\hat\mu}\bigr\rangle,\\
s_{\rm on} &= (1-2\lambda)\bigl\langle\phi_x^2\bigr\rangle+\lambda\bigl\langle\phi_x^4\bigr\rangle,
\end{aligned}
\label{eq:action_density_split}
\end{equation}
where the on-site moments follow from the tabulated single-site cumulants through $\langle\phi_x^2\rangle=\kappa^{(\mathrm{site})}_2$ and $\langle\phi_x^4\rangle=\kappa^{(\mathrm{site})}_4+3\bigl[\kappa^{(\mathrm{site})}_2\bigr]^2$ for ensembles with vanishing odd moments. At $\lambda=0.9$ the coefficient $1-2\lambda=-0.8$ makes the two terms of $s_{\rm on}$ opposite in sign and comparable in magnitude. At $\kappa=0.18$ the HMC moments give $s_{\rm on}=-0.469+0.566=0.097$, and $s_{\rm hop}=-0.164$ cancels further against $s_{\rm on}$, leaving $\langle S\rangle/V=-0.067$ as a small residue of much larger terms. Evaluating $s_{\rm on}$ with the DM and HMC moments of \Tab{tab:indist_scalar_L64_3d} gives the differences $\Delta s_{\rm on}=0.013$, $0.025$, and $0.020$ at the three couplings, against the observed action-density shifts $\Delta\langle S\rangle/V=0.013$, $0.019$, and $0.016$. The deviation is therefore carried almost entirely by $s_{\rm on}$, with a small compensating shift of the nearest-neighbor correlator in $s_{\rm hop}$. Although the single-site moments are reproduced at the few-percent level, the cancellations within $s_{\rm on}$ and between $s_{\rm on}$ and $s_{\rm hop}$ amplify this bias into the large relative deviation of $\langle S\rangle/V$. Consistent with this interpretation, the cross-$L$ ensembles of \Sec{sec:crossL}, whose single-site moments lie closer to HMC, reduce the action-density deviation to the few-percent level. The scalar-observable analysis therefore identifies two residual errors in three dimensions, an overestimate of the infrared zero-mode width and a local-moment bias most visible in the action density. The infrared component is tested more directly by the propagator in the next subsection.

\begin{figure*}[!htbp]
  \centering
  \includegraphics[width=0.5\textwidth]{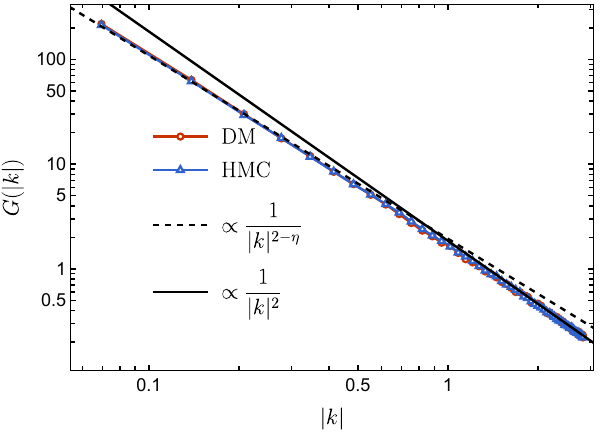}\includegraphics[width=0.5\textwidth]{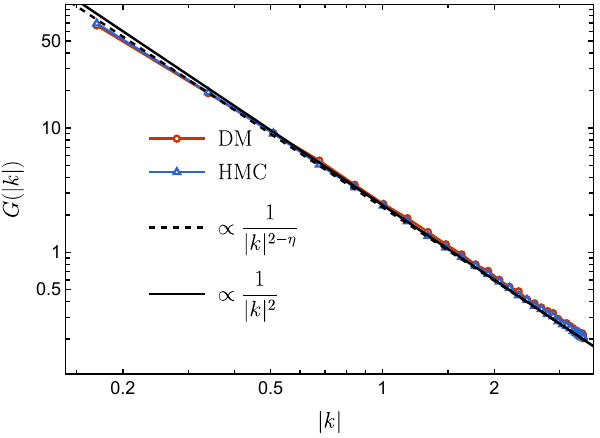}
\caption{Critical scaling of the propagator near $\kappa_c$ on log-log axes. The converged in-distribution diffusion-model ensemble, trained and sampled at the same lattice size, is compared with the HMC reference. The left panel shows two dimensions at $L=128$, $\lambda=0.022$, $\kappa=0.2705\simeq\kappa_c$, together with the critical scaling form $G(|k|)\propto |k|^{-(2-\eta)}$ for $\eta=1/4$ (dashed line) and the free-field behavior $1/|k|^2$ (solid line). The right panel shows three dimensions at $L=64$, $\lambda=0.9$, $\kappa=0.1923\simeq\kappa_c$, with $\eta\simeq0.036$. Both panels use the same line convention. Error bars are bootstrap $1\sigma$.}
  \label{fig:prop_crit_scaling}
\end{figure*}

\subsection{Momentum-space two-point function}
\label{sec:propagator_results}

\begin{figure*}[!tbp]
\centering
\includegraphics[width=\textwidth]{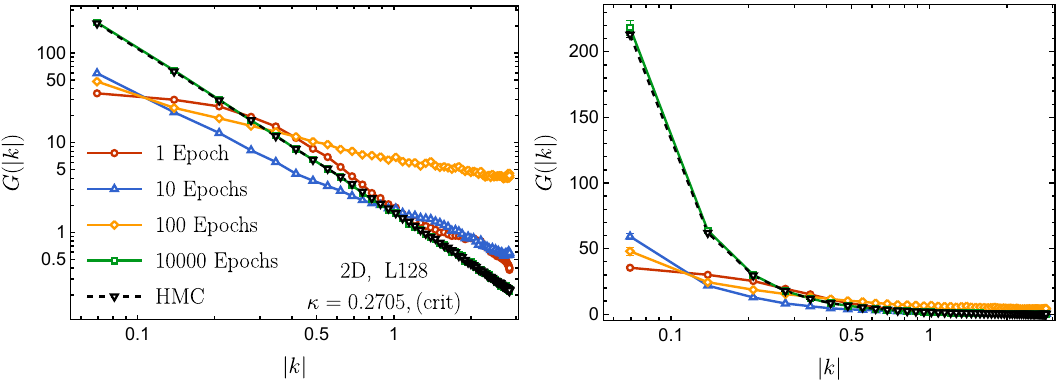}
\caption{\textbf{Near-critical regime}. Momentum-space propagator $G(|k|)$ at $\kappa=0.2705\simeq\kappa_c$ in the two-dimensional theory, $L=128$, $\lambda=0.022$. The panels show the HMC reference and diffusion-model samples from selected training epochs on double-logarithmic (left) and semi-logarithmic (right) axes.}
\label{fig:prop_2d_crit}
\end{figure*}

\begin{figure*}[!tbp]
  \centering
  \includegraphics[width=\textwidth]{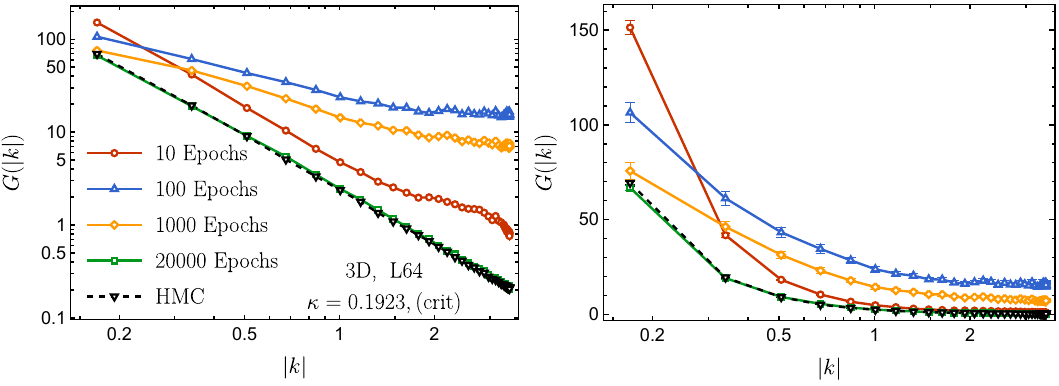}
\caption{\textbf{Near-critical regime}. Momentum-space propagator $G(|k|)$ at $\kappa=0.1923\simeq\kappa_c$ in the three-dimensional theory, $L=64$, $\lambda=0.9$. The panels show the HMC reference and diffusion-model samples from selected training epochs on double-logarithmic (left) and semi-logarithmic (right) axes.}
  \label{fig:prop_3d_crit}
  \end{figure*}

The scalar observables discussed above probe selected low-dimensional projections of the generated ensemble. We now consider the connected momentum-space two-point function, denoted by $G(|k|)$ and referred to below as the propagator. This observable tests the two-point structure of the generated field configurations mode-by-mode, and is therefore particularly sensitive to residual long-wavelength errors. It complements the single-site cumulants in \Sec{sec:indist_scalar}, which probe the single-site field distribution but do not resolve correlations between different lattice sites.

Fourier modes on the periodic lattice are labeled by,
\begin{equation}
  q_\mu = \frac{2\pi n_\mu}{L},
  \qquad
  n_\mu = 0,\ldots,L-1 .
\end{equation}
We use the corresponding lattice momentum,
\begin{equation}
  k_\mu(q)
  =
  2\sin\frac{q_\mu}{2},
  \qquad
  |k(q)|
  =
  \left(
  \sum_\mu k_\mu^2(q)
  \right)^{1/2},
  \label{eq:lattice_momentum}
\end{equation}
which is the natural momentum variable for the nearest-neighbor lattice action. The lattice Laplacian is diagonal in this variable with eigenvalue $|k|^2$, so the free-field action is quadratic in $\tilde\phi(q)$ and gives the propagator
\begin{equation}
  G_0(|k|)
  \propto
  \frac{1}{|k|^2+m_0^2},
  \label{eq:free_lattice_prop}
\end{equation}
which reduces to $1/|k|^2$ in the massless limit. In the interacting theory this Gaussian form holds only approximately. It describes the low-momentum propagator in the massive phases, up to a field normalization and a renormalized mass, whereas near the critical point the momentum dependence becomes anomalous and is no longer captured by a constant residue, as given by \Eq{eq:critical_scaling_prop}.

For each configuration we compute the Fourier amplitude,
\begin{equation}
  \tilde\phi(q)
  =
  \frac{1}{\sqrt{V}}
  \sum_x e^{-iq\cdot x}\phi_x,
  \qquad
  V=L^D .
  \label{eq:fourier_field}
\end{equation}
For non-zero momenta the constant component of the field does not contribute, and by translation invariance the connected two-point function is simply,
\begin{equation}
  G(|k(q)|)
  =
  \left\langle
  \tilde\phi(q)\tilde\phi(-q)
  \right\rangle
  =
  \left\langle
  |\tilde\phi(q)|^2
  \right\rangle ,
  \label{eq:prop_estimator}
\end{equation}
where we used the reality condition $\tilde\phi(-q)=\tilde\phi(q)^*$. The zero-momentum mode is not included in the propagator plots. It is instead characterized by the susceptibility and the zero-mode cumulants in \Sec{sec:indist_scalar}.

We restrict the comparison to diagonal non-zero momenta,
\begin{equation}
  q_1=\cdots=q_D\equiv q,
  \qquad
  q=\frac{2\pi n}{L},
  \qquad
  n=1,\ldots,\frac{L}{2},
  \label{eq:diagonal_momenta}
\end{equation}
for which,
\begin{equation}
  |k_n|
  =
  \sqrt{D}\,2\sin\frac{\pi n}{L}.
  \label{eq:diagonal_lattice_momentum}
\end{equation}
This projection gives a single one-dimensional curve without shell-averaging over momenta with different lattice artifacts. Statistical uncertainties are estimated by bootstrap resampling of configurations.

\subsubsection{Near-critical regime}
\label{sec:results_nearcritical}

At the critical point, the propagator is expected to follow the scaling form,
\begin{equation}
  G(|k|)
  \propto
  |k|^{-(2-\eta)} .
  \label{eq:critical_scaling_prop}
\end{equation}

In the two-dimensional theory at the pseudocritical point $\kappa=0.2705\simeq\kappa_c$, \Fig{fig:prop_2d_crit} shows that the early diffusion-model checkpoints substantially underestimate the infrared propagator. At intermediate training the large-$|k|$ modes instead develop a plateau above the HMC curve, visible at $100$ epochs, before the curve settles onto the reference. By the final checkpoint, the diffusion-model propagator agrees with HMC over the plotted range. The converged comparison in \Fig{fig:prop_crit_scaling} shows the expected two-dimensional Ising scaling with the anomalous dimension $\eta=1/4$, clearly separated from the free-field $1/|k|^2$ behavior.

In the three-dimensional theory at the pseudocritical point $\kappa=0.1923\simeq\kappa_c$, \Fig{fig:prop_3d_crit} shows that the early checkpoints overestimate the propagator at all $|k|$. The infrared modes converge first, and the intermediate and large $|k|$ retain a plateau above the expected power-law decay until late in training. At the final checkpoint the diffusion-model propagator matches HMC. The converged comparison in the right panel of \Fig{fig:prop_crit_scaling} follows three-dimensional Ising scaling with $\eta\simeq0.036$. Because $\eta$ is small, $|k|^{-(2-\eta)}$ stays close to the free-field $1/|k|^2$, and both the HMC and the diffusion-model points lie close to the two reference lines. The detailed evolution of the propagator with training depends on the diffusion-model formulation and on its hyperparameters. The present results use the variance-exploding SDE with $g(t)=\sigma^t$, so changing the noise schedule can alter the intermediate-epoch trajectory at the same target.

\begin{figure*}[!tbp]
  \centering
  \includegraphics[width=\textwidth]{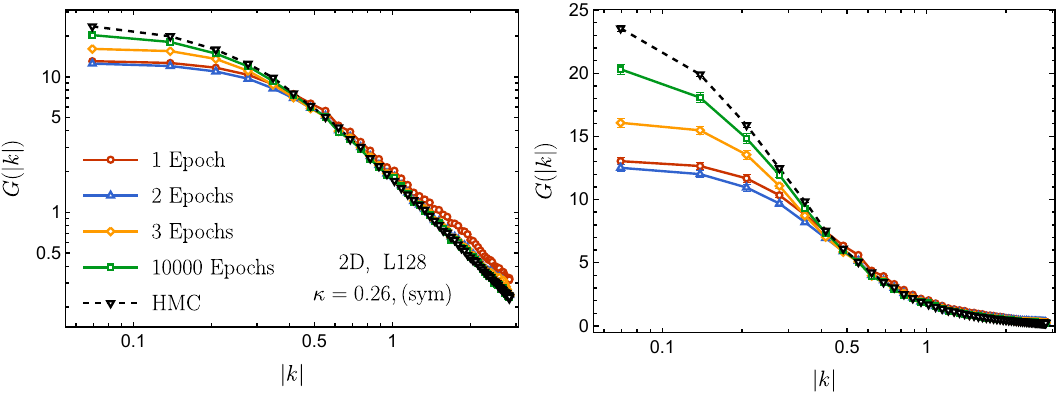}
\caption{\textbf{Symmetric phase}. Momentum-space propagator $G(|k|)$ at $\kappa=0.26$ in the two-dimensional symmetric phase, $L=128$, $\lambda=0.022$. The panels show the HMC reference and diffusion-model samples from selected training epochs on double-logarithmic (left) and semi-logarithmic (right) axes.}
  \label{fig:prop_2d_sym}
  \end{figure*}

\begin{figure*}[!tbp]
    \centering
    \includegraphics[width=\textwidth]{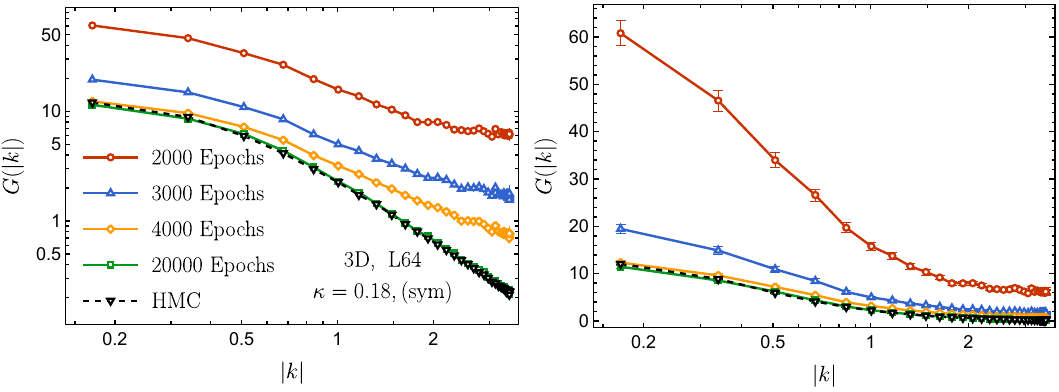}
\caption{\textbf{Symmetric phase}. Momentum-space propagator $G(|k|)$ at $\kappa=0.18$ in the three-dimensional symmetric phase, $L=64$, $\lambda=0.9$. The panels show the HMC reference and diffusion-model samples from selected training epochs on double-logarithmic (left) and semi-logarithmic (right) axes.}
    \label{fig:prop_3d_sym}
\end{figure*}

\subsubsection{Symmetric phase}
\label{sec:results_symmetric}

In the symmetric phase the correlation length is finite and the propagator develops an infrared plateau. \Fig{fig:prop_2d_sym} shows the two-dimensional result at $\kappa=0.26$. The intermediate and large $|k|$ already match HMC by $1$--$3$ epochs, while the infrared plateau is underestimated. The plateau is approached from below as training continues. At $10000$ epochs the lowest few diagonal modes are still slightly below HMC, and the rest of the propagator agrees with the reference.

\Fig{fig:prop_3d_sym} shows the three-dimensional result at $\kappa=0.18$. The early checkpoints overestimate the propagator at all $|k|$, with a flat tail at large $|k|$ instead of the expected fall-off. The overshoot decreases rapidly with training. By $4000$ epochs the diffusion-model curve matches HMC in the infrared, while a small residual remains at large $|k|$. The agreement extends to all $|k|$ by $20000$ epochs.

\subsubsection{Broken phase}
\label{sec:results_broken}

\begin{figure*}[!tbp]
\centering
\includegraphics[width=\textwidth]{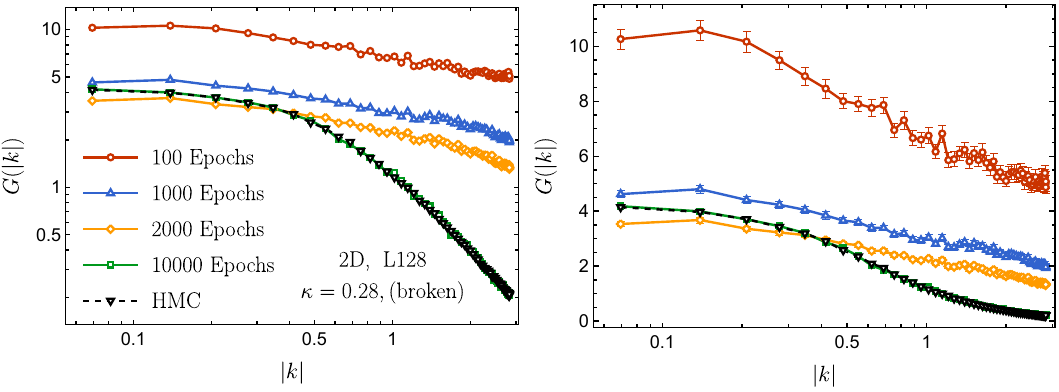}
\caption{\textbf{Broken phase}. Momentum-space propagator $G(|k|)$ at $\kappa=0.28$ in the two-dimensional broken phase, $L=128$, $\lambda=0.022$. The panels show the HMC reference and diffusion-model samples from selected training epochs on double-logarithmic (left) and semi-logarithmic (right) axes.}
\label{fig:prop_2d_broken}
\end{figure*}

In the broken phase the momentum-space propagator is infrared regulated by the finite mass gap within each ordered sector. \Fig{fig:prop_2d_broken} shows the two-dimensional result at $\kappa=0.28$. At $100$ epochs the diffusion-model curve overshoots HMC at all $|k|$, with a long flat tail at large $|k|$ in place of the expected fall-off. The overshoot decreases during training. By $10000$ epochs the propagator matches HMC across the range.
\begin{figure*}[!tbp]
  \centering
  \includegraphics[width=\textwidth]{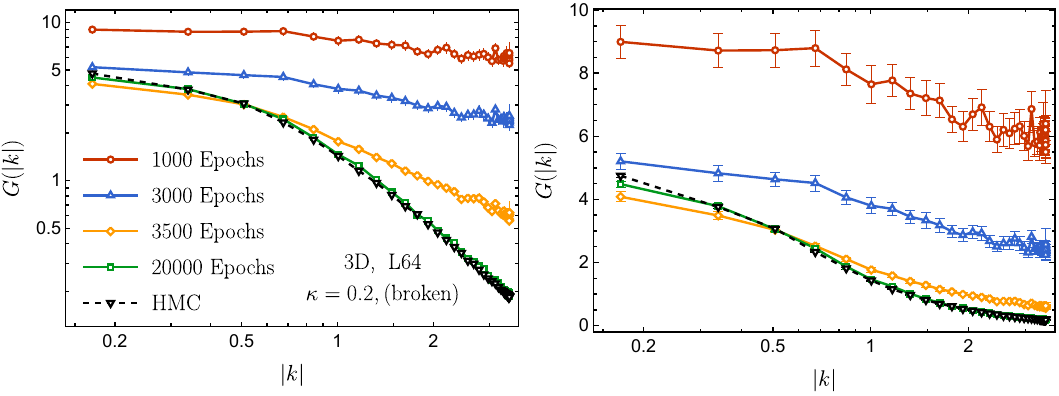}
\caption{\textbf{Broken phase}. Momentum-space propagator $G(|k|)$ at $\kappa=0.2$ in the three-dimensional broken phase, $L=64$, $\lambda=0.9$. The panels show the HMC reference and diffusion-model samples from selected training epochs on double-logarithmic (left) and semi-logarithmic (right) axes.}
  \label{fig:prop_3d_broken}
\end{figure*}

\Fig{fig:prop_3d_broken} shows the three-dimensional result at $\kappa=0.2$. At $1000$ epochs the diffusion-model curve overshoots HMC at all $|k|$, with a flat tail at large $|k|$. The overshoot decreases during training, and the propagator matches HMC by $20000$ epochs.

Across the six in-distribution tests, the converged diffusion-model propagators match HMC at all measured $|k|$. In the two-dimensional critical and symmetric cases the infrared modes are the last to converge. In the remaining cases the early checkpoints overestimate the propagator at all $|k|$, with a flat tail at large $|k|$ that retracts as training proceeds. The residual diffusion-model bias after training is concentrated in the zero-mode and action density, rather than in the finite-momentum propagator.

\section{Quality diagnostics for diffusion models}
\label{sec:diagnostics}

Having established the lattice setup in \Sec{sec:phi4_setup}, we now develop diagnostics for assessing whether a trained score-based diffusion model has learned the target distribution. We use two local diagnostics plus an HMC-referenced ESS. The direct score quality at small diffusion time (\Sec{sec:score_quality}) and the MALA acceptance rate (\Sec{sec:acceptance}) test the score network as a function approximator. The HMC-referenced MSE-based effective sample size (\Sec{sec:ess_mse_2d}) instead tests the generated DM configurations through physical observables. It reflects both score errors along the reverse trajectory and integration errors accumulated during the reverse-SDE evolution.

\subsection{Learned score network diagnostic}
\label{sec:score_quality}

We first examine the score network locally by comparing its output at small diffusion time with the negative action gradient on reference Monte Carlo configurations. This test isolates the trained network as a function approximator and is independent of the subsequent reverse-SDE sampling procedure.

The lattice action $S[\phi]$ never enters the training loss directly. Training matches the score network to the noised data distribution, and $S[\phi]$ appears only implicitly through the Boltzmann distribution $p_0\propto e^{-S[\phi]}$ of the training samples. By \Eq{eq:score_small_t}, the exact score reduces to $-\nabla_\phi S[\phi]$ in the limit $t\to 0$. If the trained network reproduces the negative action gradient in the small diffusion time limit, it has reconstructed the distribution from samples alone. This is a stronger statement about what the DM has learned than agreement on a finite set of ensemble observables.

We characterize the closeness of $s_\theta$ to $-\nabla_\phi S$ by three indicators computed over $N=512$ reference configurations $\phi$ from the training data. The cosine similarity captures the directional alignment of the two fields independent of their magnitudes. The magnitude ratio compares their overall scales independent of direction. The relative mean-squared error combines both,
\begin{equation}
\begin{aligned}
\text{cosine similarity} &= \left\langle\frac{-\,s_\theta\cdot\nabla_\phi S}{\|s_\theta\|\,\|\nabla_\phi S\|}\right\rangle,
\\
\text{relative MSE} &= \left\langle\frac{\|s_\theta+\nabla_\phi S\|^2}{\|\nabla_\phi S\|^2}\right\rangle,
\\
\text{magnitude ratio} &= \left\langle\frac{\|s_\theta\|}{\|\nabla_\phi S\|}\right\rangle,
\end{aligned}
\label{eq:cos_sim}
\end{equation}
where $s_\theta=s_\theta(\phi,t_{\text{eval}})$, $\nabla_\phi S=\nabla_\phi S[\phi]$, and $\langle\cdot\rangle$ averages over the $N=512$ reference configurations. The relative MSE vanishes only when the cosine similarity reaches unity and the magnitude ratio reaches one, so the three diagnostics resolve direction and magnitude errors separately.

\begin{figure*}[!tbp]
\centering
\includegraphics[width=\textwidth]{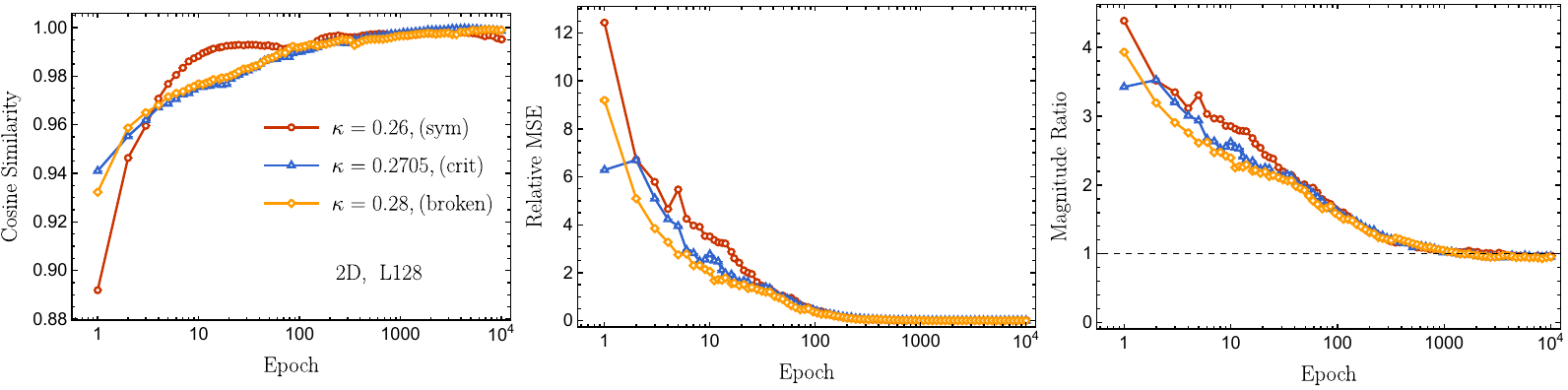}
\caption{Score-network quality diagnostic for the 2D $\phi^4$ model at $L=128$, evaluated at every saved checkpoint with EMA-averaged weights, $t_{\text{eval}}=10^{-4}$ and $N=512$ reference configurations. The left panel shows the cosine similarity rising from $\simeq 0.9$ at epoch~$1$ to $\simeq 0.998$ at late training. The center panel shows the relative MSE falling by roughly two orders of magnitude over the same window. The right panel shows the magnitude ratio decreasing from $\simeq 3$--$4$ at epoch~$1$ to $\simeq 1$ by epoch $\sim 5000$. The dashed line marks unity.}
\label{fig:score_quality_2d}
\end{figure*}

\fig{fig:score_quality_2d} shows the evolution of the three averages with training epoch. The cosine similarity reaches near-unity early, while the magnitude ratio decreases toward one over a much longer timescale. The dominant early error is therefore in the magnitude rather than the direction.

\begin{figure*}[!btp]
\centering
\includegraphics[width=\textwidth]{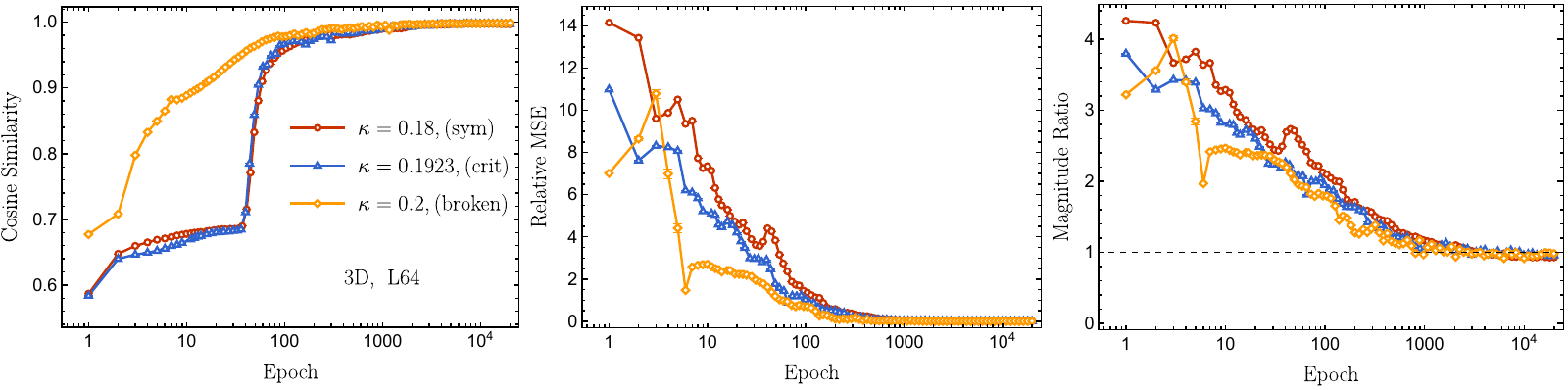}
\caption{Score-network quality diagnostic for the 3D $\phi^4$ model at $L=64$, $\lambda=0.9$, evaluated at every saved checkpoint with EMA-averaged weights, $t_{\text{eval}}=10^{-4}$ and $N=64$ reference configurations. The cosine similarity rises from $\sim 0.6$--$0.7$ at epoch~$1$ to near unity at late training, with the broken-phase curve improving earliest. The relative MSE falls from values of order $10$ to near zero over the same window. The magnitude ratio decreases from $\sim 3$--$4$ at early epochs to $\sim 1$ at late training. The dashed line marks unity.}
\label{fig:score_quality_3d}
\end{figure*}

The same diagnostic applied to the 3D models at $L=64$, $\lambda=0.9$ with $N=64$ reference configurations yields the trends shown in \fig{fig:score_quality_3d}, mirroring the 2D results. The cosine similarity reaches near-unity early while the magnitude ratio converges to one more slowly, confirming that the score network learns both direction and magnitude of $-\nabla_\phi S$ in three dimensions.

This diagnostic is evaluated only on reference configurations and therefore does not test the accumulation of errors along reverse-SDE trajectories. The trajectory-accumulated effect is tested instead by the HMC-referenced effective sample size (\Sec{sec:ess_mse_2d}), which compares DM-generated samples against HMC at the level of observables.

\subsection{MALA acceptance-rate diagnostic}
\label{sec:acceptance}

We measure how well a diffusion model has learned the target distribution by the Metropolis--Hastings (MH) acceptance rate of a Metropolis-Adjusted Langevin Algorithm (MALA)~\cite{Roberts1998} that uses the score network as proposal drift.

The update is carried out in the normalized field coordinate used by the score network. To keep notation light, hats are suppressed in this subsection, and the proposal variables are denoted by $\phi$ and $\psi$. Whenever the action appears in the Metropolis--Hastings ratio, it is evaluated on the raw lattice field obtained by the inverse linear rescaling. The constant Jacobian of this map cancels in the ratio.

From the current state $\phi_\tau$, the MALA proposal is
\begin{equation}
\psi_{\tau+1} = \phi_\tau + h\,s_\theta(\phi_\tau,\,t_{\text{mh}}) + \sqrt{2h}\,\eta,
\label{eq:mala_proposal}
\end{equation}
where $h$ is the step size, $t_{\text{mh}}$ the diffusion time at which the MH step is performed, and $\eta\sim\mathcal{N}(0,\mathbb{1})$. The associated proposal kernel is
\begin{equation}
\begin{aligned}
q(\psi|\phi) &\propto\exp\!\left[-\frac{1}{4h}\|\psi-\phi-h\,s_\theta(\phi,t_{\text{mh}})\|^2\right], \\
q(\phi|\psi) &\propto\exp\!\left[-\frac{1}{4h}\|\phi-\psi-h\,s_\theta(\psi,t_{\text{mh}})\|^2\right],
\end{aligned}
\label{eq:mala_kernel}
\end{equation}
where the second line is the reverse proposal obtained from the first by exchanging $\phi\leftrightarrow\psi$. The accept-reject step is the standard MH update,
\begin{equation}
\phi_{\tau+1} =
\begin{cases}
\psi_{\tau+1} & \text{with prob. } \min\!\bigl\{1,\,\alpha\bigr\}, \\[6pt]
\phi_\tau      & \text{otherwise},
\end{cases}
\label{eq:mala_accept}
\end{equation}
with target $p(\phi)\propto e^{-S(\phi)}$ and MH ratio,
\begin{equation}
\alpha \;=\; \frac{p(\psi_{\tau+1})\,q(\phi_\tau|\psi_{\tau+1})}{p(\phi_\tau)\,q(\psi_{\tau+1}|\phi_\tau)}.
\label{eq:mh_ratio}
\end{equation}
Substituting the Gaussian kernel \Eq{eq:mala_kernel} into \Eq{eq:mh_ratio} and using the forward-proposal identity $\psi_{\tau+1} - \phi_\tau - h\,s_\theta(\phi_\tau,t_{\text{mh}}) = \sqrt{2h}\,\eta$ from \Eq{eq:mala_proposal} gives
\begin{multline}
\log\alpha = S(\phi_\tau) - S(\psi_{\tau+1}) + \tfrac{1}{2}\|\eta\|^2 \\
- \tfrac{1}{4h}\bigl\|\sqrt{2h}\,\eta + h\bigl[s_\theta(\phi_\tau,t_{\text{mh}})+s_\theta(\psi_{\tau+1},t_{\text{mh}})\bigr]\bigr\|^2,
\label{eq:mala_logalpha}
\end{multline}
the form used in practice. The MH update targets $p\propto e^{-S}$ for any proposal drift $s_\theta$. At the chosen step size, the acceptance rate $\langle\min(1,\alpha)\rangle$ therefore probes the local quality of the learned drift and is high when $s_\theta$ is close to the target score.

The step size $h=c/L^D$ and the evaluation time $t_{\text{mh}}$ are calibrated in \App{app:mala_calibration} to maximize the late--early gap in acceptance, giving $c=0.2$ and $t_{\text{mh}}=10^{-4}$ uniformly in 2D and 3D. Reference configurations are drawn from the raw HMC ensemble and pushed through the same linear map before the trial step. At each saved checkpoint we apply a single MH step using the EMA-averaged network weights. The number of reference configurations is specified in each figure or table caption. Using a single step keeps the trials independent and avoids the warm-up bias of multi-step chains.

\begin{figure}[!htbp]
\centering
\includegraphics[width=\columnwidth]{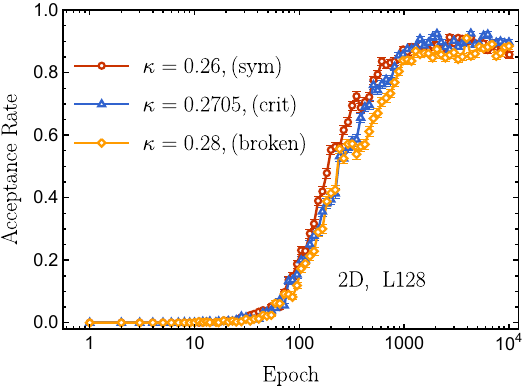}
\caption{MALA acceptance rate vs.\ training epoch for the 2D $\phi^4$ model at $L=128$, evaluated at every saved checkpoint with EMA weights, $c=0.2$ ($h=c/L^2$), $t_{\text{mh}}=10^{-4}$, $N=2048$ reference configurations and a single MH step, for three values of $\kappa$ spanning the phase diagram.}
\label{fig:acc_vs_epoch_2d}
\end{figure}

\fig{fig:acc_vs_epoch_2d} shows the resulting curves at $L=128$. The three $\kappa$ values track each other closely throughout training and saturate near $0.85$--$0.90$ by $\sim 10^3$ epochs, with no visible sensitivity to the phase.

\Tab{tab:acc_early_late} sweeps lattice size. The late-training rate exceeds $\sim 65\%$ at every $L\ge 8$ in both dimensions and at every $\kappa$ tested, and rises gently with $L$. The latter is consistent with our $h\propto L^{-D}$ scaling being more conservative than the optimum for $\phi^4$, which is expected to lie between the i.i.d.\ value $L^{-D/3}$~\cite{Roberts1998} and our choice. The lower late-training rate at $L=4$ reflects the network bottleneck at this lattice size, where the encoder collapses each axis to a single site and only about two of the three resolution levels remain active (\App{app:ncsnpp}).

\begin{table*}[t]
\centering
\caption{MALA acceptance rate of the in-distribution 2D ($\lambda=0.022$) and 3D ($\lambda=0.9$) $\phi^4$ score networks, each trained and evaluated at the same lattice size, in an early and a late training window. The early entries average over the saved checkpoints up to epoch~$50$, and the late entries average over the latest saved checkpoints, at epochs~$5000$--$10\,000$ in 2D and $9000$--$10\,000$ in 3D. All entries use $c=0.2$ ($h=c/L^D$), $t_{\text{mh}}=10^{-4}$, a single MH step and EMA weights. The 2D entries use $N=2048$ reference configurations, while the 3D entries use $N=1024$. Values are quoted as mean$(\text{SEM})$ over the binomial trials in the window, with the digit(s) in parentheses giving the $1\sigma$ uncertainty on the last displayed digit. Dashes mark $(L,\kappa)$ combinations outside the training grid.}
\label{tab:acc_early_late}
\begin{tabular}{lccccccc}
\toprule
                                    &       & $L=4$       & $L=8$       & $L=16$      & $L=32$      & $L=64$      & $L=128$     \\
\midrule
\multirow{2}{*}{3D, $\kappa=0.1923$} & early & $0.061(7)$  & $0.001(1)$  & $0.002(1)$  & $0.032(6)$  & $0.032(6)$  & ---         \\
                                    & late  & $0.371(15)$ & $0.714(14)$ & $0.801(12)$ & $0.891(10)$ & $0.911(9)$  & ---         \\
\midrule
\multirow{2}{*}{2D, $\kappa=0.26$}   & early & ---         & $0.016(1)$  & $0.019(2)$  & $0.008(2)$  & $0.008(2)$  & $0.040(2)$  \\
                                    & late  & ---         & $0.799(5)$  & $0.751(5)$  & $0.824(4)$  & $0.846(6)$  & $0.867(3)$  \\
\midrule
\multirow{2}{*}{2D, $\kappa=0.2705$} & early & ---         & $0.038(1)$  & $0.027(2)$  & $0.003(1)$  & $0.006(1)$  & $0.030(1)$  \\
                                    & late  & ---         & $0.747(7)$  & $0.837(5)$  & $0.763(6)$  & $0.835(3)$  & $0.896(4)$  \\
\midrule
\multirow{2}{*}{2D, $\kappa=0.28$}   & early & ---         & $0.117(5)$  & $0.021(2)$  & $0.005(2)$  & $0.007(1)$  & $0.022(2)$  \\
                                    & late  & ---         & $0.698(13)$ & $0.827(4)$  & $0.762(5)$  & $0.798(5)$  & $0.881(4)$  \\
\bottomrule
\end{tabular}
\end{table*}

\begin{figure}[!htbp]
\centering
\includegraphics[width=\columnwidth]{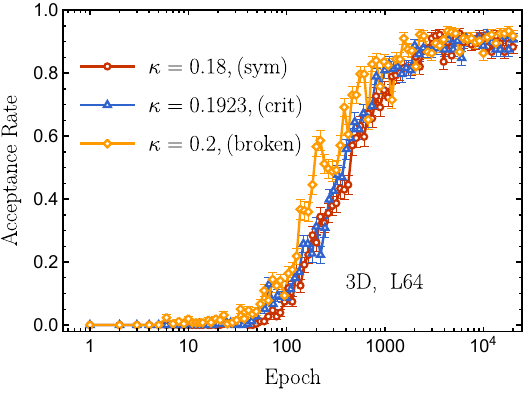}
\caption{MALA acceptance rate vs.\ training epoch for the 3D $\phi^4$ model at $L=64$, $\lambda=0.9$, evaluated at every saved checkpoint with EMA weights, $c=0.2$ ($h=c/L^3$), $t_{\text{mh}}=10^{-4}$, $N=256$ reference configurations and a single MH step, for three values of $\kappa$.}
\label{fig:acc_vs_epoch_3d}
\end{figure}

\fig{fig:acc_vs_epoch_3d} repeats the diagnostic in 3D at $L=64$, $\lambda=0.9$ with $h=c/L^3$ and the same calibrated $c=0.2$, $t_{\text{mh}}=10^{-4}$. The late-training band reproduces the 2D case.

Across both 2D and 3D and for every $L\ge 8$ in our study, the late-training acceptance rate stays above $\sim 65\%$ at all three $\kappa$ values, including the pseudocritical point. The trained score therefore provides a high-acceptance proposal for an exact local MALA correction to $p_0\propto e^{-S}$ even close to criticality, indicating that the learned local drift remains well matched to the target at the tested step size. Acceptance alone does not determine sampling efficiency, which would require autocorrelation measurements for an actual MALA chain and is quantified below for the diffusion ensembles through the effective sample size. It is nevertheless a necessary local check for an efficient corrected sampler at this step size.

\subsection{MSE-based effective sample size}
\label{sec:ess_mse_2d}
The MALA acceptance rate provides a local diagnostic of the score network. For a fixed proposal step size, it probes whether the score gives an accurate local drift. It does not measure the sampling error or residual bias of observables measured on the final DM configurations. For a Markov-chain sampler, the statistical efficiency is usually characterized by the integrated autocorrelation time and the corresponding ESS. This construction is not useful for the present DM ensemble. The DM configurations are obtained from $N_{\rm DM}$ independent Gaussian fields evolved through independent reverse-diffusion trajectories. They are therefore mutually independent and carry essentially no autocorrelation, so an autocorrelation-based ESS would be close to $N_{\rm DM}$ even when the learned reverse dynamics has a residual bias. We therefore use an HMC-referenced MSE-based effective sample size. For each observable $O$, this quantity is the number of independent target-distributed samples that would give the same mean-squared error as the finite DM estimator. The estimated MSE includes the sampling variance of the DM ensemble, the residual bias relative to the HMC reference, and the finite uncertainty of the autocorrelated HMC reference.

\begin{figure*}[!tbp]
\centering
\includegraphics[width=\textwidth]{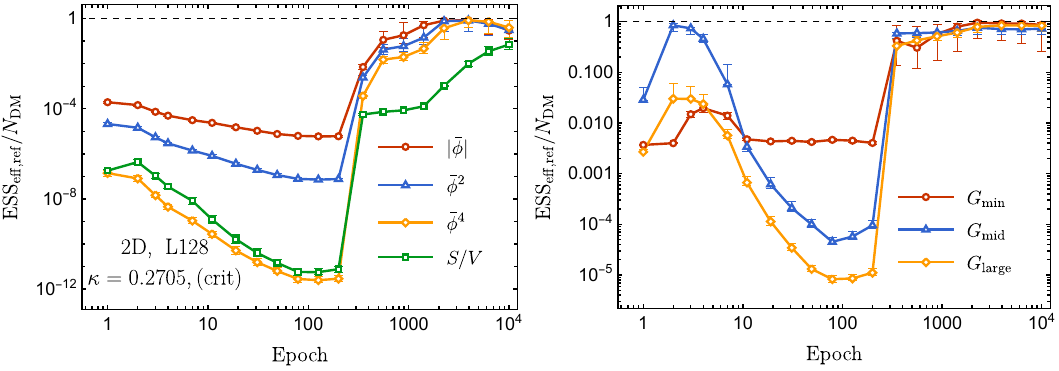}
\caption{Efficiency ratio $N_{\rm eff,ref}/N_{\rm DM}$ vs.\ training epoch for the 2D $\phi^4$ DM ensemble at $L=128$, $\kappa=0.2705$, $\lambda=0.022$, evaluated at every saved checkpoint with EMA weights. The left panel shows scalar moments $\{|\bar\phi|,\bar\phi^2,\bar\phi^4,S/V\}$. The right panel shows diagonal-propagator bands $G_{\rm min},\,G_{\rm mid},\,G_{\rm large}$ with the band $N_{\rm eff,ref}$ built as the uniform mean of per-mode $N_{\rm eff,ref}$. Markers show the full-data point estimate, and error bars show the $[16\%,\,84\%]$ bootstrap percentiles. The bootstrap uses $400$ replicas, with DM rows drawn i.i.d.\ and HMC rows drawn by moving block bootstrap with block $\lceil 2\max_k\tau_{{\rm int},k}^{\rm HMC}\rceil$. The dashed line at unity marks the i.i.d.-target reference line. The pure-shrinkage $N_{\rm eff}/N_{\rm DM}$ from \eq{eq:neff_pure} differs only by the HMC reference variance and is indistinguishable from $N_{\rm eff,ref}/N_{\rm DM}$ on this scale.}
\label{fig:ess_mse_2d}
\end{figure*}

\begin{figure*}[!tbp]
  \centering
  \includegraphics[width=\textwidth]{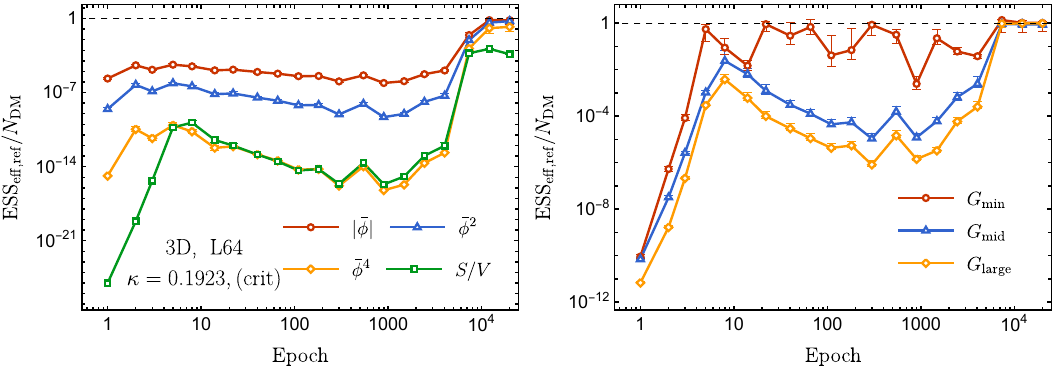}
\caption{Efficiency ratio $N_{\rm eff,ref}/N_{\rm DM}$ vs.\ training epoch for the 3D $\phi^4$ DM ensemble at $L=64$, $\kappa=0.1923$, $\lambda=0.9$. Layout and conventions are as in \fig{fig:ess_mse_2d}, with $N_{\rm DM}=64$ raw DM samples per checkpoint and $N_{\rm HMC}=2560$ HMC configurations.}
  \label{fig:ess_mse_3d}
\end{figure*}

For an observable $O$ under the target distribution $p$, let its target mean and target single-configuration variance be,
\begin{equation}
\begin{aligned}
\mu_O&=\langle O[\phi]\rangle_{\phi\sim p},\\
\sigma_{p,O}^2&=\mathrm{Var}_{\phi\sim p}(O[\phi])=
\left\langle (O[\phi]-\mu_O)^2\right\rangle_{\phi\sim p}.
\label{eq:target_mean_var}
\end{aligned}
\end{equation} 
Given $N$ independent target-distributed configurations $\phi_1,\ldots,\phi_N\sim p$, the corresponding i.i.d.\ target estimator is,
\begin{equation}
\bar O_N=\frac{1}{N}\sum_{i=1}^{N}O[\phi_i].
\label{eq:iid_target_estimator}
\end{equation}
It is unbiased, $\langle \bar O_N\rangle=\mu_O$, and its mean-squared error is,
\begin{equation}
{\rm MSE}^{\rm iid}_{O}(N)=\frac{\sigma_{p,O}^2}{N}.
\label{eq:mse_iid_target}
\end{equation}

For a fixed trained DM, let the mean and single-configuration variance of $O$ in the generated DM ensemble be,
\begin{equation}
\begin{aligned}
\mu_{{\rm DM},O}
&=
\langle O[\phi]\rangle_{\rm DM},\\
\sigma_{{\rm DM},O}^2
&=
\mathrm{Var}_{\rm DM}(O[\phi]).
\end{aligned}
\label{eq:dm_mean_var}
\end{equation}
Given $N_{\rm DM}$ independently generated raw DM configurations, the DM estimator is,
\begin{equation}
\bar O_{\rm DM}
=
\frac{1}{N_{\rm DM}}\sum_{i=1}^{N_{\rm DM}}O[\phi_i].
\label{eq:dm_estimator}
\end{equation}
Its systematic bias relative to the target mean is,
\begin{equation}
\delta_O
=
\mu_{{\rm DM},O}-\mu_O,
\label{eq:dm_bias}
\end{equation}
and its sampling variance is,
\begin{equation}
\mathrm{Var}(\bar O_{\rm DM})
=
\frac{\sigma_{{\rm DM},O}^2}{N_{\rm DM}}.
\label{eq:dm_sampling_variance}
\end{equation}
Therefore the mean-squared error of the DM estimator with respect to the exact target mean is,
\begin{equation}
{\rm MSE}_O^{\rm DM}
=
\left\langle
(\bar O_{\rm DM}-\mu_O)^2
\right\rangle_{\rm DM}
=
\delta_O^2+\mathrm{Var}(\bar O_{\rm DM}).
\label{eq:mse_dm}
\end{equation}
We define the effective sample size by matching the MSE of the DM estimator to the MSE of an i.i.d.\ estimator drawn directly from the target distribution. That is, $N_{{\rm eff},O}$ is defined by,
\begin{equation}
{\rm MSE}^{\rm iid}_{O}(N_{{\rm eff},O})
=
{\rm MSE}_O^{\rm DM}.
\label{eq:neff_matching_condition}
\end{equation}
Using Eqs.~\eqref{eq:mse_iid_target} and \eqref{eq:mse_dm}, this gives,
\begin{equation}
N_{{\rm eff},O}
=
\frac{\sigma_{p,O}^2}{\delta_O^2+\mathrm{Var}(\bar O_{\rm DM})}.
\label{eq:neff_pure}
\end{equation}
Thus $N_{{\rm eff},O}$ is the size of an i.i.d.\ target ensemble whose sample mean would have the same MSE as the DM sample mean for the observable $O$. At this point the finite uncertainty of the HMC reference has not yet been included.

In practice the exact target mean $\mu_O$ is unknown, so we compare the DM estimator against a finite HMC reference mean $\bar O_{\rm HMC}$. The squared difference of the two means mixes the genuine DM bias with the statistical fluctuations of the two finite ensembles. We separate them by subtracting the combined sampling variance from the squared difference and flooring the result at zero, so that a DM-HMC difference smaller than the sampling error is read as no bias. Inserting this estimate of $\delta_O^2$ into \eq{eq:neff_pure}, and adding to the denominator the variance of the HMC reference mean corrected for its autocorrelation, gives the reference-corrected effective sample size $N_{{\rm eff,ref},O}$ plotted in \fig{fig:ess_mse_2d} and \fig{fig:ess_mse_3d}. The explicit estimators, together with the bootstrap and the propagator-band averaging, are given in \App{app:ess_details}.

We estimate the effective sample size for the $\mathbb{Z}_2$-even observables $\{|\bar\phi|,\bar\phi^2,\bar\phi^4,S/V\}$ and for three propagator observables $G_{\rm min}$, $G_{\rm mid}$, and $G_{\rm large}$. These propagator observables correspond to the lowest non-zero momentum, middle momentum, and UV momentum. For the two-dimensional critical ensemble, we draw $N_{\rm DM}=512$ DM configurations at each training epoch shown, using reverse-SDE integration with $2000$ Euler--Maruyama steps and EMA weights.

\fig{fig:ess_mse_2d} shows $N_{{\rm eff,ref},O}/N_{\rm DM}$ for the two-dimensional theory at $L=128$, $\lambda=0.022$, and $\kappa=0.2705\simeq\kappa_c$. We show the ratio rather than $N_{{\rm eff,ref},O}$ to compare directly with $N_{\rm DM}$ independent target samples. A ratio close to one means that the $N_{\rm DM}$ samples from DM have a mean-squared error comparable to $N_{\rm DM}$ independent target samples for the same observable, up to the finite error of the HMC reference. At the beginning of training the DM estimate of $\langle O\rangle$ differs from the HMC estimate, so the estimated bias term dominates and the ratio is small. As training proceeds, the DM-HMC difference becomes comparable to the statistical errors of the two finite ensembles, so the estimated bias is set to zero. After that point the ratio measures the finite-sample precision of the DM estimator relative to independent target sampling. At the end of training, most scalar and propagator observables have ratios close to one. The remaining deviations are small compared with the many-order suppression seen at early training.

\fig{fig:ess_mse_3d} shows the same diagnostic for the three-dimensional critical ensemble at $L=64$, $\lambda=0.9$, and $\kappa=0.1923\simeq\kappa_c$. For this three-dimensional ensemble, we draw $N_{\rm DM}=64$ DM configurations at each training epoch shown. The ratios are again very small at the beginning of training and increase during training. At the end of training, the three propagator bands and the zero-mode moments $|\bar\phi|$ and $\bar\phi^2$ reach ratios close to one, while $\bar\phi^4$ ends a factor of a few below one and the action density saturates two to three orders of magnitude below unity. The suppressed action-density ratio reflects the residual $\langle S\rangle/V$ bias of the three-dimensional in-distribution $L=64$ model identified in \Sec{sec:indist_scalar}. The action density averages over the whole lattice, so its variance under the target distribution is of order $1/V$, and the MSE-matched ESS converts even a small absolute shift of its mean into a large penalty, as in the free-field analysis of \App{app:ess_free_field}. Taken together, \fig{fig:ess_mse_2d} and \fig{fig:ess_mse_3d} show that near the critical point, and on the largest lattices studied here, the final HMC-referenced ESS ratios approach one for the propagator bands and the leading zero-mode moments, while the three-dimensional action density remains limited by its residual bias. Thus, the DM framework, including the score-network architecture, the training procedure, and reverse-SDE sampling, is effective for critical sampling in both two and three dimensions, with the three-dimensional action density as the main exception.
\section{Cross-$L$ generalization}
\label{sec:crossL}

For a local field theory, the interaction carries no explicit dependence on the lattice size. Our fully convolutional score-network architecture is designed to learn a local drift that depends on the field and its neighbors. A well-trained score network can therefore apply the same local transformation at any lattice size with the same learned weights. We test this cross-$L$ generalization by training on a set of smaller lattice sizes and then sampling an unseen larger lattice size~\cite{Mate:2024huc,Zhu:2025pmw}.

\begin{figure*}[!tbp]
  \centering
  \includegraphics[width=0.5\textwidth]{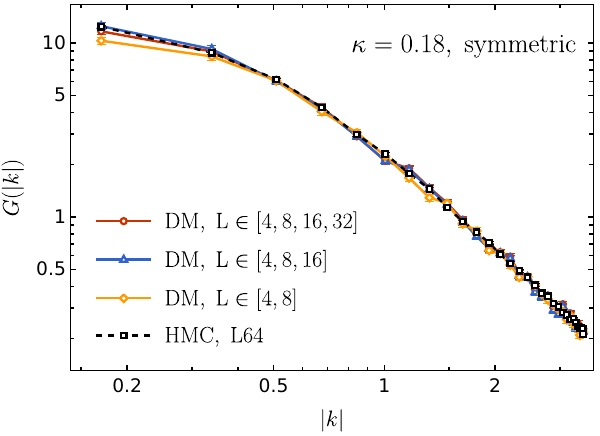}\includegraphics[width=0.5\textwidth]{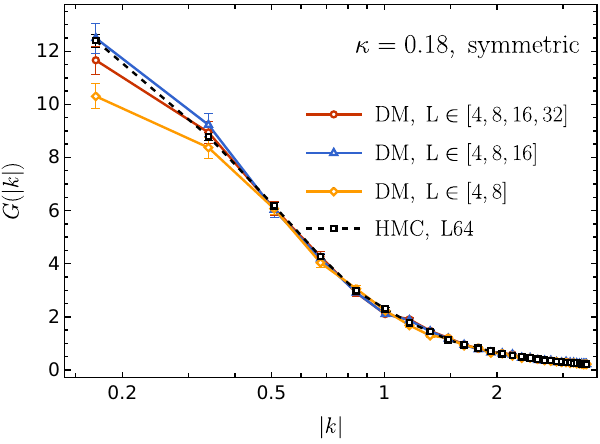}
\caption{Cross-$L$ propagator $G(|k|)$ at $L=64$ for $\kappa=0.18$ (symmetric phase), $\lambda=0.9$. Three multi-$L$ training sets ($\{4,8\}$, $\{4,8,16\}$, $\{4,8,16,32\}$) are compared against the FA-HMC--Wolff reference (dashed black) at the same lattice size. The $\{4,8\}$, $\{4,8,16\}$, and $\{4,8,16,32\}$ models use coarsely chosen noise scales $\sigma=100$, $280$, and $200$, respectively. The left panel uses log-log axes and the right panel uses log-linear axes. Errors are bootstrap $1\sigma$ over configurations.}
  \label{fig:crossL_3d_k018}
\end{figure*}
\begin{figure*}[!tbp]
  \centering
  \includegraphics[width=0.5\textwidth]{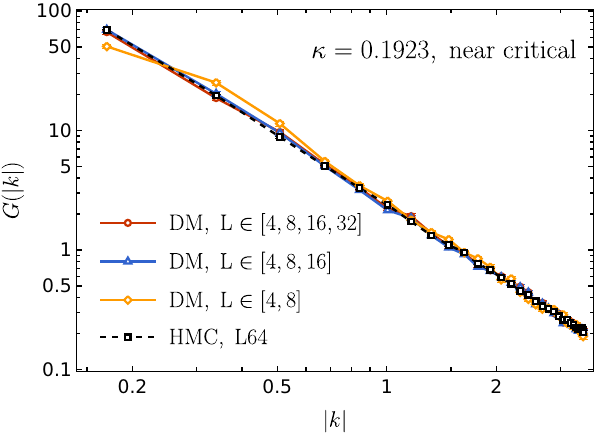}\includegraphics[width=0.5\textwidth]{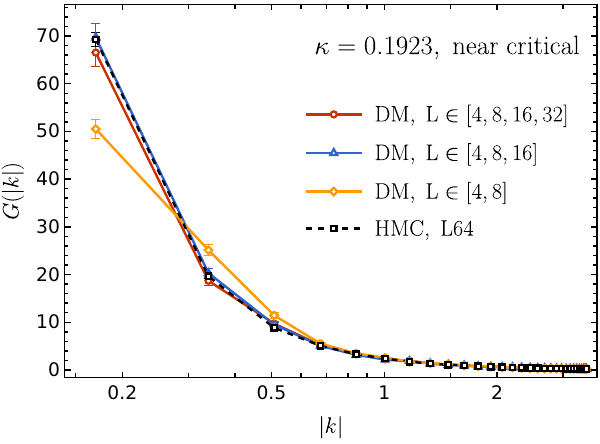}
\caption{Cross-$L$ propagator at $L=64$ for $\kappa=0.1923\simeq\kappa_c$, with the same multi-$L$ training sets as \fig{fig:crossL_3d_k018}. The $\{4,8\}$, $\{4,8,16\}$, and $\{4,8,16,32\}$ models use coarsely chosen noise scales $\sigma=100$, $280$, and $800$, respectively. The left panel uses log-log axes and the right panel uses log-linear axes.}
  \label{fig:crossL_3d_k01923_multi}
\end{figure*}
\begin{figure*}[!tbp]
  \centering
  \includegraphics[width=0.5\textwidth]{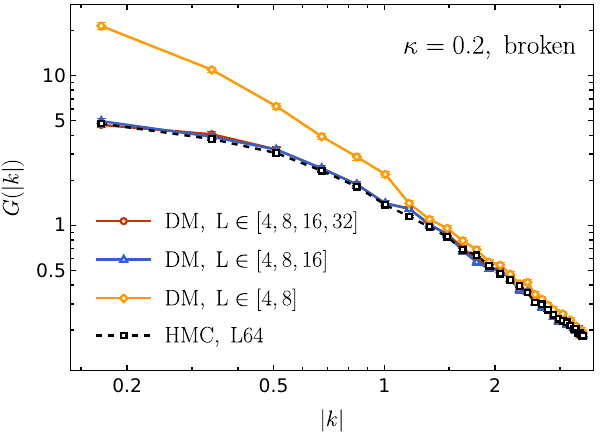}\includegraphics[width=0.5\textwidth]{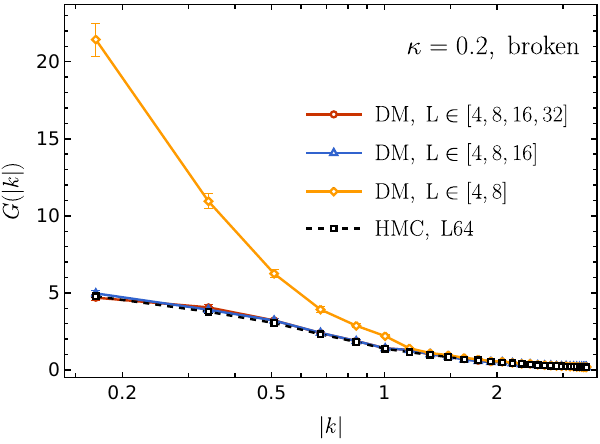}
\caption{Cross-$L$ propagator at $L=64$ for $\kappa=0.2$ (broken phase). The $\{4,8\}$, $\{4,8,16\}$, and $\{4,8,16,32\}$ models use the coarsely chosen noise scales $\sigma=100$, $280$, and $400$, respectively.}
  \label{fig:crossL_3d_k02}
\end{figure*}

We begin with the two-dimensional theory at $\lambda=0.022$. The network is trained on $L\in\{8,16,32,64\}$ and then used to sample the unseen lattice size $L=128$ at the three values of $\kappa$ already used in \Sec{sec:indist_scalar}. The cross-$L$ results in \Tab{tab:multiL_HMC_cumulants} should be read against the in-distribution $L=128$ baseline in \Tab{tab:indist_scalar_L128}, which is trained and sampled at the target size. In the symmetric phase, the cross-$L$ and in-distribution $L=128$ models both reproduce the HMC zero-mode observables, while the cross-$L$ model is closer to HMC for the action density and for the listed single-site cumulants. In the broken phase, the cross-$L$ model also improves the comparison for the susceptibility, $\chi=5.0(1)$ instead of the in-distribution $L=128$ value $\chi=5.44(18)$, and for the higher zero-mode and single-site cumulants. The main exception is the near-critical point. There the in-distribution $L=128$ model tracks the HMC susceptibility, $\chi=752(17)$ versus $749(9)$, whereas the cross-$L$ model gives $\chi=673(20)$ and underestimates the higher single-site cumulants. Thus, away from criticality, the two-dimensional cross-$L$ model does not merely remain competitive with in-distribution $L=128$ training, but gives modest improvements across many of the scalar and single-site entries listed in the tables. The leading loss is confined to the critical zero-mode.

We turn to three dimensions, where we compare the cross-$L$ ensembles with HMC through the scalar observables and then the momentum-space propagator, following the order of the in-distribution analysis in \Sec{sec:indist}. We train on the lattice size subsets $L\in\{4,8\}$, $\{4,8,16\}$ and $\{4,8,16,32\}$ at $\lambda=0.9$, and sample at the unseen lattice size $L=64$ at three values of $\kappa$ spanning the 3D Ising transition. At criticality we add single-size extrapolation baselines, each trained on a single lattice size $L\in\{4,8,16,32\}$ and extrapolated to $L=64$, to disentangle the effect of seeing multiple lattice sizes from the effect of seeing one particular largest lattice size. The noise scales $\sigma$ in \Eq{eq:ve_geometric_g} for these cross-$L$ runs are coarse choices rather than results of the calibration used for the in-distribution $L=64$ baselines. Each is fixed by applying the prescription of \Eq{eq:ve_sigma_max_prescription} to the ensembles of its own training set, so no $L=64$ configurations entered the selection.

In the symmetric phase, cross-$L$ training brings the action density and single-site cumulants much closer to HMC than the in-distribution $L=64$ baseline of \Tab{tab:indist_scalar_L64_3d}. The zero-mode sector at $\kappa=0.18$ behaves differently. The $\{4,8,16,32\}$ model overestimates $\langle|\bar\phi|\rangle$ and $\chi$ by roughly a factor of two, $0.0121(3)$ against the HMC $0.00571(9)$ and $11.3(6)$ against $5.0(2)$, and gives a Binder cumulant $U_L=0.40(2)$ against the HMC $-0.02(3)$. The smaller training sets deviate much less. These relative deviations are large only because the zero-mode observables are themselves small in the symmetric phase. In absolute terms the excess corresponds to a root-mean-square magnetization of $0.014$ against $0.007$ in the reference, and the $\chi$ excess of about $6$ is one percent of the near-critical susceptibility $\chi\simeq620$. The shift is a small absolute distortion of the magnetization distribution and has no visible effect on the finite-momentum propagator (\fig{fig:crossL_3d_k018}). At the pseudocritical point, the $\{4,8,16,32\}$ model is closer to HMC than the in-distribution $L=64$ model for $\langle|\bar\phi|\rangle$, $\chi$, the action density, the higher zero-mode cumulants, and most single-site cumulants, in line with the propagator comparison presented below.

The unfiltered comparison in \Tab{tab:multiL_HMC_cumulants_nofilter} strengthens the case for multi-$L$ training. Although the same support filters of \App{app:numerical_tables} are applied to every DM ensemble, they are needed mainly by the in-distribution $L=64$ run, whose zero-mode results degrade sharply once they are removed. At $\kappa=0.1923\simeq\kappa_c$, the largest cross-$L$ model is essentially unchanged, moving from the filtered $\chi=694(30)$ to $\chi=698(22)$ and remaining far closer to the HMC value $621(15)$ than the in-distribution $L=64$ result, which inflates to $\chi=1.31(26)\times10^3$. In the broken phase the largest cross-$L$ model changes from $\chi=9.3(6)$ to $13.1(6)$, against $4.9(1)$ in HMC, while the in-distribution $L=64$ result changes from $7.8(5)$ to $23.6(9)$. The support filters therefore do not create the observed cross-$L$ gain.

The broken phase reinforces this picture away from the zero-mode. Once $L=16$ is included, the action density and several single-site cumulants move close to the reference values, and the momentum-space propagator follows HMC over the plotted range (\fig{fig:crossL_3d_k02}). The largest cross-$L$ model also improves $\langle|\bar\phi|\rangle$ and the higher zero-mode cumulants relative to the in-distribution $L=64$ run. The susceptibility remains the main exception. With the support filters it is comparable to the in-distribution run, and without them it is substantially less distorted than the in-distribution run.

Turning to the momentum-space propagator, in the symmetric phase $\kappa=0.18$ the multi-$L$ propagators are already close to HMC over the full momentum range (\fig{fig:crossL_3d_k018}). The ultraviolet and intermediate momenta are insensitive to the training set. The visible differences are confined to the lowest momenta, where $\{4,8\}$ undershoots and the larger training sets track HMC more closely. This is the first indication that the finite momentum sector can transfer cleanly even when the largest training lattice is still up to a factor of eight smaller than the target size.

\begin{figure}[!htbp]
\centering
\includegraphics[width=0.45\textwidth]{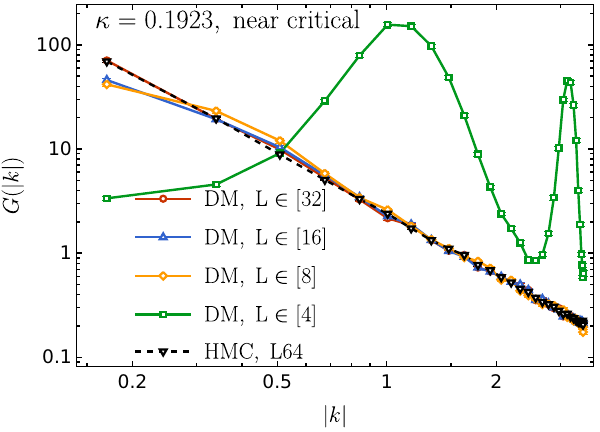}
\caption{Single-size extrapolation baselines at $\kappa=0.1923$. Networks trained on a single lattice size $L\in\{4,8,16,32\}$ and evaluated at $L=64$ degrade strongly for the smaller training sizes, while the $L=32$ model comes closest to the HMC reference. The comparison with the multi-$L$ model in \fig{fig:crossL_3d_k01923_multi} shows that the transfer is not controlled by the largest training size alone.}
\label{fig:crossL_3d_k01923_single}
\end{figure}

At the pseudocritical point $\kappa\approx\kappa_c$, the advantage of multi-$L$ training is most apparent. The $\{4,8,16,32\}$ network follows HMC from the ultraviolet down to $k_{\min}$, although it has never seen an $L=64$ configuration (\fig{fig:crossL_3d_k01923_multi}). The $\{4,8,16\}$ network is also close over most of the range, whereas the $\{4,8\}$ network misses part of the lowest momentum rise and develops a small excess at intermediate momenta. The single-size extrapolation baselines in \fig{fig:crossL_3d_k01923_single} show a different pattern. The $L=4$ model is strongly distorted, $L=8$ and $L=16$ undershoot the infrared, and the largest single training size ($L=32$) is closest to the target $L=64$. Thus, near criticality even a single training lattice can extrapolate to $L=64$, while the multi-$L$ models remain the most accurate.

In the broken phase $\kappa=0.2$, the $\{4,8\}$ network strongly overestimates the propagator at low and intermediate momenta (\fig{fig:crossL_3d_k02}), consistent with the large susceptibility in \Tab{tab:multiL_HMC_cumulants}. Adding $L=16$ removes most of this excess, and the $\{4,8,16\}$ and $\{4,8,16,32\}$ networks track HMC over the plotted momenta.

The MALA acceptance rate diagnostic of \Sec{sec:acceptance}, applied directly at the unseen lattice size $L=64$, sees the same hierarchy at the level of the score itself (\Tab{tab:acc_multiL_3d}). At epoch $50$, the acceptance is below $0.3\%$ for every training set and coupling. By epoch $10000$, it rises from $0.58$--$0.63$ for $\{4,8\}$ to $0.80$--$0.83$ for $\{4,8,16\}$ and $0.86$--$0.88$ for $\{4,8,16,32\}$. The cross-$L$ noise scales are coarse choices rather than the calibrated values used for the in-distribution $L=64$ baselines, so the acceptance rates test the multi-$L$ prescription without fine tuning. Even so, the largest cross-$L$ model reaches a MALA compatibility within a few percentage points of the in-distribution $L=64$ baseline. The test is stringent because the score is evaluated at a lattice size never seen in training, yet still provides a high-acceptance local Metropolis drift for the $L=64$ Boltzmann target.

\begin{table}[t]
\centering
\caption{MALA acceptance rate at the unseen lattice size $L=64$ for the 3D cross-$L$ models. Rows compare an early checkpoint (epoch $50$) with a late checkpoint (epoch $10000$). The diagnostic uses the same calibrated step size and diffusion time as \fig{fig:acc_vs_epoch_3d}, with $c=0.2$, $h=c/L^3$, $t_{\text{mh}}=10^{-4}$, $N=1024$ reference configurations and a single MH step. Each entry is the mean over five random seeds, with the uncertainty the standard error across seeds. None of the training sets contains $L=64$, so every entry tests the network at a lattice size it never saw during training. The listed cross-$L$ noise scales follow the pairwise-distance prescription applied to the training-lattice ensembles rather than the calibration at the target size used for the in-distribution $L=64$ baselines. The $\{4,8\}$ and $\{4,8,16\}$ models use $\sigma=100$ and $280$ at all three couplings. The $\{4,8,16,32\}$ model uses $\sigma=200$, $800$, and $400$ at $\kappa=0.18$, $0.1923$, and $0.2$, respectively.}
\label{tab:acc_multiL_3d}
\begin{tabular}{llccc}
\toprule
training set &       & $\kappa=0.18$ & $\kappa=0.1923$ & $\kappa=0.2$ \\
\midrule
\multirow{2}{*}{$\{4,8\}$}       & early & $0.0023(4)$ & $0.0010(4)$ & $0.0004(4)$ \\
                                 & late  & $0.631(10)$ & $0.578(8)$  & $0.628(8)$  \\
\midrule
\multirow{2}{*}{$\{4,8,16\}$}    & early & $0.0016(4)$ & $0.0004(2)$ & $0.0020(5)$ \\
                                 & late  & $0.803(8)$  & $0.830(6)$  & $0.826(7)$  \\
\midrule
\multirow{2}{*}{$\{4,8,16,32\}$} & early & $0.0029(7)$ & $0.0002(2)$ & $0.0014(5)$ \\
                                 & late  & $0.861(3)$  & $0.877(6)$  & $0.859(6)$  \\
\bottomrule
\end{tabular}
\end{table}

The MALA update of \Sec{sec:acceptance} can also be promoted from a diagnostic to an exact refinement of the generated ensembles. The rows labeled DM\,$\{4,8,16,32\}$+MALA in \Tab{tab:multiL_HMC_cumulants} evolve the filtered cross-$L$ ensembles with an exact composite Markov chain targeting $p_0\propto e^{-S}$, constructed in \App{app:numerical_tables}, with the Langevin drift given either by the learned score or by the exact action gradient $-\nabla S$. Both variants keep the exact action in the Metropolis test, so both are exact for $e^{-S}$. They differ in mixing. Away from criticality the two agree and reach the HMC values within errors. At $\kappa=0.1923$ they separate. The exact-gradient chain reproduces the HMC zero mode, $\langle|\bar\phi|\rangle=0.116(2)$ and $\chi=642(23)$ against $0.1165(9)$ and $621(15)$, while the score chain stalls at $\langle|\bar\phi|\rangle=0.126(2)$ and $\chi=678(26)$. The gap does not close with chain length. It reflects the residual error of the learned drift in the long-wavelength modes, which enters the acceptance and slows the near-critical zero-mode relaxation. The exact gradient carries no such error and equilibrates the zero mode at the same chain length. On the lattice this gradient is a cheap local operation, so the $(-\nabla S)$ variant is the accurate choice, while the (score) variant measures how far the learned drift alone can drive an exact local sampler.

Taken together these results show that multi-$L$ training improves the accuracy of the generated ensembles and provides a route to generation across lattice sizes. The learned score is not tied to the lattice size on which the network was trained. A fully convolutional score network trained only on smaller lattices can be evaluated at a larger unseen lattice size and used to generate configurations there. In the present tests this transfer reproduces the momentum-space propagator, improves many single-site and zero-mode observables relative to the in-distribution baselines, and yields a high-acceptance MALA drift at the target lattice size. The unfiltered comparison in \App{app:numerical_tables} gives further support to this conclusion for the most sensitive three dimensional zero-mode observables. Removing the support filters enlarges the susceptibility of the in-distribution $L=64$ sampler much more than that of the largest cross-$L$ sampler at the critical and broken couplings. The practical consequence is that expensive training at the largest lattice size~\cite{Abbott:2022zsh} is not always required. Smaller lattices can provide the training data needed to learn a score that extrapolates to larger lattices.

\section{Summary and outlook}
\label{sec:summary}

We have investigated variance-exploding score-based diffusion models as generative samplers for lattice $\phi^4$ theory in two and three dimensions, validating the generated ensembles against FA-HMC--Wolff references across the symmetric, near-critical, and broken regimes. On the trained lattice sizes, the reverse-SDE sampler reproduces the order parameter, the Binder cumulant, and the higher zero-mode cumulants, together with the joint distribution of magnetization and action density, the translation-invariant single-site cumulants, and the momentum-space propagator. The largest remaining discrepancies are infrared. In the broken phases the two $\mathbb{Z}_2$-related peaks sit at the correct magnetizations, and the leading bias is that the peaks are too wide, which shows up most directly in the susceptibility. At the three-dimensional near-critical point the in-distribution $L=64$ model shifts $\langle|\bar\phi|\rangle$, $\chi$, and the higher zero-mode cumulants toward the ordered side, while the single-site cumulants stay much closer to HMC. The three-dimensional action density is also too high, by up to about twenty percent. This shift comes from a few-percent bias in the single-site moments, acting through the cancellations at $\lambda=0.9$. The dominant limitation is therefore the zero-mode, with a small single-site moment bias that is amplified only in three dimensions, while the finite-momentum sector remains well controlled.

The diagnostics in \Sec{sec:diagnostics} probe complementary aspects of the sampler. The small-diffusion-time score test compares $s_\theta(\phi,t)$ with $-\nabla_\phi S[\phi]$ on reference Monte Carlo configurations and shows that the direction of the action gradient is learned early, while its overall magnitude converges more slowly. The MALA acceptance test uses the same score as a proposal drift for an exact Boltzmann update. It reaches high values for the converged in-distribution models and increases systematically with the amount of cross-$L$ training data at unseen lattice sizes. The HMC-referenced effective sample size combines statistical variance and residual bias into an observable-level measure of ensemble quality. Its training curves show the same convergence seen in the near-critical scalar and propagator comparisons, with both classes of observables moving from bias-dominated early checkpoints to ESS ratios close to the independent-target reference line at late checkpoints.

The cross-$L$ analysis provides the central evidence for cross-$L$ generalization. Because the score network is fully convolutional and uses circular padding, the same weights define a map on the lattice sizes used in this study, including $L=4$ and the larger multiples of eight. This structural property becomes useful when the training data contain enough finite volume information. In two dimensions, the multi-$L$ model matches or modestly improves in-distribution $L=128$ training in the symmetric and broken phases, with the main degradation confined to the near-critical zero-mode. In three dimensions, the model trained on $L\in\{4,8,16,32\}$ can be evaluated at the unseen lattice size $L=64$ and reproduces the propagator from the ultraviolet down to the lowest diagonal momentum. At the pseudocritical point it also improves several zero-mode and single-site observables relative to the in-distribution $L=64$ model. The unfiltered appendix tables show that this comparison is not an artifact of the support filters. The single-size extrapolation baselines show that the severe propagator failure occurs when the network is trained only at $L=4$ and then evaluated at $L=64$. The multi-$L$ $\{4,8\}$ model is instead a weaker but still meaningful transfer case. Its main residual error is a low momentum excess in the broken phase, visible also in the susceptibility, and adding $L=16$ removes most of this effect. In the symmetric phase the largest training set overestimates the small zero-mode observables by about a factor of two, an absolute distortion that is minute on the scale of the near-critical zero-mode fluctuations. The transfer at full accuracy is demonstrated up to a factor of two in linear size beyond the largest training lattice. Larger ratios, probed by the smaller training sets and by the single-size baselines, show increasing infrared degradation.

Several directions remain open. The present work uses a plain VE-SDE implementation in which the noise scale $\sigma$ remains a practical hyperparameter. Modern diffusion formulations such as EDM and EDM2~\cite{Karras2022EDM,Karras2024EDM2} provide improved parameterizations, schedules, and training stabilization, and applying them to the present cross-$L$ setting should be a direct way to improve the generated ensembles, while reducing the number of score evaluations per configuration from the 2000 steps used here to a few tens. Our preliminary work on EDM2 indicates that it can remove the residual zero-mode bias found here. Noise schedules inspired by renormalization-group diffusion models~\cite{Masuki:2025pfx} are another promising route, especially for the zero-mode and other near-global components that dominate the remaining bias.

A second direction is to use diffusion samples as starting points for exact Monte Carlo evolution. The generated configurations are already close to thermal equilibrium in the observables studied here. They can therefore initialize a modern HMC or HMC-like sampler, after which the exact accept-reject step removes residual diffusion-model bias while preserving much of the equilibration gain supplied by the learned sampler.

The present study concerns equilibrium field configurations. In future work the physical time direction could also be embedded into the network architecture, so that diffusion models learn dynamical evolution of the field rather than only static configurations. This would open a route to real-time~\cite{Berges:2006xc,Tan:2021zid,Tan:2024fuq} and non-equilibrium~\cite{Tan:2025bsv} field-theory applications.

Finally, the cross-$L$ results motivate applications to full QCD with dynamical fermions. The long-term goal is to train on smaller lattices where reference ensembles are cheaper, and then generate larger-volume full-QCD gauge-field configurations with the same learned local structure. Achieving this will require gauge equivariant architectures~\cite{Aarts:2026zzr}, fermion-aware conditioning~\cite{Abbott:2022zhs}, and MALA correction, but the present scalar theory shows that lattice-size extrapolation of the learned score can already be tested quantitatively.

\begin{acknowledgments}
We thank Yuto Ashida, Andreas Ipp, Kanta Masuki, Jan Pawlowski and Thomas Ranner and other DM-QFT collaboration members for insightful discussions.
We thank the DEEP-IN working group at RIKEN-iTHEMS for support in the preparation of this paper.
LW and YT are supported by the JST-BOOST Grant No.\ JPMJBY24H9. LW is also supported by the RIKEN-TRIP initiative (RIKEN-Quantum) and JSPS KAKENHI Grant No.\ 25H01560.
GA and BL are supported by STFC Consolidated Grant ST/X000648/1. 
GA is also supported by a Royal Society Leverhulme Trust Senior Research Fellowship.
BL is supported in part by the STFC Consolidator Grant No.\ ST/X00063X/1.
DH is supported by the UKRI AIMLAC CDT EP/S023992/1.
\end{acknowledgments}

\section*{Code availability}
The code accompanying this work is publicly available in the \href{https://github.com/Yangyang-Tan/DM-for-phi4}{DM-for-phi4} repository~\cite{DM-for-phi4}.

\appendix

\section{Detailed numerical tables}
\label{app:numerical_tables}

This appendix collects the large numerical tables used in \Sec{sec:indist_scalar} and \Sec{sec:crossL}. The main text keeps only the trends needed for the argument.

For \Tab{tab:indist_scalar_L128}, the HMC rows use the raw reference ensembles. The two-dimensional DM samples use epoch $10000$ reverse-SDE ensembles with $2000$ EM steps on a logarithmic time grid and $N_{\rm DM}=2048$ configurations before any augmentation. The two-dimensional HMC rows use the full reference chain of $N_{\rm HMC}=10240$ configurations. The rows labeled HMC ($N{=}2048$) use every fifth configuration of the same chain, matching $N_{\rm DM}$. The three dimensional DM samples use epoch $20000$ in-distribution $L=64$ networks with $\sigma=2000,2048,2760$ at $\kappa=0.18,0.1923,0.2$ and $2000$ EM steps on a linear grid. Before any observable is measured, each raw ensemble of $512$ DM configurations is passed through two support filters defined from the $L=64$ HMC reference at the same $\kappa$, while the HMC reference itself is never filtered. A configuration is rejected if any lattice site lies outside the HMC field range $[\phi_{\min},\phi_{\max}]$ (site support), or if its volume average $|M|=|\langle\phi\rangle_V|$ exceeds the largest $|M|$ in the HMC ensemble (magnetization support). The first criterion removes reverse-SDE trajectories that drifted into a field range never seen during training. The second removes a late training $\mathbb{Z}_2$ branch collapse artifact that drives the order parameter outside its physical range. For these in-distribution ensembles the rejected counts are $11$, $37$ and $200$ out of $512$ at $\kappa=0.18,0.1923,0.2$. The large rejection in the broken phase at $\kappa=0.2$ is dominated by the magnetization criterion, which accounts for $193$ of the $200$. The error estimates differ between the two dimensions. The two-dimensional errors are $50$ bin block jackknife over configurations, while the three dimensional errors are bootstrap $1\sigma$ over configurations using the full $N_{\rm HMC}=2560$ HMC ensemble. In both dimensions each single-site cumulant is evaluated per site and then averaged over sites, following \Eq{eq:site_cumulants_avg}. This per-site estimate carries a finite-sample bias that grows with the cumulant order. The higher single-site cumulants of the DM ensembles should therefore be read against the HMC rows at matched $N$. The common observables in \Tab{tab:indist_scalar_L128} and \Tab{tab:multiL_HMC_cumulants} are ordered identically. \Tab{tab:multiL_HMC_cumulants} additionally lists $\kappa^{(0)}_2$ and $\langle|\phi|\rangle$.

\begin{table*}[!tbp]
\centering
\caption{Scalar observables of the in-distribution $\phi^4$ diffusion model, trained and sampled at the same lattice size, compared with HMC in 2D ($L=128$, $\lambda=0.022$) and 3D ($L=64$, $\lambda=0.9$). The DM ensembles hold $N_{\rm DM}=2048$ configurations in 2D and $N_{\rm DM}=512$ in 3D, the latter before the support filters described in the text.}
\label{tab:indist_scalar_L128}\label{tab:indist_scalar_L64_3d}
\resizebox{\textwidth}{!}{%
\begin{tabular}{llcccccccccccc}
\toprule
 & & \multicolumn{4}{c}{Global / action} & \multicolumn{3}{c}{Zero-mode cumulants $\kappa^{(0)}_n$} & \multicolumn{5}{c}{Single-site cumulants} \\
\cmidrule(lr){3-6} \cmidrule(lr){7-9} \cmidrule(lr){10-14}
$\kappa$ & source & $\langle|\bar\phi|\rangle$ & $\chi$ & $\langle S\rangle/V$ & $U_L$ & $\kappa^{(0)}_4$ & $\kappa^{(0)}_6$ & $\kappa^{(0)}_8$ & $\kappa_2^{(\mathrm{site})}$ & $\kappa_4^{(\mathrm{site})}$ & $\kappa_6^{(\mathrm{site})}$ & $\kappa_8^{(\mathrm{site})}$ & $U_4^{(\mathrm{site})}$ \\
\midrule
\multicolumn{14}{l}{\textit{2D}, $L=128$, $\lambda=0.022$} \\
\midrule
$0.26$ & HMC & $0.0315(2)$ & $9.2(2)$ & $0.45120(7)$ & $0.01(2)$ & $-1(1)\!\times\!10^{-7}$ & $1(1)\!\times\!10^{-9}$ & $1(1)\!\times\!10^{-11}$ & $0.9105(2)$ & $-0.2726(3)$ & $0.602(1)$ & $-3.19(1)$ & $0.1095(1)$ \\
 & HMC ($N{=}2048$) & $0.0314(5)$ & $9.2(3)$ & $0.4513(1)$ & $0.01(3)$ & $-1(2)\!\times\!10^{-7}$ & $-2(2)\!\times\!10^{-9}$ & $5(16)\!\times\!10^{-12}$ & $0.9103(4)$ & $-0.2744(7)$ & $0.610(3)$ & $-3.35(4)$ & $0.1100(2)$ \\
       & DM  & $0.0316(5)$ & $9.32(40)$ & $0.4531(1)$ & $0.01(3)$ & $-0.6(2.2)\!\times\!10^{-7}$ & $-1.5(1.8)\!\times\!10^{-9}$ & $1.5(2.1)\!\times\!10^{-11}$ & $0.8977(4)$ & $-0.2565(6)$ & $0.550(3)$ & $-2.92(3)$ & $0.1053(2)$ \\
\midrule
$0.2705$ & HMC & $0.453(4)$ & $749(9)$ & $0.3984(1)$ & $0.489(4)$ & $-0.092(3)$ & $0.153(8)$ & $-0.55(4)$ & $1.420(1)$ & $-1.436(4)$ & $8.85(4)$ & $-121.5(7)$ & $0.2371(3)$ \\
 & HMC ($N{=}2048$) & $0.448(5)$ & $776(15)$ & $0.3984(2)$ & $0.484(6)$ & $-0.089(4)$ & $0.15(1)$ & $-0.52(5)$ & $1.419(1)$ & $-1.434(5)$ & $8.84(5)$ & $-121.8(9)$ & $0.2368(4)$ \\
         & DM  & $0.451(6)$ & $752(17)$ & $0.3998(2)$ & $0.487(6)$ & $-0.0905(45)$ & $0.149(11)$ & $-0.531(54)$ & $1.4160(13)$ & $-1.418(4)$ & $8.70(4)$ & $-119.3(8)$ & $0.2350(3)$ \\
\midrule
$0.28$ & HMC & $1.4558(2)$ & $4.22(6)$ & $0.23817(6)$ & $0.666505(2)$ & $-8.983(4)$ & $152.3(1)$ & $-5487(5)$ & $2.6899(4)$ & $-9.801(4)$ & $171.1(1)$ & $-6397(5)$ & $0.45146(5)$ \\
 & HMC ($N{=}2048$) & $1.4558(4)$ & $4.2(2)$ & $0.2383(1)$ & $0.666505(6)$ & $-8.984(9)$ & $152.3(2)$ & $-5488(11)$ & $2.6902(9)$ & $-9.807(9)$ & $171.3(2)$ & $-6415(12)$ & $0.4514(1)$ \\
       & DM  & $1.4528(4)$ & $5.44(18)$ & $0.2398(2)$ & $0.666470(7)$ & $-8.928(9)$ & $150.9(2)$ & $-5420(11)$ & $2.6860(11)$ & $-9.754(10)$ & $169.9(3)$ & $-6343(13)$ & $0.45040(9)$ \\
\midrule
\multicolumn{14}{l}{\textit{3D}, $L=64$, $\lambda=0.9$} \\
\midrule
$0.18$ & HMC & $0.00571(9)$ & $5.0(2)$ & $-0.066692(25)$ & $-0.02(3)$ & $1.71(260)\!\times\!10^{-10}$ & $-6.25(604)\!\times\!10^{-14}$ & $6.85(2151)\!\times\!10^{-18}$ & $0.58604(2)$ & $-0.40116(3)$ & $1.3731(2)$ & $-10.237(2)$ & $0.38894(1)$ \\
 & DM & $0.0062(2)$ & $6.3(3)$ & $-0.05382(19)$ & $-0.10(5)$ & $1.15(51)\!\times\!10^{-9}$ & $-4.28(167)\!\times\!10^{-13}$ & $-2.37(937)\!\times\!10^{-17}$ & $0.6003(2)$ & $-0.4246(2)$ & $1.504(1)$ & $-11.65(2)$ & $0.39068(3)$ \\
\midrule
$0.1923$ & HMC & $0.1165(9)$ & $621(15)$ & $-0.116709(65)$ & $0.516(4)$ & $-3.93(12)\!\times\!10^{-4}$ & $4.33(19)\!\times\!10^{-5}$ & $-1.03(6)\!\times\!10^{-5}$ & $0.61815(4)$ & $-0.46164(7)$ & $1.7020(4)$ & $-13.640(4)$ & $0.40231(2)$ \\
 & DM & $0.147(2)$ & $755(49)$ & $-0.09754(33)$ & $0.559(7)$ & $-0.00101(6)$ & $1.78(17)\!\times\!10^{-4}$ & $-6.77(85)\!\times\!10^{-5}$ & $0.6409(4)$ & $-0.4997(5)$ & $1.927(3)$ & $-16.23(3)$ & $0.40324(4)$ \\
\midrule
$0.2$ & HMC & $0.51623(9)$ & $4.9(1)$ & $-0.247603(52)$ & $0.666574(2)$ & $-0.14204(10)$ & $0.3028(3)$ & $-1.372(2)$ & $0.70104(4)$ & $-0.64717(7)$ & $2.8696(5)$ & $-27.624(6)$ & $0.43652(1)$ \\
 & DM & $0.5213(2)$ & $7.8(5)$ & $-0.23195(28)$ & $0.666522(9)$ & $-0.147(1)$ & $0.317(6)$ & $-1.45(5)$ & $0.7175(7)$ & $-0.6806(6)$ & $3.099(4)$ & $-30.69(6)$ & $0.43727(3)$ \\
\bottomrule
\end{tabular}%
}
\end{table*}

For \Tab{tab:multiL_HMC_cumulants}, all quantities are evaluated in the physical field space. The two-dimensional block reports raw HMC at $L=128$, $\lambda=0.022$ and cross-$L$ DM\,$\{8,16,32,64\}$ $\mathbb{Z}_2$ symmetrized ensembles. The two-dimensional DM ensembles hold $N_{\rm DM}=2048$ configurations before augmentation. The two-dimensional HMC reference is shown both at the full chain $N=10240$ and at a matched $N=2048$ subsample built from every fifth configuration. The three dimensional block reports $L=64$, $\lambda=0.9$ HMC and cross-$L$ DM ensembles. The 3D DM samples are drawn from multi-$L$ epoch $10000$ using EM sampling on a linear time grid with $2000$ steps. Each raw DM ensemble holds $N_{\rm DM}=512$ configurations and is passed through the same two support filters used for \Tab{tab:indist_scalar_L128}, here defined from the $L=64$ HMC reference at the matching $\kappa$. Across the nine multi-$L$ ensembles the rejected counts are small, at most $21$ out of $512$, with the single exception of $\{4,8,16,32\}$ at $\kappa=0.2$, where $43$ out of $512$ are rejected, dominated by the magnetization criterion. The 3D HMC reference is shown both at full $N=2560$ and at a matched $N=512$ subsample. Cumulants follow the recursion of \cite{Aarts:2024rsl}. Errors are bootstrap $1\sigma$ over configurations. The single-site cumulants are evaluated per site and then averaged over sites, following \Eq{eq:site_cumulants_avg}.

The rows labeled DM\,$\{4,8,16,32\}$+MALA refine the filtered DM ensembles with an exact Markov chain whose stationary distribution is $p_0\propto e^{-S}$. Each retained configuration, $503$, $509$ and $469$ of $512$ at $\kappa=0.18$, $0.1923$ and $0.2$, initializes an independent chain of $5000$ composite updates. One update combines a Metropolis-adjusted Langevin step, a global shift $\phi\to\phi+\delta$ every second step, with Gaussian $\delta$ of width $2.4\sqrt{\chi_{\rm HMC}/V}$ and acceptance $\min(1,e^{-\Delta S})$, and an embedded Swendsen--Wang cluster update in the Brower--Tamayo embedding every tenth step~\cite{Swendsen:1987ce,Wolff:1988uh,Brower:1989mt}, which freezes aligned nearest-neighbor bonds with probability $1-e^{-4\kappa\phi_x\phi_{x+\mu}}$ and flips each cluster sign with probability $1/2$. The shift targets the zero mode and the cluster update the $\mathbb{Z}_2$ sign sector, the two collective directions that the local step moves slowly. All three kernels obey detailed balance with respect to $e^{-S}$, so the composite chain is exact for any drift in the Langevin proposal and its output needs no support filter.

The two variants differ only in the Langevin drift. The rows labeled (score) use the learned score $s_\theta(\phi,t_{\text{mh}})$ with $t_{\text{mh}}=10^{-4}$ and $h=c/L^{3}$ at $c=3$, the same drift as the acceptance-rate diagnostic of \Sec{sec:acceptance}. The rows labeled $(-\nabla S)$ use the exact action gradient at $c=100$, a step size the accurate gradient supports. Both keep the exact action in the Metropolis test, so both are exact for $e^{-S}$. The comparison of the two chains, and the resulting interpretation of the residual score error, are discussed in \Sec{sec:crossL}.

\begin{table*}[!tbp]
\centering
\caption{Cross-$L$ scalar observables and cumulants in 2D and 3D, comparing HMC with multi-$L$ DM models. The DM ensembles hold $N_{\rm DM}=2048$ configurations in 2D and $N_{\rm DM}=512$ in 3D, the latter before the support filters described in the text. The 3D block additionally shows the largest cross-$L$ ensemble after the exact MALA refinement described in the text.}
\label{tab:multiL_HMC_cumulants}
\resizebox{\textwidth}{!}{%
\begin{tabular}{llcccccccccccccc}
\toprule
 & & \multicolumn{4}{c}{Global / action} & \multicolumn{4}{c}{Zero-mode cumulants $\kappa^{(0)}_n$} & \multicolumn{6}{c}{Single-site cumulants} \\
\cmidrule(lr){3-6} \cmidrule(lr){7-10} \cmidrule(lr){11-16}
$\kappa$ & source & $\langle|\bar\phi|\rangle$ & $\chi$ & $\langle S\rangle/V$ & $U_L$ & $\kappa^{(0)}_2$ & $\kappa^{(0)}_4$ & $\kappa^{(0)}_6$ & $\kappa^{(0)}_8$ & $\langle|\phi|\rangle$ & $\kappa_2^{(\mathrm{site})}$ & $\kappa_4^{(\mathrm{site})}$ & $\kappa_6^{(\mathrm{site})}$ & $\kappa_8^{(\mathrm{site})}$ & $U_4^{(\mathrm{site})}$ \\
\midrule
\multicolumn{16}{l}{\textit{2D}, $L=128$, $\lambda=0.022$} \\
\midrule
$0.26$ & HMC & $0.0315(2)$ & $9.2(2)$ & $0.45120(7)$ & $0.01(2)$ & $0.00156(2)$ & $-1(1)\!\times\!10^{-7}$ & $1(1)\!\times\!10^{-9}$ & $1(1)\!\times\!10^{-11}$ & $0.77415(9)$ & $0.9105(2)$ & $-0.2726(3)$ & $0.602(1)$ & $-3.19(1)$ & $0.1095(1)$ \\
 & HMC ($N{=}2048$) & $0.0314(5)$ & $9.2(3)$ & $0.4513(1)$ & $0.01(3)$ & $0.00155(4)$ & $-1(2)\!\times\!10^{-7}$ & $-2(2)\!\times\!10^{-9}$ & $5(16)\!\times\!10^{-12}$ & $0.7740(2)$ & $0.9103(4)$ & $-0.2744(7)$ & $0.610(3)$ & $-3.35(4)$ & $0.1100(2)$ \\
 & DM\,$\{8,16,32,64\}$ & $0.0321(5)$ & $9.6(3)$ & $0.4525(1)$ & $0.02(3)$ & $0.00162(5)$ & $-1(2)\!\times\!10^{-7}$ & $-2(2)\!\times\!10^{-9}$ & $2(2)\!\times\!10^{-11}$ & $0.7748(2)$ & $0.9123(4)$ & $-0.2722(8)$ & $0.609(4)$ & $-3.33(5)$ & $0.1087(2)$ \\
\midrule
$0.2705$ & HMC & $0.453(4)$ & $749(9)$ & $0.3984(1)$ & $0.489(4)$ & $0.251(3)$ & $-0.092(3)$ & $0.153(8)$ & $-0.55(4)$ & $0.9955(4)$ & $1.420(1)$ & $-1.436(4)$ & $8.85(4)$ & $-121.5(7)$ & $0.2371(3)$ \\
 & HMC ($N{=}2048$) & $0.448(5)$ & $776(15)$ & $0.3984(2)$ & $0.484(6)$ & $0.248(5)$ & $-0.089(4)$ & $0.15(1)$ & $-0.52(5)$ & $0.9949(6)$ & $1.419(1)$ & $-1.434(5)$ & $8.84(5)$ & $-121.8(9)$ & $0.2368(4)$ \\
 & DM\,$\{8,16,32,64\}$ & $0.462(4)$ & $673(20)$ & $0.4026(2)$ & $0.514(5)$ & $0.255(4)$ & $-0.100(4)$ & $0.17(1)$ & $-0.66(5)$ & $0.9849(6)$ & $1.396(1)$ & $-1.347(5)$ & $8.02(5)$ & $-106.9(9)$ & $0.2300(4)$ \\
\midrule
$0.28$ & HMC & $1.4558(2)$ & $4.22(6)$ & $0.23817(6)$ & $0.666505(2)$ & $2.1196(5)$ & $-8.983(4)$ & $152.3(1)$ & $-5487(5)$ & $1.4829(2)$ & $2.6899(4)$ & $-9.801(4)$ & $171.1(1)$ & $-6397(5)$ & $0.45146(5)$ \\
 & HMC ($N{=}2048$) & $1.4558(4)$ & $4.2(2)$ & $0.2383(1)$ & $0.666505(6)$ & $2.120(1)$ & $-8.984(9)$ & $152.3(2)$ & $-5488(11)$ & $1.4829(3)$ & $2.6902(9)$ & $-9.807(9)$ & $171.3(2)$ & $-6415(12)$ & $0.4514(1)$ \\
 & DM\,$\{8,16,32,64\}$ & $1.4542(4)$ & $5.0(1)$ & $0.2401(2)$ & $0.666475(6)$ & $2.115(1)$ & $-8.943(9)$ & $151.3(2)$ & $-5439(10)$ & $1.4817(2)$ & $2.6883(7)$ & $-9.771(6)$ & $170.4(2)$ & $-6366(9)$ & $0.45043(9)$ \\
\midrule
\multicolumn{16}{l}{\textit{3D}, $L=64$, $\lambda=0.9$} \\
\midrule
$0.18$ & HMC & $0.00571(9)$ & $5.0(2)$ & $-0.06669(2)$ & $-0.02(3)$ & $5.16(15)\!\times\!10^{-5}$ & $1.71(260)\!\times\!10^{-10}$ & $-6.25(604)\!\times\!10^{-14}$ & $6.85(2151)\!\times\!10^{-18}$ & $0.66846(2)$ & $0.58604(2)$ & $-0.40116(3)$ & $1.3731(2)$ & $-10.237(2)$ & $0.38894(1)$ \\
 & HMC ($N{=}512$) & $0.0059(2)$ & $5.3(4)$ & $-0.06665(4)$ & $-0.03(6)$ & $5.50(37)\!\times\!10^{-5}$ & $2.23(531)\!\times\!10^{-10}$ & $-1.41(130)\!\times\!10^{-13}$ & $1.81(607)\!\times\!10^{-17}$ & $0.66845(4)$ & $0.58603(5)$ & $-0.40313(9)$ & $1.3907(6)$ & $-10.496(8)$ & $0.38918(3)$ \\
 & DM\,$\{4,8\}$ & $0.0062(2)$ & $6.0(5)$ & $-0.06811(4)$ & $-0.03(6)$ & $6.09(38)\!\times\!10^{-5}$ & $3.26(700)\!\times\!10^{-10}$ & $-1.56(208)\!\times\!10^{-13}$ & $-2.21(1032)\!\times\!10^{-17}$ & $0.66526(4)$ & $0.58118(6)$ & $-0.39437(9)$ & $1.3479(6)$ & $-10.072(8)$ & $0.38702(3)$ \\
 & DM\,$\{4,8,16\}$ & $0.0072(2)$ & $6.8(5)$ & $-0.06777(4)$ & $0.08(5)$ & $7.80(44)\!\times\!10^{-5}$ & $-1.52(94)\!\times\!10^{-9}$ & $5.34(412)\!\times\!10^{-13}$ & $-5.37(278)\!\times\!10^{-16}$ & $0.66672(4)$ & $0.58361(6)$ & $-0.39810(9)$ & $1.3657(6)$ & $-10.247(8)$ & $0.38749(3)$ \\
 & DM\,$\{4,8,16,32\}$ & $0.0121(3)$ & $11.3(6)$ & $-0.06621(7)$ & $0.40(2)$ & $1.90(8)\!\times\!10^{-4}$ & $-4.33(48)\!\times\!10^{-8}$ & $4.93(85)\!\times\!10^{-11}$ & $-1.22(28)\!\times\!10^{-13}$ & $0.66934(7)$ & $0.5872(1)$ & $-0.4063(1)$ & $1.4077(8)$ & $-10.670(10)$ & $0.39058(3)$ \\
 & DM\,$\{4,8,16,32\}$+MALA (score) & $0.0058(1)$ & $4.6(2)$ & $-0.06681(4)$ & $0.08(4)$ & $5.14(21)\!\times\!10^{-5}$ & $-6.73(315)\!\times\!10^{-10}$ & $9.35(736)\!\times\!10^{-14}$ & $-3.57(330)\!\times\!10^{-17}$ & $0.66879(3)$ & $0.58649(5)$ & $-0.40415(10)$ & $1.3964(7)$ & $-10.556(8)$ & $0.38950(3)$ \\
 & DM\,$\{4,8,16,32\}$+MALA ($-\nabla S$) & $0.0055(1)$ & $4.7(3)$ & $-0.06677(4)$ & $-0.08(6)$ & $4.81(22)\!\times\!10^{-5}$ & $5.29(396)\!\times\!10^{-10}$ & $-6.48(757)\!\times\!10^{-14}$ & $-4.38(382)\!\times\!10^{-17}$ & $0.66840(3)$ & $0.58594(4)$ & $-0.4031(1)$ & $1.3906(7)$ & $-10.497(9)$ & $0.38919(3)$ \\
\midrule
$0.1923$ & HMC & $0.1165(9)$ & $621(15)$ & $-0.11671(5)$ & $0.516(4)$ & $0.0159(2)$ & $-3.93(12)\!\times\!10^{-4}$ & $4.33(19)\!\times\!10^{-5}$ & $-1.03(6)\!\times\!10^{-5}$ & $0.68999(3)$ & $0.61815(4)$ & $-0.46164(7)$ & $1.7020(4)$ & $-13.640(4)$ & $0.40231(2)$ \\
 & HMC ($N{=}512$) & $0.113(2)$ & $636(35)$ & $-0.11643(9)$ & $0.50(1)$ & $0.0151(4)$ & $-3.45(25)\!\times\!10^{-4}$ & $3.56(39)\!\times\!10^{-5}$ & $-7.89(114)\!\times\!10^{-6}$ & $0.68987(7)$ & $0.61797(10)$ & $-0.4634(1)$ & $1.7204(9)$ & $-13.94(1)$ & $0.40246(3)$ \\
 & DM\,$\{4,8\}$ & $0.0287(9)$ & $94(5)$ & $-0.10747(5)$ & $0.24(3)$ & $0.00118(6)$ & $-1.02(19)\!\times\!10^{-6}$ & $4.97(160)\!\times\!10^{-9}$ & $-5.66(248)\!\times\!10^{-11}$ & $0.68054(4)$ & $0.60565(6)$ & $-0.4369(1)$ & $1.5714(7)$ & $-12.349(9)$ & $0.39499(3)$ \\
 & DM\,$\{4,8,16\}$ & $0.064(2)$ & $445(22)$ & $-0.1180(1)$ & $0.28(2)$ & $0.0058(3)$ & $-2.84(44)\!\times\!10^{-5}$ & $7.46(193)\!\times\!10^{-7}$ & $-4.49(158)\!\times\!10^{-8}$ & $0.68417(9)$ & $0.6090(1)$ & $-0.4476(2)$ & $1.633(1)$ & $-13.00(1)$ & $0.40026(5)$ \\
 & DM\,$\{4,8,16,32\}$ & $0.100(2)$ & $694(30)$ & $-0.1125(1)$ & $0.44(1)$ & $0.0127(4)$ & $-2.13(19)\!\times\!10^{-4}$ & $1.70(23)\!\times\!10^{-5}$ & $-2.95(53)\!\times\!10^{-6}$ & $0.6914(1)$ & $0.6204(2)$ & $-0.4682(2)$ & $1.748(1)$ & $-14.23(2)$ & $0.40345(4)$ \\
 & DM\,$\{4,8,16,32\}$+MALA (score) & $0.126(2)$ & $678(26)$ & $-0.1175(1)$ & $0.528(7)$ & $0.0184(4)$ & $-5.36(27)\!\times\!10^{-4}$ & $6.90(53)\!\times\!10^{-5}$ & $-1.90(20)\!\times\!10^{-5}$ & $0.69078(6)$ & $0.61930(8)$ & $-0.4662(2)$ & $1.736(1)$ & $-14.11(1)$ & $0.40313(4)$ \\
 & DM\,$\{4,8,16,32\}$+MALA ($-\nabla S$) & $0.116(2)$ & $642(23)$ & $-0.1165(1)$ & $0.509(7)$ & $0.0158(4)$ & $-3.80(21)\!\times\!10^{-4}$ & $4.11(35)\!\times\!10^{-5}$ & $-9.53(110)\!\times\!10^{-6}$ & $0.68989(5)$ & $0.61799(7)$ & $-0.4635(2)$ & $1.7209(10)$ & $-13.94(1)$ & $0.40248(4)$ \\
\midrule
$0.2$ & HMC & $0.51623(9)$ & $4.9(1)$ & $-0.24760(4)$ & $0.666574(2)$ & $0.26651(9)$ & $-0.14204(10)$ & $0.3028(3)$ & $-1.372(2)$ & $0.74510(3)$ & $0.70104(4)$ & $-0.64717(7)$ & $2.8696(5)$ & $-27.624(6)$ & $0.43652(1)$ \\
 & HMC ($N{=}512$) & $0.5163(2)$ & $4.6(3)$ & $-0.24772(8)$ & $0.666579(6)$ & $0.2666(2)$ & $-0.1421(2)$ & $0.3030(7)$ & $-1.373(4)$ & $0.74516(6)$ & $0.70114(9)$ & $-0.6498(2)$ & $2.899(2)$ & $-28.15(3)$ & $0.43675(3)$ \\
 & DM\,$\{4,8\}$ & $0.372(3)$ & $1.05(15)\!\times\!10^{3}$ & $-0.1948(5)$ & $0.645(2)$ & $0.143(1)$ & $-0.0393(9)$ & $0.044(2)$ & $-0.105(5)$ & $0.7147(3)$ & $0.6545(5)$ & $-0.5407(8)$ & $2.181(5)$ & $-19.12(6)$ & $0.4188(2)$ \\
 & DM\,$\{4,8,16\}$ & $0.5072(2)$ & $7.6(5)$ & $-0.2464(1)$ & $0.666517(10)$ & $0.2573(2)$ & $-0.1324(2)$ & $0.2725(7)$ & $-1.192(4)$ & $0.74042(8)$ & $0.6935(1)$ & $-0.6305(2)$ & $2.754(2)$ & $-26.11(2)$ & $0.43527(3)$ \\
 & DM\,$\{4,8,16,32\}$ & $0.5188(3)$ & $9.3(6)$ & $-0.2415(1)$ & $0.66649(1)$ & $0.2692(3)$ & $-0.1449(3)$ & $0.312(1)$ & $-1.428(7)$ & $0.7503(2)$ & $0.7105(3)$ & $-0.6665(4)$ & $2.996(3)$ & $-29.23(4)$ & $0.43815(4)$ \\
 & DM\,$\{4,8,16,32\}$+MALA (score) & $0.5176(1)$ & $4.8(2)$ & $-0.24833(9)$ & $0.666576(4)$ & $0.2677(4)$ & $-0.1430(8)$ & $0.305(3)$ & $-1.38(3)$ & $0.74560(4)$ & $0.70173(6)$ & $-0.6489(2)$ & $2.878(1)$ & $-27.69(2)$ & $0.43730(3)$ \\
 & DM\,$\{4,8,16,32\}$+MALA ($-\nabla S$) & $0.5163(1)$ & $5.2(2)$ & $-0.24762(9)$ & $0.666568(4)$ & $0.2663(4)$ & $-0.1416(8)$ & $0.300(4)$ & $-1.35(3)$ & $0.74513(4)$ & $0.70110(6)$ & $-0.6471(2)$ & $2.866(1)$ & $-27.54(2)$ & $0.43689(3)$ \\
\bottomrule
\end{tabular}%
}
\end{table*}

The two support filters defined above can be switched off, so that every generated configuration is retained. \Tab{tab:indist_scalar_L64_3d_nofilter} and \Tab{tab:multiL_HMC_cumulants_nofilter} repeat the three dimensional comparisons in this unfiltered form. The HMC reference is unchanged, since it is never filtered. The action density and single-site cumulants change only slightly. The susceptibility is the sensitive observable. In the in-distribution $L=64$ run it rises from $\chi=755(49)$ to $\chi=1.31(26)\times10^3$ at the pseudocritical point and from $\chi=7.8(5)$ to $\chi=23.6(9)$ in the broken phase. In the largest cross-$L$ run, the near-critical value is essentially unchanged, while the broken phase value rises from $\chi=9.3(6)$ to $\chi=13.1(6)$. The unfiltered tables therefore sharpen the main point. Rare zero-mode tails are more severe for in-distribution $L=64$ training than for the multi-$L$ sampler. The remaining discrepancy should be associated with the global magnetization rather than with the local field distribution.

\begin{table*}[!tbp]
\centering
\caption{Same as the three dimensional block of \Tab{tab:indist_scalar_L64_3d}, but with the two DM support filters disabled, so every generated configuration is retained ($L=64$, $\lambda=0.9$). The HMC reference is identical to \Tab{tab:indist_scalar_L64_3d} because it is never filtered. Removing the filters mainly inflates the susceptibility $\chi$ at the pseudocritical point and in the broken phase, while the single-site cumulants are essentially unchanged.}
\label{tab:indist_scalar_L64_3d_nofilter}
\resizebox{\textwidth}{!}{%
\begin{tabular}{llcccccccccccc}
\toprule
 & & \multicolumn{4}{c}{Global / action} & \multicolumn{3}{c}{Zero-mode cumulants $\kappa^{(0)}_n$} & \multicolumn{5}{c}{Single-site cumulants} \\
\cmidrule(lr){3-6} \cmidrule(lr){7-9} \cmidrule(lr){10-14}
$\kappa$ & source & $\langle|\bar\phi|\rangle$ & $\chi$ & $\langle S\rangle/V$ & $U_L$ & $\kappa^{(0)}_4$ & $\kappa^{(0)}_6$ & $\kappa^{(0)}_8$ & $\kappa_2^{(\mathrm{site})}$ & $\kappa_4^{(\mathrm{site})}$ & $\kappa_6^{(\mathrm{site})}$ & $\kappa_8^{(\mathrm{site})}$ & $U_4^{(\mathrm{site})}$ \\
\midrule
$0.18$ & HMC & $0.00571(9)$ & $5.0(2)$ & $-0.066692(25)$ & $-0.02(3)$ & $1.71(260)\!\times\!10^{-10}$ & $-6.25(604)\!\times\!10^{-14}$ & $6.85(2151)\!\times\!10^{-18}$ & $0.58604(2)$ & $-0.40116(3)$ & $1.3731(2)$ & $-10.237(2)$ & $0.38894(1)$ \\
 & DM & $0.0062(2)$ & $6.2(4)$ & $-0.0540(1)$ & $-0.10(4)$ & $1.12(50)\!\times\!10^{-9}$ & $-4.23(168)\!\times\!10^{-13}$ & $-2.03(883)\!\times\!10^{-17}$ & $0.6003(1)$ & $-0.4246(2)$ & $1.503(1)$ & $-11.64(1)$ & $0.39066(3)$ \\
\midrule
$0.1923$ & HMC & $0.1165(9)$ & $621(15)$ & $-0.116709(65)$ & $0.516(4)$ & $-3.93(12)\!\times\!10^{-4}$ & $4.33(19)\!\times\!10^{-5}$ & $-1.03(6)\!\times\!10^{-5}$ & $0.61815(4)$ & $-0.46164(7)$ & $1.7020(4)$ & $-13.640(4)$ & $0.40231(2)$ \\
 & DM & $0.156(2)$ & $1.31(26)\!\times\!10^{3}$ & $-0.094(3)$ & $-0.1(4)$ & $3.04(1056)\!\times\!10^{-4}$ & $0.0012(7)$ & $3.17(240)\!\times\!10^{-4}$ & $0.644(2)$ & $-0.5010(8)$ & $1.947(7)$ & $-16.41(7)$ & $0.400(2)$ \\
\midrule
$0.2$ & HMC & $0.51623(9)$ & $4.9(1)$ & $-0.247603(52)$ & $0.666574(2)$ & $-0.14204(10)$ & $0.3028(3)$ & $-1.372(2)$ & $0.70104(4)$ & $-0.64717(7)$ & $2.8696(5)$ & $-27.624(6)$ & $0.43652(1)$ \\
 & DM & $0.5274(3)$ & $23.6(9)$ & $-0.2313(2)$ & $0.66623(2)$ & $-0.1541(9)$ & $0.342(4)$ & $-1.61(3)$ & $0.7247(4)$ & $-0.6922(7)$ & $3.169(5)$ & $-31.47(7)$ & $0.43754(4)$ \\
\bottomrule
\end{tabular}%
}
\end{table*}

\begin{table*}[!tbp]
\centering
\caption{Same as the three dimensional block of \Tab{tab:multiL_HMC_cumulants}, but with the two DM support filters disabled, so every generated configuration is retained ($L=64$, $\lambda=0.9$). The HMC rows are identical to \Tab{tab:multiL_HMC_cumulants} because the reference is never filtered. The action density and single-site cumulants are stable. The main differences are in the susceptibility of the largest training set in the symmetric and broken phases, while the near-critical entry is stable.}
\label{tab:multiL_HMC_cumulants_nofilter}
\resizebox{\textwidth}{!}{%
\begin{tabular}{llcccccccccccccc}
\toprule
 & & \multicolumn{4}{c}{Global / action} & \multicolumn{4}{c}{Zero-mode cumulants $\kappa^{(0)}_n$} & \multicolumn{6}{c}{Single-site cumulants} \\
\cmidrule(lr){3-6} \cmidrule(lr){7-10} \cmidrule(lr){11-16}
$\kappa$ & source & $\langle|\bar\phi|\rangle$ & $\chi$ & $\langle S\rangle/V$ & $U_L$ & $\kappa^{(0)}_2$ & $\kappa^{(0)}_4$ & $\kappa^{(0)}_6$ & $\kappa^{(0)}_8$ & $\langle|\phi|\rangle$ & $\kappa_2^{(\mathrm{site})}$ & $\kappa_4^{(\mathrm{site})}$ & $\kappa_6^{(\mathrm{site})}$ & $\kappa_8^{(\mathrm{site})}$ & $U_4^{(\mathrm{site})}$ \\
\midrule
$0.18$ & HMC & $0.00571(9)$ & $5.0(2)$ & $-0.06669(2)$ & $-0.02(3)$ & $5.16(15)\!\times\!10^{-5}$ & $1.71(260)\!\times\!10^{-10}$ & $-6.25(604)\!\times\!10^{-14}$ & $6.85(2151)\!\times\!10^{-18}$ & $0.66846(2)$ & $0.58604(2)$ & $-0.40116(3)$ & $1.3731(2)$ & $-10.237(2)$ & $0.38894(1)$ \\
 & HMC ($N{=}512$) & $0.0059(2)$ & $5.3(4)$ & $-0.06665(4)$ & $-0.03(6)$ & $5.50(37)\!\times\!10^{-5}$ & $2.23(531)\!\times\!10^{-10}$ & $-1.41(130)\!\times\!10^{-13}$ & $1.81(607)\!\times\!10^{-17}$ & $0.66845(4)$ & $0.58603(5)$ & $-0.40313(9)$ & $1.3907(6)$ & $-10.496(8)$ & $0.38918(3)$ \\
 & DM\,$\{4,8\}$ & $0.0062(1)$ & $6.0(4)$ & $-0.06809(4)$ & $-0.09(7)$ & $6.14(29)\!\times\!10^{-5}$ & $1.07(78)\!\times\!10^{-9}$ & $6.19(1824)\!\times\!10^{-14}$ & $-1.80(151)\!\times\!10^{-16}$ & $0.66526(3)$ & $0.58117(4)$ & $-0.39426(9)$ & $1.3469(6)$ & $-10.059(8)$ & $0.38701(3)$ \\
 & DM\,$\{4,8,16\}$ & $0.0073(2)$ & $6.8(4)$ & $-0.06777(4)$ & $0.09(4)$ & $7.87(33)\!\times\!10^{-5}$ & $-1.64(72)\!\times\!10^{-9}$ & $5.45(306)\!\times\!10^{-13}$ & $-5.19(209)\!\times\!10^{-16}$ & $0.66672(3)$ & $0.58362(4)$ & $-0.39807(9)$ & $1.3655(6)$ & $-10.243(8)$ & $0.38748(3)$ \\
 & DM\,$\{4,8,16,32\}$ & $0.0127(2)$ & $13.6(6)$ & $-0.06620(7)$ & $0.33(2)$ & $2.12(7)\!\times\!10^{-4}$ & $-4.45(44)\!\times\!10^{-8}$ & $5.41(77)\!\times\!10^{-11}$ & $-1.39(27)\!\times\!10^{-13}$ & $0.66935(5)$ & $0.58722(8)$ & $-0.4062(1)$ & $1.4070(8)$ & $-10.658(10)$ & $0.39057(3)$ \\
\midrule
$0.1923$ & HMC & $0.1165(9)$ & $621(15)$ & $-0.11671(5)$ & $0.516(4)$ & $0.0159(2)$ & $-3.93(12)\!\times\!10^{-4}$ & $4.33(19)\!\times\!10^{-5}$ & $-1.03(6)\!\times\!10^{-5}$ & $0.68999(3)$ & $0.61815(4)$ & $-0.46164(7)$ & $1.7020(4)$ & $-13.640(4)$ & $0.40231(2)$ \\
 & HMC ($N{=}512$) & $0.113(2)$ & $636(35)$ & $-0.11643(9)$ & $0.50(1)$ & $0.0151(4)$ & $-3.45(25)\!\times\!10^{-4}$ & $3.56(39)\!\times\!10^{-5}$ & $-7.89(114)\!\times\!10^{-6}$ & $0.68987(7)$ & $0.61797(10)$ & $-0.4634(1)$ & $1.7204(9)$ & $-13.94(1)$ & $0.40246(3)$ \\
 & DM\,$\{4,8\}$ & $0.0286(6)$ & $93(4)$ & $-0.10747(5)$ & $0.24(2)$ & $0.00117(4)$ & $-9.85(131)\!\times\!10^{-7}$ & $4.65(108)\!\times\!10^{-9}$ & $-5.07(164)\!\times\!10^{-11}$ & $0.68054(3)$ & $0.60565(4)$ & $-0.43695(10)$ & $1.5714(7)$ & $-12.349(9)$ & $0.39499(3)$ \\
 & DM\,$\{4,8,16\}$ & $0.064(1)$ & $443(16)$ & $-0.11798(9)$ & $0.27(2)$ & $0.0057(2)$ & $-2.69(32)\!\times\!10^{-5}$ & $6.79(141)\!\times\!10^{-7}$ & $-3.90(113)\!\times\!10^{-8}$ & $0.68416(6)$ & $0.60897(9)$ & $-0.4475(2)$ & $1.633(1)$ & $-13.00(1)$ & $0.40025(5)$ \\
 & DM\,$\{4,8,16,32\}$ & $0.100(2)$ & $698(22)$ & $-0.1125(1)$ & $0.44(1)$ & $0.0126(4)$ & $-2.08(15)\!\times\!10^{-4}$ & $1.63(19)\!\times\!10^{-5}$ & $-2.77(43)\!\times\!10^{-6}$ & $0.69144(8)$ & $0.6204(1)$ & $-0.4682(2)$ & $1.748(1)$ & $-14.23(2)$ & $0.40345(4)$ \\
\midrule
$0.2$ & HMC & $0.51623(9)$ & $4.9(1)$ & $-0.24760(4)$ & $0.666574(2)$ & $0.26651(9)$ & $-0.14204(10)$ & $0.3028(3)$ & $-1.372(2)$ & $0.74510(3)$ & $0.70104(4)$ & $-0.64717(7)$ & $2.8696(5)$ & $-27.624(6)$ & $0.43652(1)$ \\
 & HMC ($N{=}512$) & $0.5163(2)$ & $4.6(3)$ & $-0.24772(8)$ & $0.666579(6)$ & $0.2666(2)$ & $-0.1421(2)$ & $0.3030(7)$ & $-1.373(4)$ & $0.74516(6)$ & $0.70114(9)$ & $-0.6498(2)$ & $2.899(2)$ & $-28.15(3)$ & $0.43675(3)$ \\
 & DM\,$\{4,8\}$ & $0.372(2)$ & $1.05(11)\!\times\!10^{3}$ & $-0.1948(5)$ & $0.644(2)$ & $0.142(1)$ & $-0.0391(8)$ & $0.044(1)$ & $-0.103(4)$ & $0.7147(2)$ & $0.6545(4)$ & $-0.5406(8)$ & $2.180(5)$ & $-19.12(6)$ & $0.4187(2)$ \\
 & DM\,$\{4,8,16\}$ & $0.5072(2)$ & $7.6(4)$ & $-0.2464(1)$ & $0.666517(7)$ & $0.2571(4)$ & $-0.1319(7)$ & $0.270(3)$ & $-1.18(2)$ & $0.74042(7)$ & $0.6935(1)$ & $-0.6305(2)$ & $2.754(2)$ & $-26.10(2)$ & $0.43527(3)$ \\
 & DM\,$\{4,8,16,32\}$ & $0.5202(2)$ & $13.1(6)$ & $-0.2417(1)$ & $0.66642(1)$ & $0.2704(4)$ & $-0.1459(8)$ & $0.314(3)$ & $-1.44(3)$ & $0.7509(1)$ & $0.7115(2)$ & $-0.6683(4)$ & $3.006(3)$ & $-29.33(4)$ & $0.43828(4)$ \\
\bottomrule
\end{tabular}%
}
\end{table*}

\section{Reference ensemble generation with the FA-HMC--Wolff algorithm}
\label{app:fahmc_wolff}

Reference configurations are generated with the FA-HMC--Wolff algorithm, a Markov chain that combines Fourier-accelerated Hybrid Monte Carlo with Wolff cluster reflections.  Each production cycle consists of one FA-HMC trajectory followed by $n_{\rm W}$ Wolff reflections, with $n_{\rm W}=5$ in the production runs and $n_{\rm W}=10$ for the near-critical ensembles.  The HMC part updates the continuous amplitudes of the scalar field across momentum scales, while the cluster reflections improve tunnelling of the global $\mathbb{Z}_2$ mode at the pseudocritical point and in the broken phase.  Configurations used in the comparisons are recorded only after warm-up, during which the Fourier metric and the HMC step size are adapted.

For the HMC update we introduce an auxiliary real momentum field $\pi$, with one component $\pi_x$ conjugate to each lattice variable $\phi_x$, and use the auxiliary Hamiltonian,
\begin{equation}
\begin{aligned}
H[\phi,\pi]&=S[\phi]+K[\pi],\\
K[\pi]&=\frac{1}{2}\sum_{q}
\widetilde\pi(q)^*M^{-1}(q)\widetilde\pi(q) ,
\end{aligned}
\label{eq:fahmc_hamiltonian}
\end{equation}
where the sum runs over the lattice Fourier modes $q_\mu=2\pi n_\mu/L$ with $n_\mu=0,\ldots,L-1$, and $\widetilde\pi(q)$ is the Fourier amplitude of $\pi$ in the normalization of \Eq{eq:fourier_field}. The chain samples the joint distribution proportional to $e^{-H[\phi,\pi]}=e^{-S[\phi]}\,e^{-K[\pi]}$. The momenta enter only through the Gaussian factor $e^{-K[\pi]}$, so the marginal distribution of $\phi$ remains the target Boltzmann distribution, and $\pi$ is refreshed from this Gaussian at the beginning of every trajectory. The metric $M$ is positive and diagonal in the lattice Fourier basis.  At the beginning of warm-up it is initialized by the free-field form,
\begin{equation}
M_0(q)=|k(q)|^2+m_0^2,
\label{eq:fourier_mass}
\end{equation}
with the lattice momentum $k_\mu(q)=2\sin(q_\mu/2)$ defined in \Eq{eq:lattice_momentum}, so that $M_0$ is the inverse of the free lattice propagator in \Eq{eq:free_lattice_prop}. The bare mass keeps the zero-mode entry of the metric positive, and the production runs use $m_0^2=10^{-2}$.
This is the form used in the implementation, with the two-dimensional code storing the independent real-FFT half-grid.  The refreshment of $\pi$ is implemented by drawing a white Gaussian field, transforming it to momentum space, multiplying by $M^{1/2}$, and transforming back.  The kinetic energy and the field update use the corresponding multiplication by $M^{-1}$.

The Hamiltonian trajectory is integrated with a reversible leapfrog scheme.  The field update uses $M^{-1}\pi$, evaluated by FFT, and the momentum updates use the negative derivative of the action in \Eq{eq:phi4_action}.  In coordinate space this derivative is evaluated from the forward and backward nearest neighbors of each site together with the on-site quadratic and quartic terms.  The code omits the additive constant $\lambda V$ in $\sum_x\lambda(\phi_x^2-1)^2$, since this constant cancels from the Hamiltonian difference.  A proposed field $\phi'$ is accepted with probability $\min(1,e^{-\Delta H})$, where $\Delta H=H[\phi',\pi']-H[\phi,\pi]$, and a rejection restores the field at the beginning of the trajectory.

Two adaptations are applied during warm-up and then frozen.  The step size $\varepsilon$ is tuned by dual averaging with target acceptance $\delta=0.75$.  The number of leapfrog steps is independently randomised for each trajectory in the integer range from $0.8n_{\rm steps}$ to $1.2n_{\rm steps}$, reducing resonances of the integrator.  The Fourier metric is updated in a sequence of adaptation windows of increasing length.  For a window containing $N_w$ stored configurations we estimate the mode-wise propagator,
\begin{equation}
G_w(q)=\frac{1}{N_w}\sum_{i=1}^{N_w}\bigl|\widetilde\phi_i(q)\bigr|^2,
\label{eq:fahmc_window_propagator}
\end{equation}
with $\widetilde\phi_i(q)$ the Fourier amplitude \Eq{eq:fourier_field} of the $i$-th stored configuration.  The next metric is the inverse of a prior-mixed covariance estimate, using the measured $G_w$ with weight $1-\alpha$ and the previous covariance $M^{-1}$ with weight $\alpha=0.05$.  After each metric update the dual-averaging tuning of $\varepsilon$ is restarted from half the current step size.  Windows with too few stored configurations are skipped.  The production chain uses the final metric and final step size without further adaptation, giving a fixed Metropolis-corrected HMC transition kernel.

The Wolff reflection is applied after each FA-HMC proposal.  For fixed magnitudes $|\phi_x|$, the signs $\sigma_x={\rm sign}(\phi_x)$ define an embedded nearest-neighbor Ising model with coupling $J_{xy}=2\kappa |\phi_x\phi_y|$.  Starting from a uniformly chosen seed site, the cluster is grown through same-sign nearest-neighbor pairs.  The bond probability used in the code is
\begin{equation}
P_{\rm bond}(x,y)=1-\exp\!\left[-4\kappa\phi_x\phi_y\right],
\qquad \phi_x\phi_y>0 .
\end{equation}
After no further bonds are activated, all fields in the cluster are reflected, $\phi_z\mapsto-\phi_z$.  The on-site part of the action is invariant under this reflection, and the bond probability above is the Wolff probability for the embedded Ising signs.  The reflection is therefore rejection-free and leaves the same Boltzmann measure invariant.

The production kernel is the composition of one FA-HMC kernel and $n_{\rm W}$ Wolff kernels.  Since both kernels preserve the target distribution proportional to $e^{-S[\phi]}$, their composition preserves it as well.  The two-dimensional implementation uses periodic nearest-neighbor tables and real FFTs.  The three-dimensional reference chains use the same construction with nearest neighbors and Fourier modes in three spatial directions.  Autocorrelation times quoted elsewhere in the paper are measured after discarding the warm-up configurations.

\section{Finite-size determination of the critical hopping}
\label{app:kappac}

This appendix records the finite-size analysis used to choose the near-critical hopping parameters in the main text.  The zero-mode observables used here are the order parameter, susceptibility and Binder cumulant defined in \Sec{sec:indist_scalar}.  The order parameter and susceptibility plots locate the finite-volume crossover and show its sharpening as the lattice size grows.  The susceptibility peak gives a useful check on the transition window.  The central estimates quoted below are obtained from Binder-cumulant crossings.

The limiting values of the Binder cumulant provide useful orientation.  It approaches zero in the symmetric phase when the zero-mode distribution is approximately Gaussian, and it approaches \(2/3\) deep in the ordered phase.  At a continuous transition the curves for different lattice sizes intersect, up to finite-size corrections, near a critical Binder value fixed by the dimension, boundary conditions, and aspect ratio.  We use the observed sequence of pairwise intersections only as an empirical finite-size estimator of \(\kappa_c\).  No correction-to-scaling fit is performed.

\subsection{Two-dimensional theory at $\lambda=0.022$}
\label{app:kappac_2d}

\begin{figure}[!htbp]
\centering
\includegraphics[width=\columnwidth]{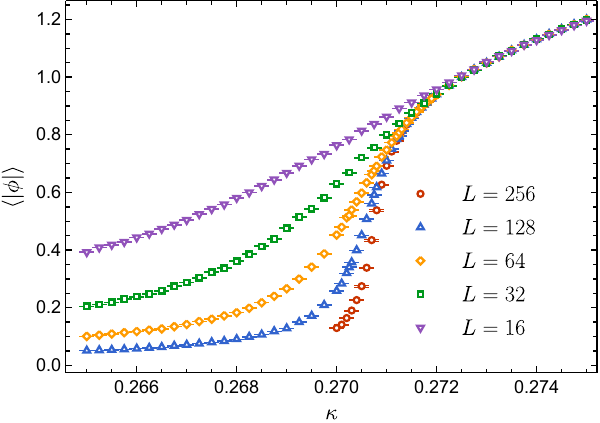}
\caption{Order parameter $\langle|\bar\phi|\rangle$ as a function of $\kappa$ in two dimensions at $\lambda=0.022$.  The curves show $L=16,32,64,128,256$.  The crossover sharpens as the lattice size increases.}
\label{fig:orderparameter_2d}
\end{figure}

The two-dimensional scan uses $L=16,32,64,128,256$.  The order parameter in \fig{fig:orderparameter_2d} changes from a broad finite-volume crossover at small $L$ to a steep rise at the largest volumes.  The transition window is centered near $\kappa\simeq0.271$, and the production point $\kappa=0.2705$ used in the main text lies within this finite-volume critical region.

\begin{figure}[!htbp]
\centering
\includegraphics[width=\columnwidth]{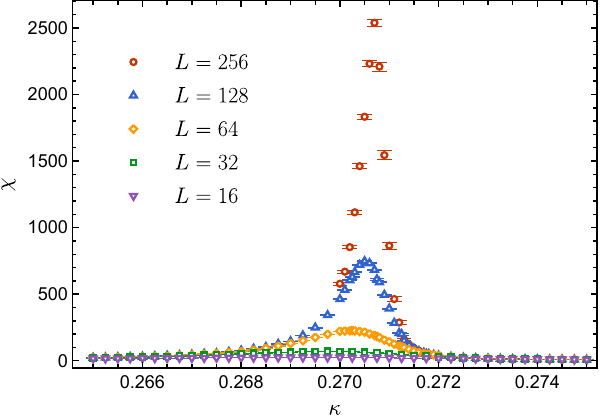}
\caption{Magnetic susceptibility $\chi_L$ as a function of $\kappa$ in two dimensions at $\lambda=0.022$.  The peak height grows strongly with $L$ and the maximum indicates the critical region.}
\label{fig:chi_2d_1}
\end{figure}

The susceptibility in \fig{fig:chi_2d_1} shows the same transition window.  Its peak grows rapidly with $L$, as expected from finite-size scaling in the Ising universality class.  This provides an auxiliary check on the interval in which the Binder crossings should be analyzed.

\begin{figure}[!htbp]
\centering
\includegraphics[width=\columnwidth]{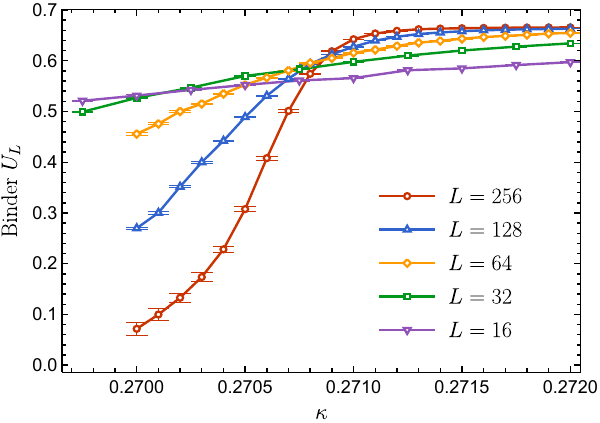}
\caption{Binder cumulant $U_L$ as a function of $\kappa$ in two dimensions at $\lambda=0.022$.  The intersections of the curves determine the finite-size sequence of crossing estimates.}
\label{fig:binder_2d}
\end{figure}

\begin{figure}[!htbp]
\centering
\includegraphics[width=\columnwidth]{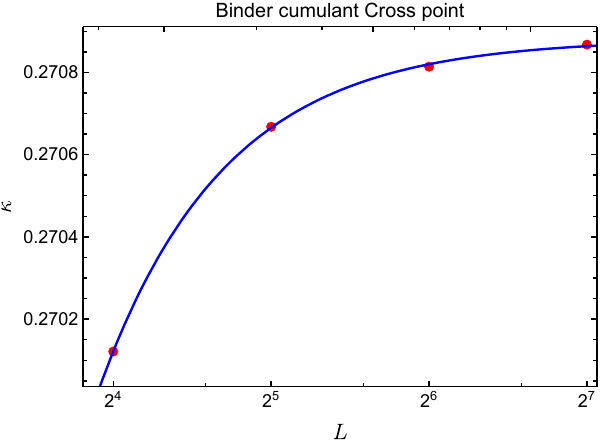}
\caption{Finite-size Binder-crossing estimates in two dimensions.  The pairwise intersections approach $\kappa_c\simeq0.27088$ at the largest volumes.  The solid curve is a guide to the eye and is not used in the estimate.}
\label{fig:bindercross_2d}
\end{figure}

The Binder cumulants in \fig{fig:binder_2d} provide the critical-point estimate.  The crossings of successive lattice sizes lie in a narrow range.  Treating these intersections as finite-volume estimates, the pairwise analysis in \fig{fig:bindercross_2d} gives,
\begin{equation}
\kappa_c(\lambda=0.022)\simeq0.27088 .
\label{eq:kappa_c}
\end{equation}
This is the Binder-crossing estimate of the two-dimensional critical hopping.  The validation runs at $\kappa=0.2705$ are slightly below this estimate and remain in the same finite-volume critical window.

\subsection{Three-dimensional theory at $\lambda=0.9$}
\label{app:kappac_3d}

\begin{figure}[!htbp]
\centering
\includegraphics[width=\columnwidth]{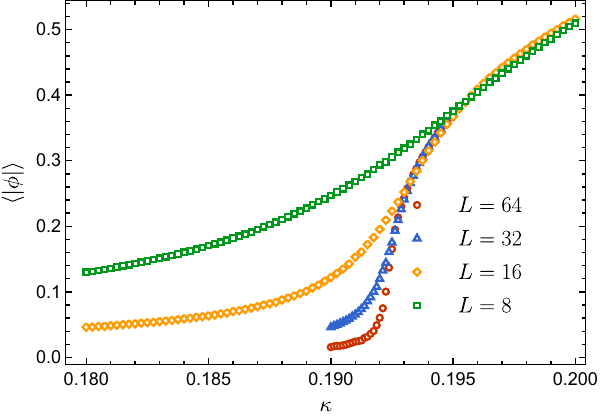}
\caption{Order parameter $\langle|\bar\phi|\rangle$ in the three-dimensional theory at $\lambda=0.9$.  The scan includes $L=8,16,32,64$.}
\label{fig:orderparameter_3d}
\end{figure}

The three-dimensional scan uses $L=8,16,32,64$.  The order parameter in \fig{fig:orderparameter_3d} rises from the symmetric regime to the broken regime over a window centered near $\kappa\simeq0.192$.  The broad crossover at $L=8$ is replaced by a much sharper change at $L=64$, consistent with the growth of the correlation length.

\begin{figure}[!htbp]
\centering
\includegraphics[width=\columnwidth]{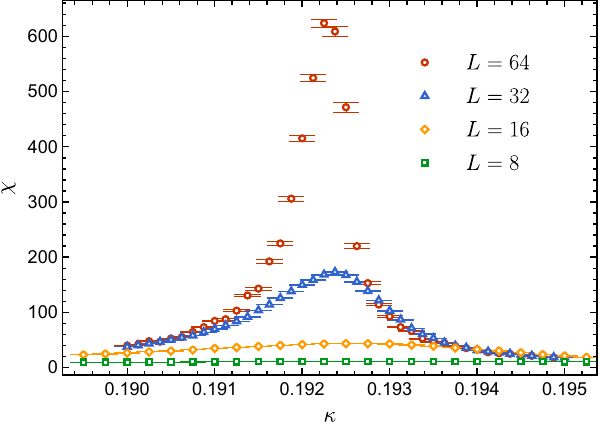}
\caption{Magnetic susceptibility $\chi_L$ as a function of $\kappa$ in three dimensions at $\lambda=0.9$.  The growth and narrowing of the peak mark the finite-volume critical region.}
\label{fig:chi_3d}
\end{figure}

The susceptibility in \fig{fig:chi_3d} peaks in the same region.  The peak height increases from $L=8$ to $L=64$, and the largest-volume maximum lies close to the Binder crossing quoted below.  The peak positions were located by weighted quadratic fits to the points around each maximum.  The fitted peak heights obey the expected three-dimensional Ising finite-size trend.  A log-log fit of $\chi_{\rm peak}(L)$ over $L=16,32,64$ gives $\gamma/\nu=1.946(42)$, compatible with the Ising value $\gamma/\nu=1.9642$.  The $L=8$ point is left out of this exponent fit because it is visibly pre-asymptotic.  We use this scaling behavior as a check on the HMC reference ensembles and on the critical window.  The quoted critical hopping below remains the Binder-crossing estimate.

\begin{figure}[!htbp]
\centering
\includegraphics[width=\columnwidth]{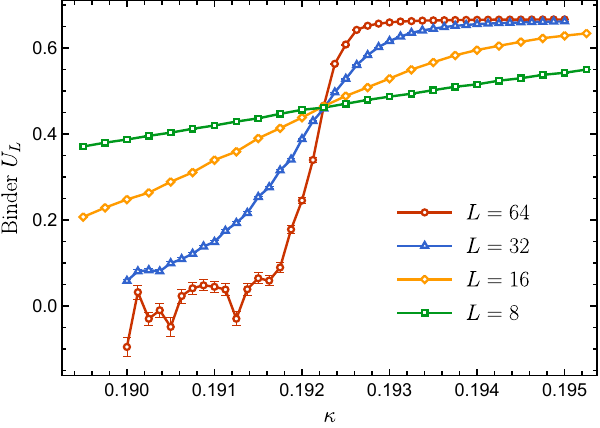}
\caption{Binder cumulant $U_L$ as a function of $\kappa$ in three dimensions at $\lambda=0.9$.  The crossing of the $L=32$ and $L=64$ curves gives $\kappa_c\simeq0.19225$.}
\label{fig:binder_3d}
\end{figure}

The Binder cumulants in \fig{fig:binder_3d} cross near $\kappa=0.192$.  The smaller lattices carry visible finite-size corrections, so the crossing of the two largest volumes is used for the central value.  We obtain,
\begin{equation}
\kappa_c(\lambda=0.9)\simeq0.19225 .
\label{eq:kappa_c_3d}
\end{equation}
The corresponding crossing value of $U_L$ lies between the symmetric and broken limits, as expected for a critical Binder cumulant on a finite periodic lattice.  Using the same central $\kappa_c$ and the three-dimensional Ising exponent $\nu=0.6296$, the Binder curves collapse in the central scaling window when plotted against $(\kappa-\kappa_c)L^{1/\nu}$, with a crossing value near $U_L^\ast\simeq0.4655$.  This collapse provides a visual check that the largest-volume crossing is already in the universal scaling regime.  No additional fit parameter is extracted from the collapse.  The diffusion-model runs at $\kappa=0.1923$ therefore sit inside this finite-volume critical window.

\section{Calibration of the MALA acceptance diagnostic}
\label{app:mala_calibration}

The MALA acceptance-rate diagnostic in \Sec{sec:acceptance} uses the learned score only as a proposal drift for an exact Metropolis--Hastings update with target density $p_0(\phi)\propto e^{-S[\phi]}$. The diagnostic therefore contains two auxiliary choices. The first is the overall step-size coefficient $c$ in $h=c/L^D$. The second is the diffusion time $t_{\text{mh}}$ at which the score is evaluated inside the proposal. These parameters set the sensitivity of the acceptance rate to the quality of the learned score, while the target distribution remains fixed by the exact MH ratio.

For a given pair $(c,t_{\text{mh}})$ we measure the acceptance rate using EMA-averaged score weights and a single MH trial from each reference configuration. The early-training acceptance rate is averaged over the checkpoints in epochs $0$--$50$, while the late-training rate is averaged over checkpoints in epochs $6000$--$10000$. We define the separation between these two windows by,
\begin{equation}
\Delta A(c,t_{\text{mh}})
= A_{\rm late}(c,t_{\text{mh}})-A_{\rm early}(c,t_{\text{mh}}) .
\label{eq:mala_diagnostic_gap}
\end{equation}
A useful diagnostic choice gives a large positive $\Delta A$ while keeping the late-time acceptance away from the trivial limits of zero and one. All calibration sweeps include the three representative values of $\kappa$ used in the main text, giving a common parameter choice across the symmetric, near-critical and broken regimes.

The usual asymptotic MALA scaling for an i.i.d.\ product target has limited relevance for the present lattice ensembles. For a $d$-dimensional product density $\pi(x)=\prod_i f(x_i)$, Roberts and Rosenthal~\cite{Roberts1998} found an optimal proposal variance scaling as $d^{-1/3}$, with limiting acceptance rate $0.574$. In an interacting lattice field theory, the components are coupled by the nearest-neighbor term in the action, and the drift used here is an approximate score evaluated at a nonzero diffusion time. Direct sweeps show that the product-target scaling gives moves that are too large for the present purpose. We instead use the volume-normalized family,
\begin{equation}
 h = \frac{c}{L^D} .
\label{eq:mala_volume_scaled_h}
\end{equation}
With this parametrization, the stochastic part of the proposal has expected squared norm $\bigl\langle\,\|\sqrt{2h}\,\eta\|^2\,\bigr\rangle=2c$, independent of the number of lattice sites. The dimensionless coefficient $c$ then controls a comparable global proposal size as $L$ and $D$ are varied.

\begin{figure*}[!tbp]
\centering
\includegraphics[width=\textwidth]{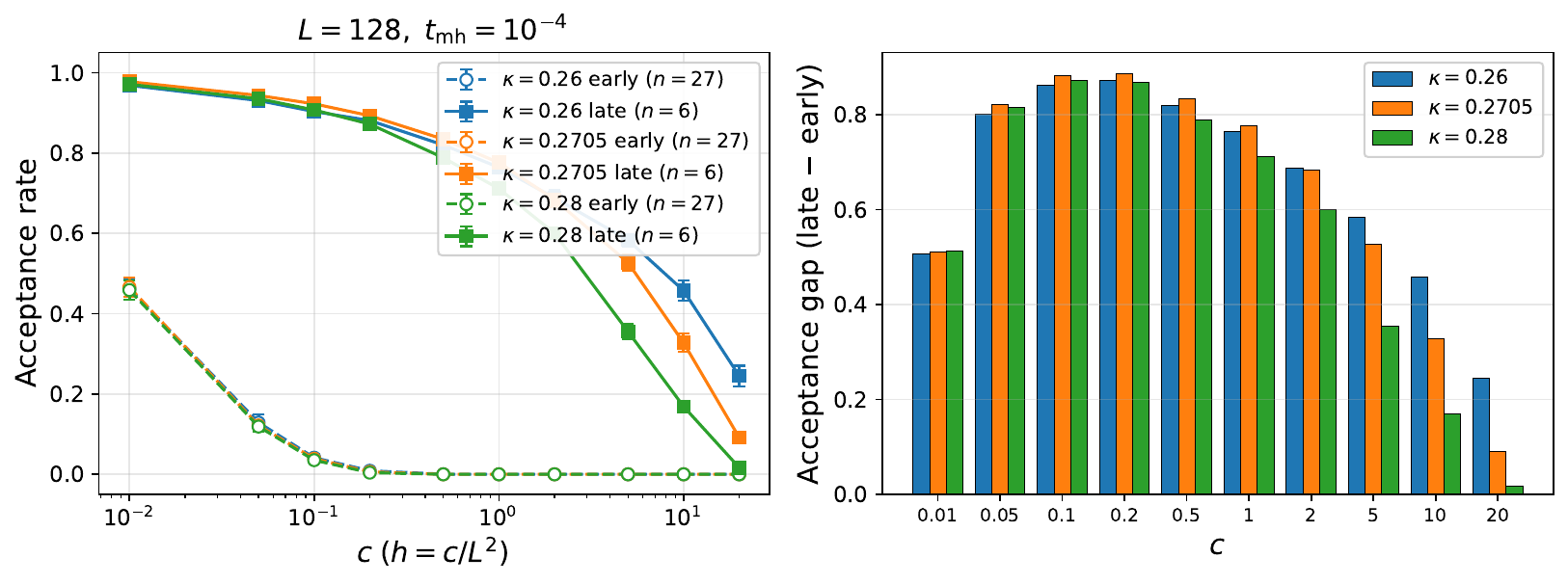}
\caption{Two-dimensional calibration of the MALA step size $h=c/L^2$ at $L=128$, with $t_{\text{mh}}=10^{-4}$, $N=2048$ reference configurations, a single MH step and EMA-averaged score weights. The acceptance rates are averaged separately over the early checkpoint window at epochs $0$--$50$ and the late window at epochs $6000$--$10000$. The right panel shows the diagnostic gap $\Delta A$ in \Eq{eq:mala_diagnostic_gap}. The common maximum over the three values of $\kappa$ is broad and includes the value $c=0.2$ used in the main text.}
\label{fig:calibrate_h}
\end{figure*}

The two-dimensional step-size sweep at $L=128$ is shown in \fig{fig:calibrate_h}. At very small $c$, both early and late checkpoints have high acceptance, and the acceptance rate has little resolving power. At large $c$, the proposal becomes too aggressive and the late-time acceptance falls. The separation $\Delta A$ has a broad maximum around $c=0.1$--$0.2$ for all three values of $\kappa$. We use $c=0.2$, which lies in this common maximum and leaves the late-training acceptance in an informative range.

\begin{figure*}[!tbp]
\centering
\includegraphics[width=\textwidth]{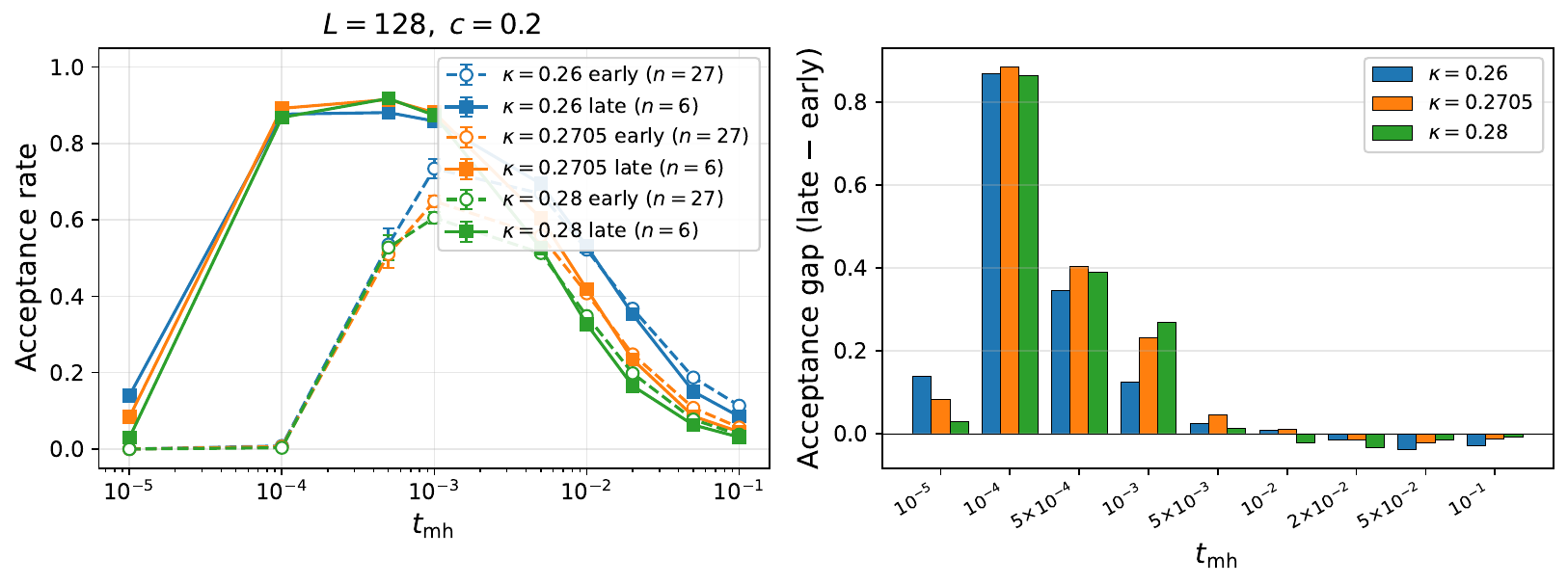}
\caption{Two-dimensional calibration of the score-evaluation time $t_{\text{mh}}$ at $L=128$, with $c=0.2$, $N=2048$ reference configurations, a single MH step and EMA-averaged score weights. The acceptance gap peaks at $t_{\text{mh}}=10^{-4}$ for all three values of $\kappa$. For $t_{\text{mh}}\gtrsim 2\times10^{-2}$ the gap becomes negative because the proposal drift is then governed by the score of the noised distribution $p_t$, while the MH correction still targets $p_0$.}
\label{fig:calibrate_tmh}
\end{figure*}

With $c$ fixed to $0.2$, the diffusion-time sweep in \fig{fig:calibrate_tmh} selects $t_{\text{mh}}=10^{-4}$. This small value keeps the score close to the physical score of $p_0$. As $t_{\text{mh}}$ is increased, the network evaluates the score of the noised distribution $p_t$, while the MH correction still uses the action of the original ensemble. The acceptance gap consequently decreases towards zero for $t_{\text{mh}}\gtrsim 5\times10^{-3}$ and becomes negative for $t_{\text{mh}}\gtrsim 2\times10^{-2}$. This sign change makes large $t_{\text{mh}}$ unsuitable for the diagnostic, even though such times can be useful along the reverse diffusion trajectory.

\begin{figure*}[!tbp]
\centering
\includegraphics[width=\textwidth]{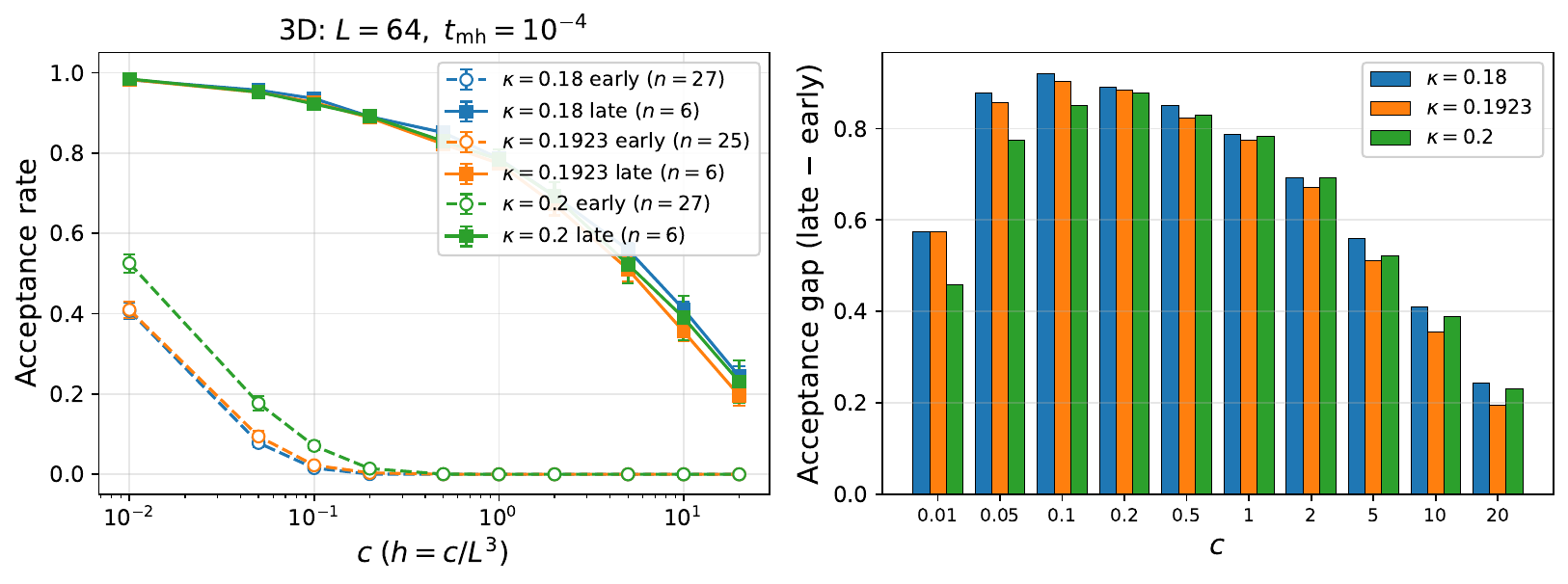}
\caption{Three-dimensional calibration of the MALA step size $h=c/L^3$ at $L=64$, $\lambda=0.9$, with $t_{\text{mh}}=10^{-4}$, a single MH step and EMA-averaged score weights. The calibration sweeps use $N = 256$ configurations per point, which suffices to resolve the acceptance gap. The averaging windows and the definition of the diagnostic gap are the same as in \fig{fig:calibrate_h}. The value $c=0.2$ again lies in the common maximum over the three representative couplings.}
\label{fig:calibrate_h_3d}
\end{figure*}

\begin{figure*}[!tbp]
\centering
\includegraphics[width=\textwidth]{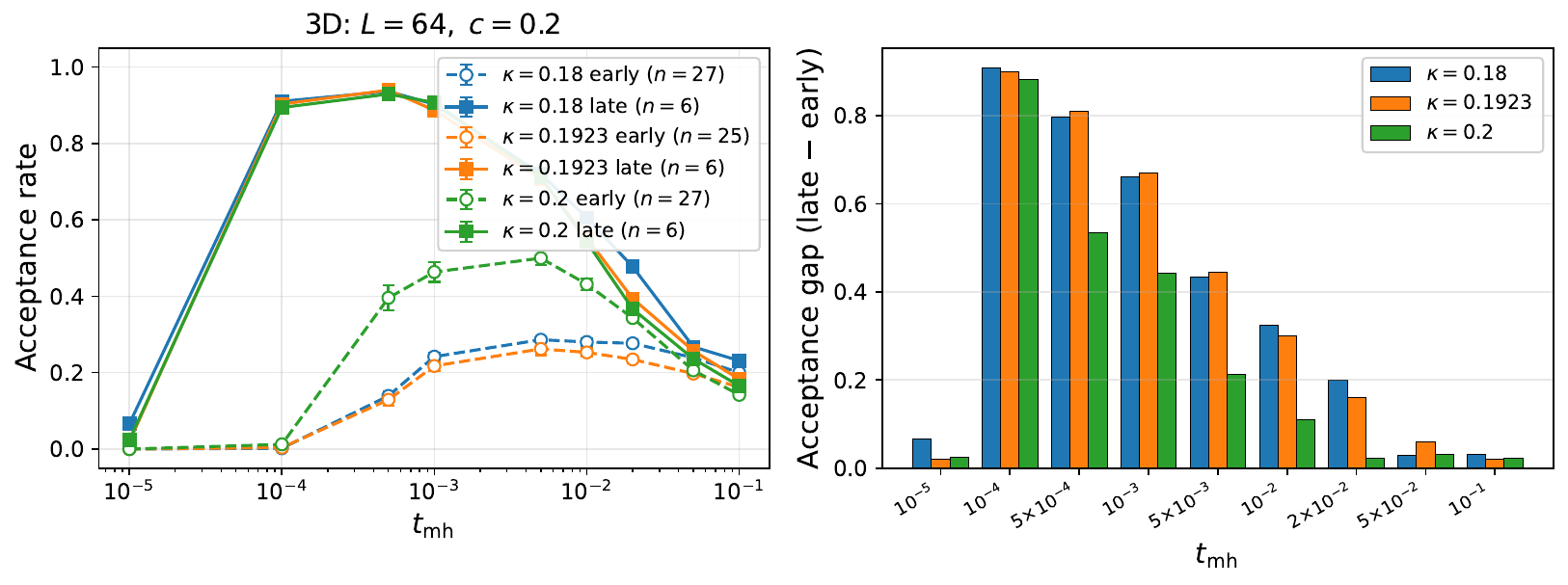}
\caption{Three-dimensional calibration of $t_{\text{mh}}$ at $L=64$, with $c=0.2$ or $h=c/L^3$, $N=256$ reference configurations, a single MH step and EMA-averaged score weights. The combined plot uses $t_{\text{mh}}$ on the horizontal axis so that the three runs with $\sigma=2000$, $2048$ and $2760$ can be compared directly. The maximum of the diagnostic gap occurs at $t_{\text{mh}}=10^{-4}$, as in two dimensions.}
\label{fig:calibrate_tmh_3d}
\end{figure*}

The same calibration in three dimensions is shown in \fig{fig:calibrate_h_3d} and \fig{fig:calibrate_tmh_3d}. The step-size sweep at $L=64$ again favors $c=0.2$ within a broad maximum of $\Delta A$. The diffusion-time sweep again selects $t_{\text{mh}}=10^{-4}$. For the trained three-dimensional noise scales $\sigma=2000$, $2048$ and $2760$, this corresponds to an injected noise scale of approximately $10^{-2}$ in the normalized field variable. The same calibrated pair $(c,t_{\text{mh}})=(0.2,10^{-4})$ is therefore used for all MALA diagnostics in both two and three dimensions, including the cross-$L$ table in \Sec{sec:crossL}, which differs only in the noise scales $\sigma$ stated in its caption.

\section{NCSN++ score network architecture and training}
\label{app:ncsnpp}

\begin{figure*}[!tbp]
\centering
\includegraphics[width=\textwidth]{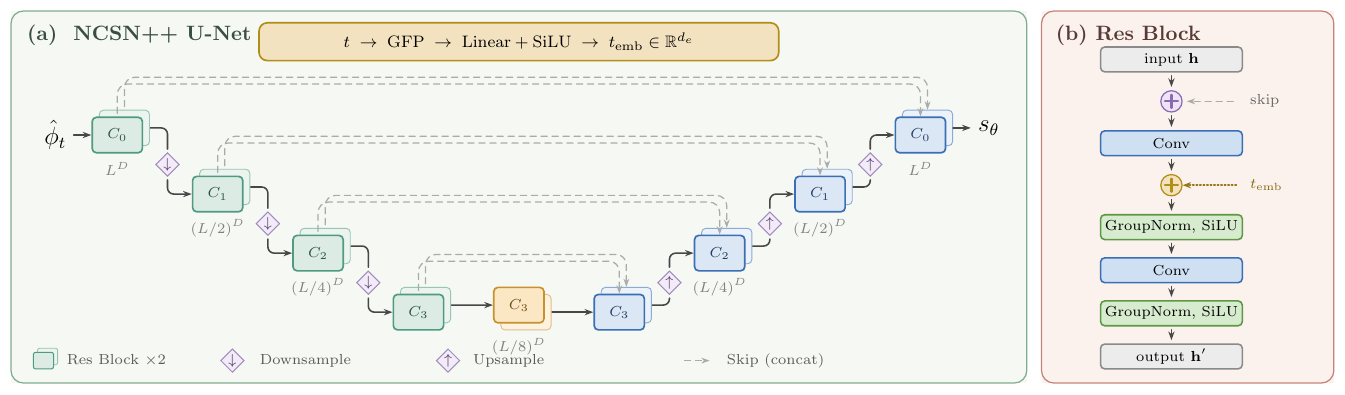}
\caption{Modified NCSN++ score network used for the lattice $\phi^4$ ensembles. Panel (a) shows the U-Net map from a noised field on a periodic $L^D$ lattice to a score field on the same lattice.  The encoder applies three stride two downsamplings.  For the larger lattices that are multiples of eight, this reduces the spatial resolution to $(L/8)^D$.  The smallest training lattice $L=4$ is handled by the same map, with its bottleneck reduced to a single site (\App{app:ncsnpp_implementation}).  The decoder restores the original resolution using features stored at the corresponding encoder scales.  The fully convolutional map is therefore evaluated with shared weights on all lattice sizes used in the cross-$L$ tests.  The scalar diffusion time $t$ is encoded by Gaussian Fourier projection and injected into every block.  The final output is divided by the VE marginal width $\Sigma(t)$.  Panel (b) shows the time conditioned convolutional block.  It follows the two convolution layout and time conditioning of the NCSN++ ResBlock, with the internal residual addition omitted for training stability.  All spatial convolutions use circular padding, matching the periodic boundary conditions of the lattice theory.}
\label{fig:architecture}
\end{figure*}

This appendix describes the score-network ansatz used in the diffusion-model runs.  The starting point is the NCSN++ architecture of the score-based SDE formulation in Ref.~\cite{Song2021SDE}, building on the earlier noise conditional score network construction of Ref.~\cite{SongErmon2019}. In its original setting, NCSN++ is a noise conditioned U-Net used to approximate the score along a diffusion trajectory.  We use the same architectural idea for lattice $\phi^4$ ensembles, with modifications adapted to periodic scalar fields.  \Fig{fig:architecture} summarizes the network, with the U-Net map in panel (a) and the time-conditioned convolutional block in panel (b).  All spatial convolutions use circular padding.  The two-dimensional and three dimensional networks share a dimension independent implementation. The residual addition inside the nominal ResBlocks is omitted in the production networks, while the skip concatenations between encoder and decoder are kept. This last choice is empirical.  When the internal residual addition was kept, the denoising score matching loss often became unstable after a few tens to about one hundred epochs.  Omitting this addition gave stable optimization in all production runs.

\subsection{Score-network ansatz}
\label{app:ncsnpp_implementation}

The learned map is $s_\theta(\hat{\phi},t)$, the score in the normalized field coordinate used by the neural network.  At zero diffusion time, the score in the raw field variable is fixed by the lattice action.  Since $p_0(\phi)\propto e^{-S[\phi]}$,
\begin{equation}
  s_y(\phi,0)=-\frac{\partial S}{\partial \phi_y}.
\end{equation}
For the lattice action in \Eq{eq:phi4_action}, this derivative gives
\begin{equation}
-\frac{\partial S}{\partial\phi_y}
=
2\kappa\sum_{\mu=1}^{D}
\left(\phi_{y+\hat\mu}+\phi_{y-\hat\mu}\right)
-2\phi_y-4\lambda\phi_y\left(\phi_y^2-1\right) .
\end{equation}
This expression motivates a local and translation-equivariant ansatz.  At small $t$, the exact score is controlled by on-site and nearest-neighbor fields.  At larger $t$, the Gaussian convolution defining $p_t$ in \Eq{eq:ve_marginal_distribution} couples this local structure to broader collective variations of the field distribution.  A convolutional U-Net provides local kernels at each resolution together with a multiscale path for long-wavelength modes.

Before training, the raw field is linearly rescaled to the interval $[-1,1]$,
\begin{equation}
  \hat{\phi}=2\,\frac{\phi-\phi_{\min}}{\phi_{\max}-\phi_{\min}}-1,
  \qquad \hat{\phi}\in[-1,1],
\end{equation}
where $\phi_{\min}$ and $\phi_{\max}$ are computed from the corresponding training ensemble and stored in the checkpoint.  During training a mini-batch is represented as
\begin{equation}
  \hat{\phi}\in\mathbb{R}^{B\times1\times L^D}.
\end{equation}
Here $B$ is the mini-batch size.  The factor $1$ denotes the single scalar-field channel, and $L^D$ denotes the number of lattice sites.  In the implementation, the last factor is stored as the usual $D$-index lattice array.  The network returns a tensor with the same lattice content, representing the score in the normalized field coordinate.  Generated configurations are transformed back to the original field variable before any observable is measured.

The architectural choices are tied to the lattice problem.  Circular padding implements the periodic boundary condition and avoids artificial boundary sites. The absence of dense layers over lattice sites makes the network a local functional with shared weights across the lattice.  This is the structural basis of the cross-$L$ tests in \Sec{sec:crossL}.  The encoder--decoder hierarchy combines the local zero-time structure above with the soft modes that become important close to criticality.  It is best regarded as a practical multiscale parametrization of the score.

The 2D and 3D networks use the same topology with different channel widths.  The 2D runs use $(16,32,64,128)$ and the 3D runs use $(10,20,40,80)$.  These widths were chosen so that the 2D and 3D score networks have comparable size after replacing each $3\times3$ convolution by a $3\times3\times3$ convolution.  The resulting parameter counts are about $2.64\times10^6$ in 2D and $2.87\times10^6$ in 3D.  This makes the comparison across dimensions a test at nearly fixed neural network capacity.  It also shows that a score network of roughly the same scale can treat both 2D and 3D scalar theories without a large increase in parameter count or training time.  Each resolution level contains two time conditioned convolutional blocks, with two further blocks at the bottleneck.  The decoder receives the two stored encoder features from the matching resolution by channel concatenation.  Self attention is disabled in all production runs, and long distance information is carried by the convolutional hierarchy.  The global $\mathbb{Z}_2$ symmetry is learned from the symmetrized training ensembles.

The same map is applied at every lattice size in the multi-$L$ training set, including the smallest lattice $L=4$, without any size specific code path.  Each encoder stage halves the spatial extent with a stride two convolution of kernel size three and unit circular padding.  The output extent of one such step is $\lfloor (L-1)/2\rfloor+1$.  For $L$ a multiple of eight the three steps give the $(L/8)^D$ bottleneck quoted above.  At $L=4$ the extent instead follows $4\to2\to1\to1$.  The first two downsamplings already reduce each axis to a single site, and the third acts trivially.  The bottleneck is then a single lattice site.  Circular padding stays well defined on a length one axis because the required one site pad equals the axis length.  The decoder does not assume a fixed factor of two upsampling.  Each upsampling stage restores the spatial shape stored at the matching encoder scale, so the collapsed $1\to1\to2\to4$ chain is inverted exactly and the output returns to the input lattice $L=4$.  In effect the network uses about two of its three resolution levels at $L=4$, with a bottleneck receptive field that already spans the whole lattice.

The scalar diffusion time $t$ is an input to the network because the score changes along the noising trajectory.  We represent this single scalar by fixed sine and cosine features,
\begin{equation}
\begin{split}
  r(t)=(&\sin(2\pi t W_1),\ldots,\sin(2\pi t W_{128}),\\
        &\cos(2\pi t W_1),\ldots,\cos(2\pi t W_{128}))
        \in\mathbb{R}^{256}.
\end{split}
\label{eq:fourier_time_embedding_app}
\end{equation}
The frequencies $W_i$ are drawn once from ${\cal N}(0,30^2)$ before training and then kept fixed.  The vector $r(t)$ is a Fourier-feature representation of the diffusion time.  It gives the following learned layers access to both slowly and rapidly varying functions of $t$.  A learned linear layer and a SiLU nonlinearity then produce a 256-dimensional time embedding.  Inside each convolutional block, this embedding is projected to the current channel width and added after the first convolution.

Let $u_\theta(\hat{\phi},t)$ denote the raw U-Net output.  The score returned by the model is
\begin{equation}
  s_\theta(\hat{\phi},t)=\frac{u_\theta(\hat{\phi},t)}{\Sigma(t)},
\end{equation}
where $\Sigma(t)$ is the VE marginal standard deviation in \Eq{eq:ve_sigma_schedule}.  With this convention, $u_\theta$ learns an $O(1)$ denoising direction and the explicit factor $1/\Sigma(t)$ restores the scale of $\nabla_{\hat{\phi}}\log p_t(\hat{\phi})$.

\subsection{Training objective and optimization}
\label{app:ncsnpp_training}

All networks are trained with the VE denoising score-matching objective in \Eq{eq:dsm_loss_ve}.  For a mini-batch of normalized fields $\hat{\phi}$, we sample $t\sim U(10^{-5},1)$ and $\eta\sim{\cal N}(0,\mathbb{1})$, form
\begin{equation}
  \hat{\phi}_t=\hat{\phi}+\Sigma(t)\eta,
\end{equation}
and minimize,
\begin{equation}
{\cal L}_{\rm batch}
=
\frac{1}{B}\sum_{b=1}^{B}
\left\|
\Sigma(t_b)s_\theta(\hat{\phi}^{(b)}_{t_b},t_b)+\eta_b
\right\|_2^2 .
\label{eq:batch_loss_app}
\end{equation}
Here $B$ is the mini-batch size.  The norm is taken over all lattice sites and the single field channel.  The optimizer uses this single scalar objective.  The UV, mid and IR histories below are diagnostics computed from the same per-sample losses by grouping training examples according to diffusion time.

The in-distribution 2D production runs use $L=128$, $\lambda=0.022$, and $10240$ training configurations at $\kappa=0.26,0.2705,0.28$, with noise scales $\sigma=300,450,640$, respectively.  The corresponding 3D runs use $L=64$, $\lambda=0.9$, and $2560$ configurations at $\kappa=0.18,0.1923,0.2$, with $\sigma=2000,2048,2760$.  Mini-batch sizes are $128$ in 2D and $64$ in 3D.  The 2D models are trained for $10000$ epochs and the 3D models for $20000$ epochs. Optimization uses Adam with learning rate $10^{-3}$ and no weight decay.

Training is run with PyTorch graph compilation through \texttt{torch.compile} and mixed bfloat16 precision.  This combination substantially reduces wall-clock time while preserving the stability of the score-matching optimization.  Among the low-precision formats we tested, bfloat16 was the lowest precision that remained stable.  Float16 made the training loss unstable.  An exponential moving average of the network parameters is maintained from the start of training.  The EMA decay is $0.9999$ for the 2D runs and for the 3D $\kappa=0.18,0.2$ runs.  The saved 3D near-critical production run uses decay $0.999$.  Unless stated otherwise, all sampling and diagnostics use the EMA-averaged weights.  Production samples are generated by the reverse-SDE Euler--Maruyama sampler with $2000$ steps.  The 2D runs use a logarithmic time grid and the 3D in-distribution runs use a linear time grid.

\subsection{Diffusion-time loss diagnostics in three dimensions}
\label{app:training_loss_3d}

\begin{figure*}[!tbp]
\centering
\includegraphics[width=\textwidth]{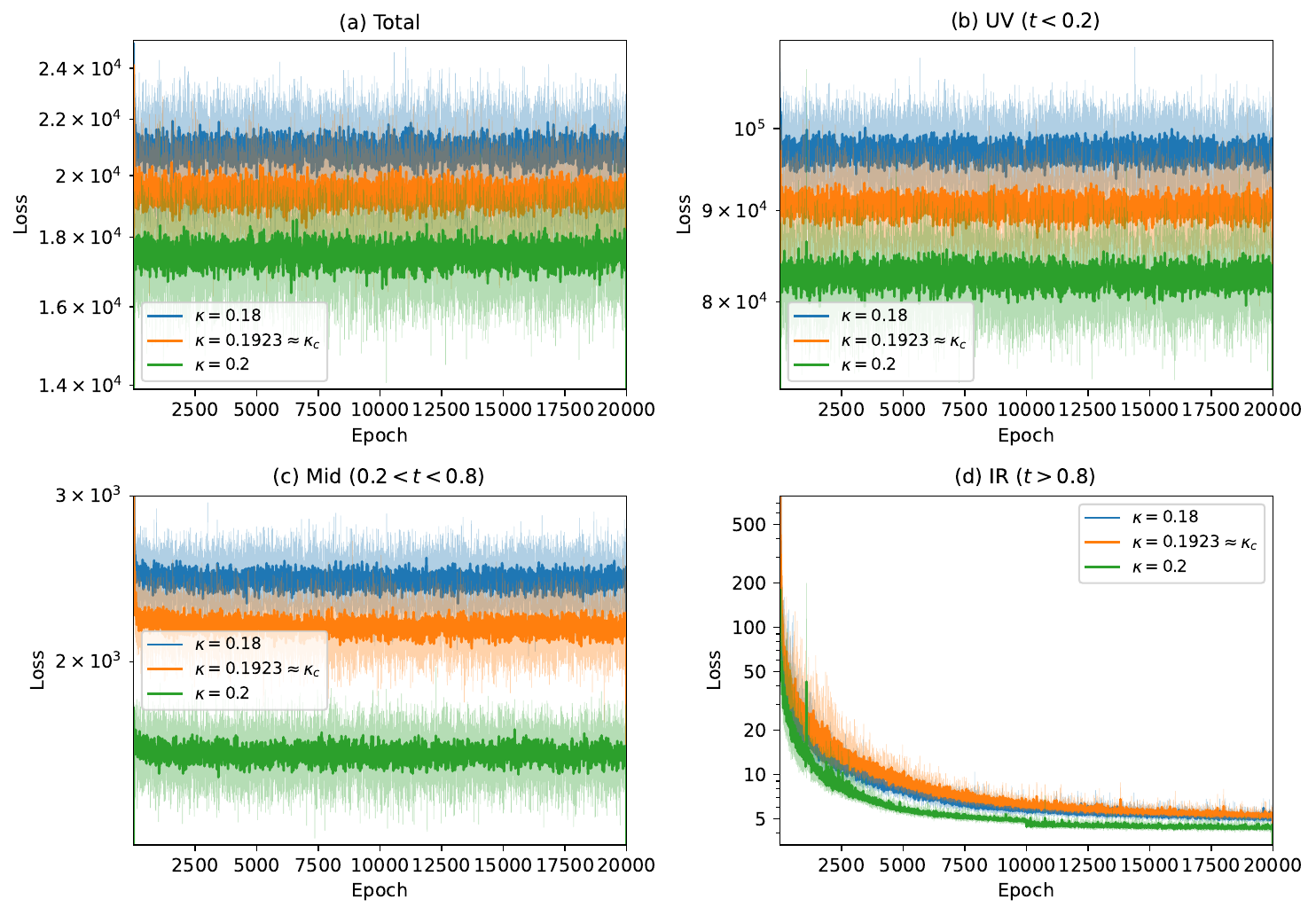}
\caption{Diffusion-time decomposition of the denoising score-matching loss for the three-dimensional runs at $L=64$, $\lambda=0.9$.  In each panel the per-sample loss $\ell_b$ of \Eq{eq:per_sample_loss_app}, averaged over the indicated range of diffusion time, is plotted against training epoch for $\kappa=0.18$ (symmetric), $\kappa=0.1923\simeq\kappa_c$ (near-critical), and $\kappa=0.2$ (broken).  Panel (a) is the total loss, the mini-batch average over $t\sim U(10^{-5},1)$.  Panels (b), (c), and (d) restrict the average to the UV ($t<0.2$), mid ($0.2\le t<0.8$), and IR ($t>0.8$) bins, in order of increasing VE noise scale $\Sigma(t)$.  The UV bin sets the scale of the total loss and saturates early, while the IR bin, shown on a logarithmic vertical axis, keeps decreasing by more than an order of magnitude before reaching a plateau of order five.  Lighter traces show the unsmoothed per-epoch loss and the darker curves a moving average over epochs.}
\label{fig:loss_3d_L64}
\end{figure*}

The curves in \fig{fig:loss_3d_L64} decompose the recorded denoising score-matching loss according to the diffusion time sampled for each training configuration.  This decomposition is useful because the VE noise scale $\Sigma(t)$ fixes the resolution of the denoising problem.  For each sample we record,
\begin{equation}
\ell_b(t_b)=
\left\|
\Sigma(t_b)s_\theta(\hat{\phi}^{(b)}_{t_b},t_b)+\eta_b
\right\|_2^2 .
\label{eq:per_sample_loss_app}
\end{equation}
The total curve is the mini-batch average of $\ell_b$.  The other three curves are conditional averages over samples whose diffusion times fall in
\begin{equation}
  10^{-5}\le t<0.2,
  \qquad
  0.2\le t<0.8,
  \qquad
  0.8\le t\le 1 .
\end{equation}
We label these bins UV, mid and IR according to the denoising resolution.  This classification depends only on $t$, or equivalently on $\Sigma(t)$.  No Fourier projection is involved.

The runs shown in \fig{fig:loss_3d_L64} have $L=64$ and $\lambda=0.9$, with $\kappa=0.18$, $\kappa=0.1923\simeq\kappa_c$, and $\kappa=0.2$.  Panels (a), (b), and (c) show that the total loss, the UV-bin loss, and the mid-bin loss reach a plateau within the early part of training.  Their ordering is the same in the three panels.  The symmetric run at $\kappa=0.18$ has the largest loss, the broken run at $\kappa=0.2$ has the smallest loss, and the near-critical run lies between them.  The UV-bin loss is of order $8\times 10^4$ to $10^5$, whereas the mid-bin loss is of order $10^3$.  The total loss therefore mainly follows the UV-bin loss.

Panel (d) shows a qualitatively different behavior.  The IR-bin loss continues to decrease after the total, UV-bin, and mid-bin losses have nearly saturated, and approaches values of order five at late epochs.  The near-critical run remains slightly above the two off-critical runs, while the broken run gives the smallest IR loss.  This behavior shows why the decomposition in diffusion time, or equivalently in noise scale, is a useful diagnostic.  Small diffusion times give the largest contribution to the total loss because the noisy fields remain close to the original configurations, where local fluctuations are only weakly smoothed.  At large diffusion times, most local structure of the field has been washed out by the noise.  The target score is then largely fixed by the Gaussian noise kernel in \Eq{eq:ve_sde_kernel}, which is a simpler object than the score near the original data distribution.  This explains why the IR-bin loss is much smaller than the UV-bin loss.  The remaining decrease of the IR-bin loss shows that the network is still learning the large-scale structure of the lattice-field distribution.  A total-loss curve alone would therefore obscure this continued learning in the large-noise regime, since its value is dominated by the UV contribution.

\section{Estimator for the HMC-referenced effective sample size}
\label{app:ess_details}

This appendix gives the estimator behind \fig{fig:ess_mse_2d} and \fig{fig:ess_mse_3d}. The main text writes the effective sample size in terms of the exact target mean and variance. Here we replace those by quantities measured from the finite DM and HMC ensembles, and we account for the fact that the HMC configurations are correlated along the Markov chain.

We need three ingredients for each observable $O$, and for each propagator mode treated as an observable. The first is the single-configuration variance of $O$, which sets the numerator of the effective sample size. We estimate it from the HMC ensemble, $\widehat\sigma_{p,O}^2=\widehat\sigma_{{\rm HMC},O}^2$. The second is the uncertainty of the DM mean, the usual error on a mean of $N_{\rm DM}$ independent draws,
\begin{equation}
\widehat V_{\rm DM}=\frac{\widehat\sigma_{{\rm DM},O}^2}{N_{\rm DM}} .
\end{equation}
The third is the uncertainty of the HMC mean. Because the HMC configurations are correlated along the chain rather than independent, this variance is larger than the independent-sample value $\widehat\sigma_{{\rm HMC},O}^2/N_{\rm HMC}$ by the factor $2\widehat\tau_{{\rm int},O}^{\rm HMC}$,
\begin{equation}
\widehat V_{\rm HMC}=\frac{2\widehat\tau_{{\rm int},O}^{\rm HMC}\,\widehat\sigma_{{\rm HMC},O}^2}{N_{\rm HMC}} .
\label{eq:vhmc_app}
\end{equation}
Here $N_{\rm HMC}$ is the chain length, and the autocorrelation time comes from the normalized autocorrelation function $\widehat\rho_O(t)$,
\begin{equation}
\widehat\tau_{{\rm int},O}^{\rm HMC}(W)=\frac{1}{2}+\sum_{t=1}^{W}\widehat\rho_O(t) .
\end{equation}
We cut the sum at the first window $W$ with $W\ge 5\widehat\tau_{{\rm int},O}^{\rm HMC}(W)$, and use that value in \eq{eq:vhmc_app}. This correction matters most near the critical point, where the long-wavelength modes of the chain decorrelate slowly.

We do not know the exact target mean, so we estimate the DM bias from the difference of the DM and HMC sample means. The squared difference,
\begin{equation}
D_O^2=(\bar O_{\rm DM}-\bar O_{\rm HMC})^2,
\end{equation}
overestimates the squared bias $\delta_O^2$, because the two sample means each fluctuate. On average it exceeds $\delta_O^2$ by the combined sampling variance $\widehat V_{\rm DM}+\widehat V_{\rm HMC}$, so we subtract this variance and floor the result at zero,
\begin{equation}
\widehat{\delta_O^2}=\max\!\left[D_O^2-\widehat V_{\rm DM}-\widehat V_{\rm HMC},\,0\right] .
\label{eq:delta2_shrink}
\end{equation}
The floor sets $\widehat{\delta_O^2}=0$ whenever the DM-HMC difference is smaller than the combined sampling error, so that an unresolved difference counts as no bias. The quantity plotted in the figures is then
\begin{equation}
\widehat N_{{\rm eff,ref},O}=\frac{\widehat\sigma_{p,O}^2}{\widehat{\delta_O^2}+\widehat V_{\rm DM}+\widehat V_{\rm HMC}}=\frac{\widehat\sigma_{p,O}^2}{\max\!\left[D_O^2,\,\widehat V_{\rm DM}+\widehat V_{\rm HMC}\right]} .
\label{eq:neff_ref_app}
\end{equation}
When a bias is resolved the denominator is the squared DM-HMC difference, and $\widehat N_{{\rm eff,ref},O}$ is limited by that bias. When none is resolved it reduces to $\widehat\sigma_{p,O}^2/(\widehat V_{\rm DM}+\widehat V_{\rm HMC})$, set by the sampling errors alone.

We obtain error bars by a bootstrap that resamples the configurations $400$ times. Because the DM configurations are independent, each resample draws $N_{\rm DM}$ of them at random with replacement. The HMC configurations are correlated along the chain, so we instead resample them in contiguous blocks of length $\lceil 2\widehat\tau_{{\rm int},O}^{\rm HMC}\rceil$, which preserves the autocorrelation. The autocorrelation time is measured once on the full chain and held fixed during the resampling. We quote the full-data value as the central estimate, and the $16\%$ and $84\%$ points of the resampled distribution as the error bar. We use these points rather than the average of the resamples because $N_{\rm eff}$ varies as one over an estimated error and develops a long tail toward large values when \eq{eq:delta2_shrink} is near zero.

We include only $\mathbb{Z}_2$-even observables. The signed mean $\bar\phi$ has exact target value zero by symmetry, so its HMC estimate is just finite-chain noise about zero, and including it would only probe the residual symmetry breaking of the ensembles rather than a physical bias. The scalar observables used in \fig{fig:ess_mse_2d} and \fig{fig:ess_mse_3d} are $\{|\bar\phi|,\bar\phi^2,\bar\phi^4,S/V\}$.

For the propagator we apply the same estimator separately to each diagonal momentum defined in \Sec{sec:propagator_results}. For mode $n$ the observable is,
\begin{equation}
G_n(\phi)=|\widetilde\phi(q_n)|^2,
\qquad
q_{n,1}=\cdots=q_{n,D}=\frac{2\pi n}{L},
\end{equation}
with $|k_n|=\sqrt{D}\,2\sin(\pi n/L)$ and $n=1,\ldots,L/2$. We compute $\widehat N_{{\rm eff,ref},n}$ for each mode and then average over a band of modes with equal weight,
\begin{equation}
\widehat N_{{\rm eff,ref},{\cal B}}=\frac{1}{|{\cal B}|}\sum_{n\in{\cal B}}\widehat N_{{\rm eff,ref},n} .
\end{equation}
Averaging the per-mode $\widehat N_{{\rm eff,ref},n}$ this way, rather than first averaging $G_n$ over the band, keeps every momentum on an equal footing. A direct average of $G_n$ would be dominated by the smallest momenta, where the propagator is largest close to criticality. The three bands are $n=1$ for $G_{\rm min}$, $n\in[L/8,L/4)$ for $G_{\rm mid}$, and $n\in[3L/8,L/2]$ for $G_{\rm large}$. Within a band the bootstrap uses the same configuration indices for all modes, and the HMC block length is set by the slowest mode in the band, $\lceil 2\max_{n\in{\cal B}}\widehat\tau_{{\rm int},n}^{\rm HMC}\rceil$.

\subsection{Free-field check of the ESS estimator}
\label{app:ess_free_field}

An independent check of the MSE-matched ESS estimator is obtained in the free-field limit, where the target distribution and the VE marginals are known exactly. We take the free-field limit of \Eq{eq:phi4_action} by setting $\lambda=0$,
\begin{equation}
S_0[\phi]
=
\sum_x
\left[
\phi_x^2
-
2\kappa\sum_{\mu=1}^{D}\phi_x\phi_{x+\hat\mu}
\right].
\label{eq:free_phi4_action}
\end{equation}
This action can be written as,
\begin{equation}
S_0[\phi]
=
\frac{1}{2}
\sum_{x,y}\phi_x (C^{-1})_{xy}\phi_y,
\label{eq:free_action_C}
\end{equation}
with,
\begin{equation}
(C^{-1})_{xy}
=
2\delta_{xy}
-
2\kappa
\sum_{\mu=1}^{D}
\left(
\delta_{x+\hat\mu,y}
+
\delta_{x-\hat\mu,y}
\right),
\label{eq:Cinv_phi4_free}
\end{equation}
where periodic boundary conditions are understood. The corresponding target distribution is,
\begin{equation}
p_0(\phi)
=
\frac{1}{(2\pi)^{V/2}\det(C)^{1/2}}
\exp\left[-\frac{1}{2}\phi^T C^{-1}\phi\right],
\label{eq:p0_free}
\end{equation}
where $V=L^D$. The covariance is diagonal in the Fourier basis defined in \Eq{eq:fourier_field}. Its eigenvalues are,
\begin{equation}
C_q^{-1}
=
2\left(
1-2\kappa\sum_{\mu=1}^{D}\cos q_\mu
\right)
=
2\left(1-2D\kappa+\kappa\hat q^2\right),
\label{eq:Cq_inv_free}
\end{equation}
with,
\begin{equation}
\hat q^2=4\sum_{\mu=1}^{D}\sin^2\frac{q_\mu}{2}.
\end{equation}
Equivalently,
\begin{equation}
C_{xy}=\frac{1}{V}\sum_q e^{iq\cdot(x-y)}C_q .
\label{eq:Cxy_free}
\end{equation}
The Gaussian distribution is normalizable for $\kappa<1/(2D)$, which keeps the zero-mode variance finite and positive.

For the VE forward process used in this work,
\begin{equation}
\phi_t=\phi_0+\Sigma(t)\eta,
\qquad
\eta\sim {\cal N}(0,\mathbb{1}),
\end{equation}
with,
\begin{equation}
\Delta_t\equiv \Sigma^2(t)
=
\frac{\sigma^{2t}-1}{2\ln\sigma}
\label{eq:Delta_t_free}
\end{equation}
for $g(t)=\sigma^t$. Since the forward kernel is Gaussian, the marginal distribution remains Gaussian,
\begin{equation}
p_t(\phi)
=
\frac{1}{(2\pi)^{V/2}\det(C+\Delta_t\mathbb{1})^{1/2}}
\exp\left[-\frac{1}{2}\phi^T(C+\Delta_t\mathbb{1})^{-1}\phi\right].
\label{eq:pt_free}
\end{equation}
Thus the diffusion process changes the covariance according to,
\begin{equation}
C_q(t)=C_q+\Delta_t .
\label{eq:Ct_free}
\end{equation}
This relation gives an analytic trajectory from the target distribution at $t=0$ to the broad VE distribution at large $t$.

At a fixed diffusion time, let $N_{\rm DM}$ independent configurations be drawn from $p_t$ and used to estimate the target expectation value at $t=0$. With $\langle\cdots\rangle_t$ denoting the average over $p_t$, the MSE-matched effective sample size is,
\begin{equation}
\frac{N_{{\rm eff},O}(t)}{N_{\rm DM}}
=
\frac{{\rm Var}_{p_0}(O)}
{N_{\rm DM}\left(\langle O\rangle_t-\langle O\rangle_0\right)^2+{\rm Var}_{p_t}(O)} .
\label{eq:free_neff_general}
\end{equation}
This is the same MSE-matching definition as in \Eq{eq:neff_pure}, specialized to a case where all averages are known analytically and no HMC reference is needed.

For the zero-mode observables, $\bar\phi=V^{-1}\sum_x\phi_x$ is Gaussian with variance $C_0(t)/V$, where $C_0(t)=C_{q=0}(t)$. The three $\mathbb{Z}_2$-even observables used in the main analysis have
\begin{align}
\frac{N_{{\rm eff},|\bar\phi|}(t)}{N_{\rm DM}}
&=
\frac{\left(1-\frac{2}{\pi}\right)C_0}
{N_{\rm DM}\frac{2}{\pi}\left(\sqrt{C_0(t)}-\sqrt{C_0}\right)^2+
\left(1-\frac{2}{\pi}\right)C_0(t)},
\label{eq:free_neff_absm}
\\
\frac{N_{{\rm eff},\bar\phi^2}(t)}{N_{\rm DM}}
&=
\frac{2C_0^2}
{N_{\rm DM}\left(C_0(t)-C_0\right)^2+2C_0(t)^2},
\label{eq:free_neff_m2}
\\
\frac{N_{{\rm eff},\bar\phi^4}(t)}{N_{\rm DM}}
&=
\frac{96C_0^4}
{9N_{\rm DM}\left(C_0(t)^2-C_0^2\right)^2+96C_0(t)^4}.
\label{eq:free_neff_m4}
\end{align}
The powers of $V$ cancel in these ratios because the target and noised zero-mode variances scale in the same way with the lattice volume.

For the free action density,
\begin{equation}
\frac{S_0[\phi]}{V}
=
\frac{1}{2V}\phi^T C^{-1}\phi .
\end{equation}
Gaussian Wick contractions give
\begin{align}
\left\langle \frac{S_0}{V}\right\rangle_t
&=
\frac{1}{2V}{\rm Tr}\left[C^{-1}(C+\Delta_t\mathbb{1})\right]
=
\frac{1}{2}+\Delta_t,
\label{eq:free_action_mean}
\\
{\rm Var}_{p_t}\left(\frac{S_0}{V}\right)
&=
\frac{1}{2V^2}
\sum_q\left[1+\Delta_t C_q^{-1}\right]^2 .
\label{eq:free_action_var}
\end{align}
In the last step we used ${\rm Tr}\,C^{-1}=2V$ for the free action in \Eq{eq:free_phi4_action}. Equation \eqref{eq:free_neff_general} then gives,
\begin{equation}
\frac{N_{{\rm eff},S_0/V}(t)}{N_{\rm DM}}
=
\frac{1/(2V)}
{N_{\rm DM}\Delta_t^2+\frac{1}{2V^2}\sum_q\left[1+\Delta_t C_q^{-1}\right]^2} .
\label{eq:free_neff_action}
\end{equation}

For a single propagator mode $G_q[\phi]=|\widetilde\phi(q)|^2$,
\begin{equation}
\langle G_q\rangle_t=C_q(t),
\qquad
{\rm Var}_{p_t}(G_q)=\alpha_q C_q(t)^2,
\label{eq:free_Gq_moments}
\end{equation}
where,
\begin{equation}
\alpha_q=
\begin{cases}
1, & q\neq -q\;({\rm mod}\;2\pi),\\
2, & q=-q\;({\rm mod}\;2\pi).
\end{cases}
\end{equation}
The corresponding ESS ratio is,
\begin{equation}
\frac{N_{{\rm eff},G_q}(t)}{N_{\rm DM}}
=
\frac{\alpha_q C_q^2}
{N_{\rm DM}\Delta_t^2+\alpha_q C_q(t)^2} .
\label{eq:free_neff_Gq}
\end{equation}
For a momentum band ${\cal B}$ we follow the same mode-wise convention as in the interacting analysis,
\begin{equation}
\frac{N_{{\rm eff},{\cal B}}(t)}{N_{\rm DM}}
=
\frac{1}{|{\cal B}|}
\sum_{q\in{\cal B}}
\frac{\alpha_q C_q^2}
{N_{\rm DM}\Delta_t^2+\alpha_q C_q(t)^2} .
\label{eq:free_neff_Gband}
\end{equation}

All expressions above give $N_{{\rm eff},O}(0)/N_{\rm DM}=1$, since $\Delta_t=0$ and $p_t=p_0$ at $t=0$. At large diffusion time the added VE noise dominates $C_q(t)$ over a growing range of modes, while the bias term in \Eq{eq:free_neff_general} also increases. The analytic ratios therefore decrease away from unity as the marginal distribution moves away from the target. The free-field calculation thus verifies that the estimator returns the independent-target value for $p_t=p_0$ and identifies the loss of efficiency when the sample distribution is broadened by VE noise.

\subsection{Likelihood ESS in the free-field limit}
\label{app:ess_free_field_likelihood}

The diagnostics used in the main text are based on an MSE-matched, observable-wise ESS. Flow-based lattice samplers~\cite{Albergo:2021vyo,Gerdes:2022eve, Kanwar:2024ujc,Bauer:2024byr,Nicoli:2020njz} instead quantify efficiency through reweighting. A configuration drawn from the proposal $p_t$ is corrected toward the target $p_0$ by the importance weight $w=p_0/p_t$, and the quality of this correction is characterized by a likelihood-based effective sample size. The free-field limit is the one setting in which both $p_0$ (\Eq{eq:p0_free}) and $p_t$ (\Eq{eq:pt_free}) are known in closed form, so the likelihood ESS can be evaluated exactly and compared with the MSE-matched estimator of \Eq{eq:free_neff_general}.

At a fixed diffusion time the configurations are drawn from $p_t$. We denote the sample size by $N_{\rm samp}$, and the importance weight is,
\begin{equation}
w_t(\phi)=\frac{p_0(\phi)}{p_t(\phi)}\,,
\label{eq:free_likelihood_weight}
\end{equation}
and the likelihood ESS~\cite{Albergo:2021vyo} is defined as,
\begin{equation}
\frac{N_{\rm eff}^{\rm like}(t)}{N_{\rm samp}}
=\frac{\langle w_t\rangle_t^2}{\langle w_t^2\rangle_t}\,.
\label{eq:free_likelihood_ess_ratio}
\end{equation}
Both densities are normalized, so the weight has unit mean and a second moment fixed by its variance,
\begin{equation}
\langle w_t\rangle_t=1 ,
\qquad
\langle w_t^2\rangle_t=1+{\rm Var}_t(w_t)\ge 1 .
\label{eq:free_weight_moments}
\end{equation}
The likelihood ESS is simplified to the inverse second moment of the weights,
\begin{equation}
\frac{N_{\rm eff}^{\rm like}(t)}{N_{\rm samp}}
=\frac{1}{\langle w_t^2\rangle_t},
\qquad
\langle w_t^2\rangle_t=\int d\phi\,\frac{p_0(\phi)^2}{p_t(\phi)} .
\label{eq:free_likelihood_ess_def}
\end{equation}
By construction the ratio lies in $(0,1]$ and reaches unity only when the weights are constant, that is when $p_t=p_0$. Its finite-sample (Kish) estimator is,
\begin{equation}
\frac{\widehat N_{\rm eff}^{\rm like}}{N_{\rm samp}}
=\frac{\left(\sum_{i=1}^{N_{\rm samp}}w_t(\phi_i)\right)^2}
{N_{\rm samp}\sum_{i=1}^{N_{\rm samp}}w_t(\phi_i)^2},
\qquad \phi_i\sim p_t .
\label{eq:free_likelihood_ess_estimator}
\end{equation}

For Gaussian $p_0$ and $p_t$ the second moment is also a Gaussian integral. Writing the covariance of $p_t$ as $C_t\equiv C+\Delta_t\mathbb{1}$ (\Eq{eq:Ct_free}),
\begin{equation}
\frac{p_0(\phi)^2}{p_t(\phi)}
=\frac{\det(C_t)^{1/2}}{(2\pi)^{V/2}\det(C)}
\exp\!\left[-\frac{1}{2}\phi^T\!\left(2C^{-1}-C_t^{-1}\right)\phi\right] .
\label{eq:free_weight_square_density}
\end{equation}
The integral converges provided $2C^{-1}-C_t^{-1}$ is positive definite. In the Fourier basis its eigenvalues are $2/C_q-1/C_q(t)$. Since $C_q(t)=C_q+\Delta_t\ge C_q>0$, one has $2/C_q-1/C_q(t)\ge 1/C_q>0$ for every mode, so the condition holds at all diffusion times, and,
\begin{equation}
\langle w_t^2\rangle_t
=\frac{\det(C_t)^{1/2}}{\det(C)}
\det\!\left(2C^{-1}-C_t^{-1}\right)^{-1/2},
\label{eq:free_w2_matrix}
\end{equation}
so that,
\begin{equation}
\frac{N_{\rm eff}^{\rm like}(t)}{N_{\rm samp}}
=\frac{\det(C)}{\det(C_t)^{1/2}}
\det\!\left(2C^{-1}-C_t^{-1}\right)^{1/2} .
\label{eq:free_likelihood_ess_matrix}
\end{equation}

Since $C$ and $C_t$ are simultaneously diagonal in the Fourier basis, the determinant factorizes over lattice momenta. With $C_q(t)=C_q+\Delta_t$,
\begin{equation}
\frac{N_{\rm eff}^{\rm like}(t)}{N_{\rm samp}}
=\prod_q\frac{\sqrt{C_q\,[\,2C_q(t)-C_q\,]}}{C_q(t)}
=\prod_q\frac{\sqrt{C_q\,(C_q+2\Delta_t)}}{C_q+\Delta_t}\,.
\label{eq:free_likelihood_ess_ve}
\end{equation}
For numerical work the equivalent logarithmic form is more stable, since the product over the lattice momenta in \Eq{eq:free_likelihood_ess_ve} can underflow at large volume,
\begin{equation}
\log\frac{N_{\rm eff}^{\rm like}(t)}{N_{\rm samp}}
=\sum_q\left[
\frac{1}{2}\log\!\left(1+\frac{2\Delta_t}{C_q}\right)
-\log\!\left(1+\frac{\Delta_t}{C_q}\right)
\right] .
\label{eq:free_likelihood_ess_log}
\end{equation}

\begin{figure}
\centering
\includegraphics[width=\linewidth]{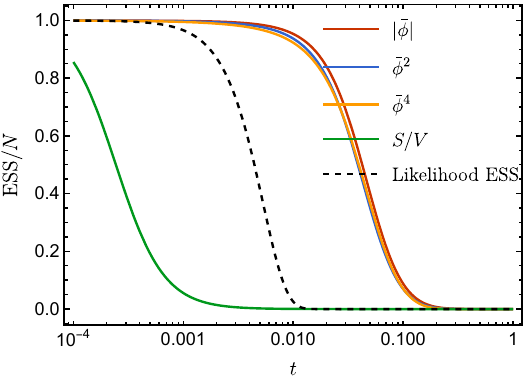}
\caption{Effective sample size ratios in the free theory at $\kappa=0.1$ on an $L=128$ two-dimensional lattice, for sample size $N_{\rm samp}=512$. Solid curves show the MSE-matched ESS of \Eq{eq:free_neff_general} for the observables $|\bar\phi|$, $\bar\phi^2$, $\bar\phi^4$, and the action density $S/V$. The dashed curve shows the likelihood ESS of \Eq{eq:free_likelihood_ess_log}, which is independent of $N_{\rm samp}$.}
\label{fig:ess_compare}
\end{figure}

Figure~\ref{fig:ess_compare} compares the MSE-matched ESS of the four observables with the likelihood ESS for the same free theory at $\kappa=0.1$ on an $L=128$ lattice in two dimensions, at sample size $N_{\rm samp}=512$. Each ratio equals unity at $t=0$ and decays to zero as $t$ increases and the VE marginal broadens. The likelihood ESS falls between the action-density curve and the three zero-mode curves and follows the same decay in $t$. The ordering reflects how each observable responds to the added noise. The action density averages over the whole lattice, so its variance under $p_0$ is of order $1/V$, while the noise shifts its mean by $\langle S/V\rangle_t-\langle S/V\rangle_0=\Delta_t$, which does not shrink with the volume. The bias term $N_{\rm samp}\Delta_t^2$ therefore overtakes the variance at $\Delta_t\sim(VN_{\rm samp})^{-1/2}$. For the zero-mode observables the mean shift and the variance carry the same power of the volume, so the volume drops out of this balance and the curves drop at $\Delta_t\sim N_{\rm samp}^{-1/2}$. The likelihood ESS is independent of $N_{\rm samp}$ and drops at $\Delta_t\sim V^{-1/2}$, which lies between the other two scales whenever $1\lesssim N_{\rm samp}\lesssim V$. The likelihood ESS therefore stays bracketed by the MSE-matched curves from $N_{\rm samp}\simeq1$, where it meets the action-density curve, to $N_{\rm samp}\simeq V$, where it meets the zero-mode curves, so the comparison at $N_{\rm samp}=512$ is representative rather than fine-tuned. This agreement supports the MSE-matched ESS as a measure of sampling efficiency.

The two ESS estimation methods differ in computational cost. The MSE-matched ESS needs only low-order moments of a chosen observable, which are cheap to estimate from samples. The likelihood ESS instead needs the ratio $p_0/p_t$ on each configuration, and its sample estimator becomes unreliable at large volume because the importance weights grow highly non-uniform, so a stable estimate would require a number of configurations that grows exponentially with the volume. The dashed curve in Fig.~\ref{fig:ess_compare} is therefore taken from the analytic expression in \Eq{eq:free_likelihood_ess_log}, which is available only because the free theory is Gaussian. For a trained diffusion model the density is not known in closed form and can be obtained only from the learned score by the costly probability flow ODE, so the likelihood ESS is far more expensive to evaluate than the MSE-matched ESS in almost any case.

The two ESS also respond differently to the sample size. The likelihood ESS ratio is independent of $N_{\rm samp}$, while the MSE-matched ratio in \Eq{eq:free_neff_general} depends on $N_{\rm samp}$ through the bias term $N_{\rm samp}(\langle O\rangle_t-\langle O\rangle_0)^2$. When the sampler carries a systematic bias $\langle O\rangle_t\neq\langle O\rangle_0$, the variance of the estimator falls with $N_{\rm samp}$ while the bias does not, so the mean squared error stops at the squared bias $(\langle O\rangle_t-\langle O\rangle_0)^2$. Matching this error to that of an unbiased estimator on the target, the effective sample size $N_{{\rm eff},O}$ approaches the finite value ${\rm Var}_{p_0}(O)/(\langle O\rangle_t-\langle O\rangle_0)^2$ and no longer grows with $N_{\rm samp}$. Drawing more configurations then lowers the ratio $N_{{\rm eff},O}/N_{\rm samp}$. The MSE-matched ESS thus reports that a systematic bias cannot be removed by increasing statistics, while the $N_{\rm samp}$-independent likelihood ESS does not provide that information.

\section{Non-conservative part of the learned score}
\label{app:nonconservative}

\begin{figure}[!tbp]
\centering
\includegraphics[width=\columnwidth]{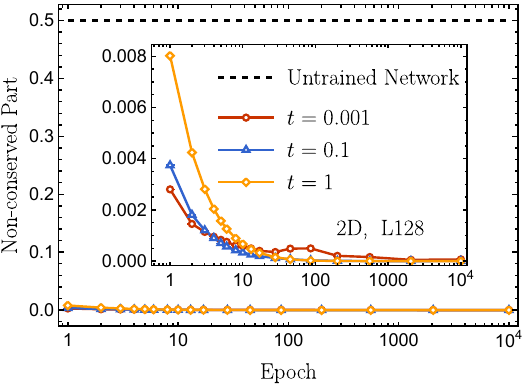}
\caption{Non-conservative fraction $R(t)$ of \Eq{eq:nonconserved_fraction} against training epoch, at $L=128$ and $\kappa=0.28$ in the broken phase, for diffusion times $t=0.001$, $0.1$, and $1$. The dashed line marks the random value $R=\tfrac12$ reached by an untrained network. The inset magnifies the trained-network curves. The fraction decreases with training and stays well below the random value at every epoch.}
\label{fig:nonconserved_2d}
\end{figure}

The exact score is the gradient of the log-density, $s(\phi,t)=\nabla_\phi\log p_t(\phi)$, so its Jacobian is the Hessian of $\log p_t$ and is symmetric. A finite score network $s_\theta$ is a generic vector field on configuration space and carries no such constraint, as noted below \Eq{eq:score_small_t}~\cite{Aarts:2025lpi}. This appendix quantifies how far the trained networks deviate from a gradient field.

Writing $\partial_j\equiv\partial/\partial\phi_j$, with $i$ and $j$ labeling the lattice sites, the Jacobian of the learned score at fixed diffusion time is $J_{ij}=\partial_j s_{\theta,i}(\phi,t)$, with $s_{\theta,i}$ the component of the score at site $i$. The learned score admits the Helmholtz decomposition
\begin{equation}
s_\theta=\nabla\Phi+r,\qquad \nabla\cdot r=0,
\end{equation}
into a conservative part $\nabla\Phi$ and a divergence-free remainder $r$ that carries the non-conservative content. On the $V$-dimensional configuration space a divergence-free field can be written as $r_i=\partial_j\mathcal{A}_{ij}$, and $\mathcal{A}_{ij}=-\mathcal{A}_{ji}$ an antisymmetric tensor potential, the analogue of $r=\nabla\times\mathcal{A}$ in three dimensions. The antisymmetric part of the Jacobian is then
\begin{equation}
A_{ij}\equiv\tfrac12\bigl(J_{ij}-J_{ji}\bigr)
=\tfrac12\bigl(\nabla\times s_\theta\bigr)_{ij}
=\tfrac12\bigl(\nabla\times r\bigr)_{ij}
=\tfrac12\,\nabla^2\mathcal{A}_{ij}.
\label{eq:antisym_curl}
\end{equation}
Here $(\nabla\times v)_{ij}\equiv\partial_j v_i-\partial_i v_j$ is the curl on configuration space and $\nabla^2=\partial_k\partial_k$ the corresponding Laplacian. The conservative part drops out because $\nabla\times\nabla\Phi=0$. The last equality holds in the gauge $\partial_i\mathcal{A}_{jk}+\partial_j\mathcal{A}_{ki}+\partial_k\mathcal{A}_{ij}=0$, which the residual freedom in $\mathcal{A}$ always permits.

The antisymmetric part vanishes if and only if the score is conservative. The matrix $A$ is therefore the complete measure of the deviation from conservativeness, and we quantify it by the non-conservative fraction
\begin{equation}
R(t)=\frac{\lVert J-J^\top\rVert_F^2}{4\,\lVert J\rVert_F^2}=\frac{\lVert A\rVert_F^2}{\lVert J\rVert_F^2}.
\label{eq:nonconserved_fraction}
\end{equation}
The symmetric and antisymmetric parts of $J$ are orthogonal in the Frobenius inner product, so $R(t)$ lies between $0$ and $1$ and measures the fraction of $\lVert J\rVert_F^2$ carried by the non-conservative component. A conservative field gives $R=0$. The value $R=1$ requires a purely antisymmetric Jacobian and hence a solenoidal field, though a generic solenoidal field also carries a symmetric part and gives $R<1$. A matrix with independent zero-mean random entries gives $R\to\tfrac12$ in large dimension, which sets the scale of an untrained network.

We evaluate $R(t)$ on HMC configurations noised to diffusion time $t$ with the VE kernel $\phi_t=\phi_0+\Sigma(t)\eta$, so the score is probed on the noised marginal $p_t$ seen during training and not on reverse-generated samples. The Frobenius norms in \Eq{eq:nonconserved_fraction} are estimated with a Hutchinson estimator. For random probe vectors $v$ with unit covariance one has $\langle\lVert Jv\rVert^2\rangle_v=\lVert J\rVert_F^2$, and the same identity applied to $J-J^\top$ gives the numerator. Each probe requires one Jacobian-vector product $Jv$ and one vector-Jacobian product $J^\top v$ through the network, so the full Jacobian is never formed. The reported values average over configurations and probe vectors.

\Fig{fig:nonconserved_2d} shows $R(t)$ against training epoch at $L=128$ and $\kappa=0.28$ for the diffusion times $t=0.001$, $0.1$, and $1$. An untrained network sits at the random value $R\simeq\tfrac12$. Training lowers the fraction by orders of magnitude, reaching $R\sim10^{-5}$ at late epochs. The three curves cross during training, and the late-epoch residual is largest at the smallest diffusion time. This ordering follows from the training objective. The weighting $\omega(t)=\Sigma^2(t)$ in \Eq{eq:dsm_loss_ve} is smallest at small $t$, so the score near the data manifold receives the weakest training signal and remains the least constrained. As a check of the estimator, the exact drift $-\nabla_\phi S$ is a gradient by construction and returns $R\sim10^{-15}$, the floor set by floating-point and estimator noise.

The trained score therefore carries a small but nonzero non-conservative part, orders of magnitude below the random value and orders of magnitude above the conservative floor. A nonzero $R$ means the score is not the gradient of any scalar, so no log-density can be reconstructed from it by path-independent integration. The reverse samplers \Eq{eq:reverse_sde} and \Eq{eq:prob_flow_ode} are unaffected, since they use the score only as a drift field, and any residual bias in the sampled distribution is removed by the MALA correction of \Sec{sec:acceptance}. The fraction $R$ tracks non-conservativeness and not accuracy. A conservative but inaccurate score still gives $R=0$, so this diagnostic complements the score-quality tests of \Sec{sec:score_quality}.

\bibliography{ref-lib}

\end{document}